%% file: integrability.tex
\documentclass[12pt, utf8]{article}
\usepackage[body={17cm, 23cm, centered}]{geometry}
\usepackage[english]{babel}
\usepackage[utf8]{inputenc}

\usepackage{amsmath,amssymb,amsthm,amscd,cite,comment,amsfonts,indentfirst,color,setspace,bbold,arydshln}
\usepackage{mathtext,marvosym,textcomp}
\usepackage[makeroom]{cancel}
\usepackage{mathtools}
\usepackage{slashed}
\usepackage{bbm}

\usepackage[debug,pageanchor=false]{hyperref}
\hypersetup{colorlinks=true,linktocpage,breaklinks,
	urlcolor=blue,
	linkcolor=red,
	citecolor=blue
}

\usepackage[vcentermath]{youngtab}

\usepackage[mathmode,centertableaux]{ytableau}

\usepackage{tikz}
\usetikzlibrary{arrows}
\usetikzlibrary{shapes.misc}
\usepackage{tikz-cd}

\tikzstyle{format} = [on grid]

\tikzset{cross/.style={cross out, draw=black, minimum size=2*(#1-\pgflinewidth), inner sep=0pt, outer sep=0pt},
	cross/.default={5pt}}
\usepackage[ normalem]{ulem}
\usepackage{tocloft}

\numberwithin{equation}{section}

\selectfont

\input{Defs.tex}

\begin{document}
\renewcommand{\refname}{\begin{center}References\end{center}}
	
\begin{titlepage}
		
	\vfill
	\begin{flushright}

	\end{flushright}
		
	\vfill
	
	\begin{center}
		\baselineskip=16pt
		{\Large \bf 
		    Integrability structures in string theory
		}
		\vskip 1cm
			Kirill Gubarev$^{a,b}$\footnote{\texttt{kirill.gubarev@phystech.edu}}, Edvard T. Musaev$^{a}$\footnote{\texttt{musaev.et@phystech.edu}}
		\vskip .3cm
		\begin{small}
			{\it 
			    $^a$Moscow Institute of Physics and Technology, \\ Institutskii per. 9, Dolgoprudny, 141700, Russia\\
			    $^b$National Research Centre ``Kurchatov Institute'', 123182, Moscow, Russia
			}
		\end{small}
	\end{center}
		
	\vfill 
	\begin{center} 
		\textbf{Abstract}
	\end{center} 
	\begin{quote}
         This review is a collection of various methods and observations relevant to structures in three-dimensional systems similar to those responsible for integrability of two-dimensional systems. Particular focus is given to Nambu structures and loop variables naturally appearing in membrane dynamics. While reviewing each topic in more details we emphasize connections between them and speculate on possible relations to membrane integrability.
	\end{quote} 
	\vfill
	\setcounter{footnote}{0}
\end{titlepage}
	
\clearpage
\setcounter{page}{2}
	
\tableofcontents

\section{Introduction}

\label{sec:intro}

Most generally speaking a dynamical system is said to be integrable when the number of degrees of freedom required to describe its dynamics is twice smaller than the dimension of the phase space. Speaking differently the system has a set of conserved charges allowing to integrate dynamical equations along the corresponding directions. While this understanding of integrability applies most directly to mechanical systems with finite phase space, for field theories the above becomes too vague and more strict criteria have to be introduced. In particular one speaks about classical integrability in the Liouville sense when an infinite set of conserved charges can be generated by making use of the Lax pair approach. Definitely not the simplest however relevant to the present discussion example of a classically integrable field theory system is the Green-Schwarz superstring on the AdS${}_5\times \SS^5$ background, that is a two-dimensional supersymmetric sigma-model. The notion of integrability naturally extends to quantum systems, where it means existence of an infinite set of commuting operators, one of which can be set as the Hamiltonian. This set can be generated by quantum Lax operators constructed using algebraic Bethe ansatz and Yang-Baxter relations, thermodynamic Bethe ansatz or by constructing quantum spectral curve. There are plenty of introductory lectures and reviews explaining the notion of quantum integrability and these approaches ( see e.g. \cite{Slavnov:2018lec,Schafer-Nameki:2010qho,Staudacher:2010jz,Bajnok:2010ke,Levkovich-Maslyuk:2016kfv}), here we will be mainly focused at structures responsible for classical integrability.

The narrative of the present review develops from integrability of the Green-Schwarz superstring on AdS${}_5\times \SS^5$, its deformations preserving integrability and the algebraic structures that emerge in this procedure. First integrability of this system has been observed in \cite{Bena:2003wd} by explicit construction of a flat Lax connection, that generates an infinite set of conserved currents. In particular the interest to integrability of the string on specific backgrounds advertised in this work was an extension of the AdS/CFT duality beyond the correspondence between weakly coupled strings and strongly coupled gauge theories. Strong indication of integrability of the string is the conjectured integrability of its holographic partner the $\mc{N}=4$ $d=4$ super Yang-Mills theory. This in turn was initially based on the observation that the large N dilatation operator of the gauge theory, when restricted to the sector of operators built out of scalars, can be regarded as the Hamiltonian of an integrable quantum spin chain \cite{Minahan:2002ve}. The string on AdS${}_5\times \SS^5$ is known to belong to a family of integrable sigma models that can be obtained via its deformation. In particular one finds the so-called Lunin-Maldacena $\b$-deformed background \cite{Lunin:2005jy} dual to certain Leigh-Strassler deformation of $\mc{N}=4$ $d=4$ SYM \cite{Leigh:1995ep}. Integrability of the string on the LM background have been shown in \cite{Frolov:2005ty,Frolov:2005dj} (non-integrability for complex deformations has been shown in \cite{Giataganas:2013dha}). For a review of the interplay between AdS/CFT correspondence and integrability see \cite{Beisert:2010jr}. When speaking about the open Green-Schwarz superstring one is able to find boundary  conditions that preserve integrability \cite{Dekel:2011ja}, allowing to speak about integrable configurations of D-branes. In particular one finds a D3-D5-brane system dual to  a defect CFT known to be integrable \cite{Linardopoulos:2021rfq}, and the D2-D4-brane system dual to ABJM theory in the presence of a half-BPS domain wall also known to be integrable \cite{Linardopoulos:2022wol}. When uplifted to M-theory the latter gives an M2-M5-system strongly suggesting in favor of a possibility to define membrane integrability. Note however the works \cite{Giataganas:2017guj,Giataganas:2019xdj}, where non-integrability of string motion on certain D-brane backgrounds have been shown. These results could either restrict the possibility of defining integrable structures for D-branes, or indicate that a better choice of variables must be made.

A more systematic approach to generating integrable two-dimensional sigma-models is based on Yang-Baxter deformations of sigma-models developed in \cite{Klimcik:2002zj} for principal chiral models and in \cite{Delduc:2013fga} for sigma-models on symmetric spaces. In the work \cite{Delduc:2013qra} an integrable Yang-Baxter deformation of the AdS${}_5\times \SS^5$ superstring has been presented, that gave an important impulse towards development of the methods described in the present review. The deformation is parametrized by a matrix $r$ that is a solution to the classical Yang-Baxter equation (CYBE) \cite{Kawaguchi:2014qwa,vanTongeren:2015soa}
\begin{equation}
    r^{b_1[a_1}r^{a_2|b_2|}f_{b_1b_2}{}^{a_3]}=0,
\end{equation}
where $f_{ab}{}^c$ denote structure constants of the (super)algebra\footnote{Note that  the deformation of \cite{Delduc:2013qra} does not uniquely define the deformed theory and a freedom in defining $r$-matrix remains. In \cite{Hoare:2018ngg} a unimodular $\h$-deformation of AdS${}_5\times \SS^5$ satisfying inhomogeneous (modified) YB equation has been constructed and shown to generate a solution to supergravity equations.}. The natural question of whether the deformed sigma-model can be interpreted as a superstring on a supergravity background has found its answer in \cite{Arutyunov:2015qva} where the corresponding background has been presented and shown to violate equations of D=10 supergravity. The proper set of equations satisfied by the background, currently referred to as ABF, has been found in \cite{Arutyunov:2015mqj} and is now referred to as the generalized supergravity \cite{Wulff:2016tju}. We will not cover such generalizations of supergravity equations here, interested reader can find more details in the review \cite{Orlando:2019his} and references therein. Important for us here is the result of \cite{Araujo:2017jap,Araujo:2017jkb} where a rule to perform Yang-Baxter deformations for a general background beyond coset spaces has been formulated. The rule is based on the open-closed string map and more conveniently can be formulated as a local O(10,10) transformation generated by a bi-vector $\beta = r^{a_1a_2} k_{a_1} \wedge k_{a_2}$, where $k_a=k_a{}^\m \dt_\m$ is a set of Killing vectors of the initial background. The bi-vector $\b^{\m\n}$ plays the role of the non-commutative parameter of the corresponding open-closed string map. In \cite{Bakhmatov:2018apn,Bakhmatov:2018bvp} it has been shown that for the deformed background to satisfy equations of supergravity it is sufficient to impose classical Yang-Baxter equation on $r^{a_1a_2}$ and the so-called unimodularity condition $r^{a_1a_2}f_{a_1a_2}{}^{b}=0$ first discovered in \cite{Borsato:2016ose}. Breaking the latter gives solutions to equations of generalized supergravity. An important side comment here is that precisely the classical Yang-Baxter matrix $r$ can be used to generate Poisson brackets of an integrable system, given a pair of Lax operators (see Section \ref{sec:lax}).

Written in the form of a linear O(10,10) transformations Yang-Baxter deformations of 10D backgrounds (2D $\s$-model) can be naturally generalized to deformations of 11D backgrounds (3D $\s$-model), that has been done in \cite{Bakhmatov:2019dow,Bakhmatov:2020kul,Gubarev:2020ydf}. The corresponding generalization of the classical Yang-Baxter equation (gCYBE) has been presented in \cite{Malek:2019xrf,Sakatani:2019zrs} following an algebraic approach based on the so-called exceptional Drinfeld algebra, generalizing classical Drinfeld double\footnote{Earlier this condition has been found in \cite{Bakhmatov:2019dow} in the form of a condition enough for R-flux to vanish. However, since the condition derived was sufficient rather than necessary certain terms could have been added and the final form of gCYBE equation still had to be determined.}. The check that gCYBE (together with unimodularity constraint) is enough for a deformation to generate a solution has been performed in \cite{Gubarev:2020ydf} for general backgrounds. In the case of 11D backgrounds a deformation is parametrized by a tri-vector $\W=\r^{a_1a_2a_3}k_{a_1}\wedge k_{a_2} \wedge k_{a_3}$, which now requires an object $\r^{a_1a_2a_3}$ with three indices rather than a matrix $r^{a_1a_2}$. At the moment no interpretation of gCYBE as a classical limit of an equation similar to quantum YB equation is known, although certain attempts have been done in \cite{Malek:2020hpo} to construct the quantum equation by hands and in \cite{Bakhmatov:2020kul} to speculate around Zamolodchikov tetrahedron equation. Despite that, a side comment similar to the one above can be made: a Nambu bracket of a dynamical system can be generated using $\r^{a_1a_2a_3}$ and a triple of Lax operators. 

Naturally a set of questions arises here. To what extent this system is integrable? Can Liouville integrability be formulated for three-dimensional field-theoretical systems? Does tri-vector deformation preserve integrability of the 2D $\s$-model in the usual sense? This review aims at collecting and describing in a uniform language several attempts to make sense of three-dimensional integrability and of integrability of Nambu-Poisson systems, as these seems to have close relation to the algebraic structures arising in tri-vector deformations of M-theory backgrounds. We will try to emphasize these relations in each case and speculate on possible further developments. The text is structured as follows. In this section the standard approach to Liouville integrability using Lax pairs and Lax connection is briefly reviewed mainly to introduce notations and to make the text self-contained. In Section \ref{sec:nambuall} Nambu systems and approaches to their integrability are discussed. As a particular example relevant for 3D integrability we focus at the KP hierarchy.  In Section \ref{sec:10d} we briefly review integrability of the superstring mainly focusing at integrable deformations and their interpretation as a Poisson-Lie T-duality. Section \ref{sec:11d} describes tri-vector deformations and related algebraic structures and contains a description of several approaches to integrability of the 3D membrane. We describe approaches to membrane dynamics based on loop algebra variables, which seem natural and emphasize that the deformation tensor $\W^{a_1a_2a_3}$  can be interpreted as loop non-commutativity parameter for membranes, while the deformation map has the same form as the open-closed membrane map. At the end of Section \ref{sec:11d} we discuss possible relations to tetrahedron equation of Zamolodchikov and a particular way to define Wilson surface in terms of loops. Finally, Section \ref{sec:disc} presents the results and observations discussed in the main text in the form of lists to present the picture in a more clear way, if even possible.

\subsection{Liouville integrability}
\label{sec:liouville}

Let us start with a brief reminder of the standard Lax pair approach to classical Liouville integrability of mechanical systems. The presentation below mainly follows \cite{deleeuw2013itp}, however the same can be found in any review or textbook on integrable systems. Start with a dynamical system defined by a set of equations of motion $\dot{x}^i = f^i(x)$. Given the Hamiltonian $H$ of the system the equations of motion can be written as
\begin{equation}
    \begin{aligned}
         \dot{q}^i = \fr{\dt H}{\dt p_i}, && \dot{p}_i = -\fr{\dt H }{\dt q^i}.
    \end{aligned}
\end{equation}
Equivalently $\dot{p}_i=\{H,p_i\}$, $\dot{q}^i=\{H,q^i\}$ where  the Poisson bracket for a pair of arbitrary functions $f(p,q)$ and $g(p,q)$ of dynamical variables is defined as usual as
\begin{equation}
    \{f,g\}=\fr{\dt f}{\dt q}\fr{\dt g}{\dt p}- \fr{\dt g}{\dt q}\fr{\dt f}{\dt p}.
\end{equation}
Suppose the system has non-trivial integrals of motion defined as $\dot{I}_i=0$, or equivalently
\begin{equation}
    \{ H,I_i \}=0.
\end{equation}
Each integral of motion $I$ allows to turn to the so-called action-angle variables and completely integrate dynamical equations for a pair of coordinates $(p,q)$. Hence, if the number of integrals of motion is equal to the total number of degrees of freedom, the system is completely integrable. To summarize, a system with $2n$-dimensional phase space is said to be Liouville integrable if it possesses $n$ integrals of motion $I_i$ all of which are in involution, i.e. $\{I_i,I_j\}=0$. 

Let us consider action-angle variables in more details, for which take a single integral of motion given by a function $F(p,q)=f=$ const. Solve this equation for $p$ and define a 1-form
\begin{equation}
    \a = p dq,
\end{equation}
where $p$ is understood as a function $p=p(q,f)$ of the coordinate $q$ and the integral of motion $f$. Then the 2-form $\w=d\a=dp\wedge dq$ defines a symplectic form on the phase space. Define now the action $S$ as
\begin{equation}
    S[q,f]=\int_{q_0}^q \a= \int_{q_0}^q p(q,f) dq.
\end{equation}
By construction the action is a function of two variables $q$ and $f$ and one may define new dynamical variables by writing
\begin{equation}
    \begin{aligned}
        p=\fr{\dt S}{\dt q}, && \y=\fr{\dt S}{\dt f},
    \end{aligned}
\end{equation}
where the first equality is straightforward and the second simply defines the angle variable $\y$. Such defined transformation from the variables  $(p,q)$ to the action-angle variables $(f,\y)$ is canonical, i.e. preserves the symplectic form. Indeed, considering
\begin{equation}
    0\equiv d^2 S = dp\wedge dq - d f \wedge d \y,
\end{equation}
one finds that $\w$ does not change.

Dynamics of the system becomes particularly simple in terms of such defined action-angle variables allowing to explicitly solve equations of motion for the integral $I=F(p,q)$. Indeed, we write
\begin{equation}
    \begin{aligned}
    \dot{f}&=\{H,f\}=0, \\ \dot{\y} &= \{H,\y\}=\fr{\dt H}{\dt f}=\w(f)=\mathrm{const}.
    \end{aligned}
\end{equation}
These can be easily solved to give
\begin{equation}
    \begin{aligned}
        f&=\mathrm{const}, \\
        \y&=\w(f)\,t+\y_0,
    \end{aligned}
\end{equation}
i.e. the angle variable $\psi$ corresponding to the action given by the integral of motion $f$ evolves linearly with time. Given $n$ integrals of motion the above procedure can be repeated for all $n$ pairs of variables $(p_i,q^i)$ allowing to solve equations of motion completely in terms of the corresponding action-angle variables $(f_i, \y^i)$. This is basically the explicit manifestation of integrability of the system: linear evolution and no chaos. 
 
As the simplest example illustrating the above general principle consider one-dimensional harmonic oscillator $H=\fr12(p^2 + \w^2 q^2)$. Hamiltonian equations of motion read  
\begin{equation}
    \begin{aligned}
        \dot{p}=\w^2 q, && \dot{q} = - p.
    \end{aligned}
\end{equation}
This system has only one integral of motion that is the energy  $H(p,q)=E$. Solving this equation we get for $p$ the following
\begin{equation}
    p = \sqrt{E- \w^2 q^2}.
\end{equation}
The action is then $S= \int_0^q \sqrt{E- \w^2 z^2}dz$ and the angle variable then reads
\begin{equation}
    \begin{aligned}
        \y = \fr{\dt S}{\dt E}= \int_0^q \fr{dz}{2\sqrt{E-\w^2 z^2}} = \fr{1}{2\w}\arctan \left[\fr{\w q}{p}\right].
    \end{aligned}
\end{equation}
Hence, we have
\begin{equation}
    \begin{aligned}
        \fr{1}{2\w}\arctan \left[\fr{\w q}{p}\right]&= \fr{1}{2}t,\\
        p^2 + \w^2 q^2&=E,
    \end{aligned}
\end{equation}
 which gives the standard solution $q=\sin \w t$, $p=-\w \cos \w t$.

\subsection{Lax pair}
\label{sec:lax}

Evidently the above procedure is not algorithmic as deals with solving differential equations at various steps, for which reason one would like to formulate an approach that allows to generate integrals of motion from a single expression and thus guarantee integrability in the considered sense. Such an approach is known as the Lax-Zakharov-Shabat formalism and starts with an assumption that we have managed to write equations of motion for a dynamical system in the form
\begin{equation}
    \label{eq:lax0}
    \dot{L} = [L,M],
\end{equation}
where $L,M$ are some matrices, which might additionally depend on some (spectral) parameter(s) $u$. Then, integrals of motion can be generated by simply taking trace of various matrix powers of $L$, i.e.
\begin{equation}
    F_k:= \Tr L^k \quad \Longrightarrow \quad \dot{F}_k=0. 
\end{equation}
The pair of matrices $L,M$ is referred to as the Lax pair. Altogether, the above equations imply that dependence on time for these matrices is given by
\begin{equation}
    \begin{aligned}
        L(t)&=g(t)L(0)g(t)^{-1}, \\
        M(t)&=\dot{g}(t)g(t).
    \end{aligned}
\end{equation}
Roughly speaking if we have managed to find a Lax pair for a dynamical system, i.e. to rewrite its equations of motion as above, the system is integrable. Certainly, this step is no more algorithmic as the standard action-variables method, however once the Lax pair is found, integrals of motion are generated automatically. For example for one-dimensional harmonic oscillator the Lax pair can be chosen as follows
\begin{equation}
    \begin{aligned}
        L= p\s_3 + \w q \s_1 =
        \begin{bmatrix}
            p & \w q \\
            \w q & -p
        \end{bmatrix}, &&
        M=-\fr{i}{2}\w \s_2 = \begin{bmatrix}
                0 & -\fr12 \w \\
                \fr12 \w & 0
            \end{bmatrix},
    \end{aligned}
\end{equation}
where $\s_1,\s_2,\s_3$ are the standard Pauli matrices.
The only integral of motion is then $H=\fr14 \Tr L^2$.

To check whether the integrals $\{F_i\}$ are in involution, one has to ensure $\{F_i,F_j\}=0$. For further discussion it is convenient to consider the Lax pair as the starting point and to generate a Poisson bracket for the system using classical Yang-Baxter $r$-matrix. For that start with a matrix $L\in \gl(d)$ and define the Poisson bracket as
\begin{equation}
    \label{eq:poisson}
    \{L_1,L_2\} = [r_{12},L_1] -[r_{21},L_2],
\end{equation}
where $[\,,]$ is the usual commutator in the algebra $\gl(d)$ and the matrices $L_{1,2}$ are defined as
\begin{equation}
    \begin{aligned}
        L_1 & = L\otimes \mathbbm{1}, \\
        L_2 & = \mathbbm{1}\otimes L. \\
    \end{aligned}
\end{equation}
In other words, both $L_1$ and $L_2$ belong to $\gl(d)\otimes \gl(d)$. Such defined bracket is antisymmetric by construction and it must additionally  satisfy the Jacobi equation, that is equivalent to the (modified) classical Yang-Baxter equation for the matrix $r$. Given a pair of vectors $X,Y$ with a defined action of $\gl(d)$ the mCYBE can be conveniently written as
\begin{equation}
    [r(X),r(Y)]-r([r(X),Y]+ [X,r(Y)]) = -c^2[X,Y],
\end{equation}
where $c$ is an arbitrary number with $c=0$ corresponding to the usual (homogeneous) CYBE. It is straightforward to check that the integrals $F_i$ are indeed in involution with respect to such defined Poisson bracket. Certainly this bracket defines the same evolution as \eqref{eq:lax0}:
\begin{equation}
    \begin{aligned}
        &\fr{dL}{dt}=\{F_k,L\} = [M_k,L],&& M_k = -k \Tr_1[L_1{}^{k-1}r_{21}],
    \end{aligned}
\end{equation}
where the subscript indicates that the trace is take w.r.t the first factor. A theorem states that eigenvalues of the Lax matrix $L$ (the conserved quantities $F_k$) are in involution if and only if there exists a function $r_{21}$ that satisfies CYBE and defines the Poisson bracket as above. For details of the theorem see the lectures \cite{deleeuw2013itp}, here we only mention that the $r$-matrix is a natural attribute of a classical integrable system.

As a final note in this subsection let us provide some expressions in an explicitly chose basis  $\{T^i{}_j\}=\mbox{bas}\,\gl(d)$. For the $r$-matrix $r \in  \gl(d)\wedge \gl(d)$ we then have the following component form
\begin{equation}
    \begin{aligned}
        r & =r^{i_1}{}_{j_1}{}^{i_2}{}_{j_2}T^{j_1}{}_{i_1}\wedge T^{j_2}{}_{i_2}.
    \end{aligned}
\end{equation}
The equation \eqref{eq:poisson} then becomes
\begin{equation}
    \{L^{i_1}{}_{j_1},L^{i_2}{}_{j_2}\} = r^{i_1}{}_{k_1}{}^{i_2}{}_{j_2}L^{k_1}{}_{j_1}-L^{i_1}{}_{k_1}r^{k_1}{}_{j_1}{}^{i_2}{}_{j_2}-r^{i_1}{}_{j_1}{}^{i_2}{}_{k_2}L^{k_2}{}_{j_2}+L^{i_2}{}_{k_2}r^{i_1}{}_{j_1}{}^{k_2}{}_{j_2}.
\end{equation}
We observe that classical $r$-matrix has naturally two pairs of indices each acting on a linear space.  Let us illustrate the above by the usual example of harmonic oscillator whose matrices $L$ and $M$ have been presented previously. Classical $r$ matrix can be written in the following form
\begin{equation}
    \begin{aligned}
        r&= \fr1q \Big(F \otimes E - E\otimes F\Big)\\
        & = \fr1q \begin{bmatrix}
            0 & 0 \\
            1 & 0
        \end{bmatrix}\otimes 
        \begin{bmatrix}
            0 & 1 \\
            0 & 0
        \end{bmatrix}-
        \fr1q \begin{bmatrix}
            0 & 1 \\
            0 & 0
        \end{bmatrix}\otimes 
        \begin{bmatrix}
            0 & 0 \\
            1 & 0
        \end{bmatrix},
    \end{aligned}
\end{equation}
where $F$ and $E$ are generators of  $\sl(2)$. To derive the corresponding Poisson bracket we calculate on the one hand
\begin{equation}
    [r, L\otimes {\bf 1} + {\bf 1} \otimes L] = -\w (\s_3 \otimes \s_1 - \s_1 \otimes \s_3),
\end{equation}
and on the other hand
\begin{equation}
    \{L\otimes {\bf 1}, {\bf 1}\otimes L\} = \w \{p,q\} \s_3 \otimes \s_1 + \w \{q,p\} \s_1 \otimes \s_3.
\end{equation}
Comparing the two we have $\{q,p\}=1$.

\subsection{Quantum Yang-Baxter equation}
\label{sec:qcybe}

For a quantum system integrability basically means the same as above: existence of an infinite set of conserved charges $Q_s$, commuting with each other. When speaking about quantum systems one is mainly interested in deriving its full spectrum, which is usually simple for free theories and becomes an incredibly complicated problem for interacting systems. For integrable quantum models powerful methods have been developed to compute spectrum: algebraic and coordinate Bethe ansatz, thermodynamic Bethe ansatz, spectral curve. Of particular relevance to the present discussion is the approach of thermodynamic Bethe ansatz that allows to compute spectrum of an integrable quantum system using scattering data and Yang-Baxter equation. Let us focus on the latter referring reader to the reviews \cite{Slavnov:2018lec,Schafer-Nameki:2010qho,Staudacher:2010jz,Bajnok:2010ke,Levkovich-Maslyuk:2016kfv} for more detailed description of other methods in application to AdS/CFT integrability and spin-chain models.

\begin{figure}[ht]
\centering
\begin{tikzpicture}

\draw[thick, red] (0.5,-3.5) -- (3,-1);
\draw[thick, blue] (0,-2.85) -- (4,-3.5);
\draw[thick] (1.5,-1) -- (3.5,-4);

\draw (5,-2.5) node (eq) {$=$};

\draw[thick, red] (7,-3.5) -- (9.5,-1);
\draw[thick, blue] (6,-1.35) -- (10,-2);
\draw[thick] (6.5,-1) -- (8.5,-4);

\draw (0.2, -3.7) node {1};
\draw (0, -2.5) node {2};
\draw (1.2, -0.8) node {3};

\draw (6.7, -3.7) node {1};
\draw (6, -1.7) node {2};
\draw (6.7, -0.8) node {3};

\draw[fill=black] (1,-3) circle (0.07);
\draw[fill=black] (2.1,-1.9) circle (0.07);
\draw[fill=black] (3.07,-3.35) circle (0.07);

\draw[fill=black] (6.8,-1.47) circle (0.07);
\draw[fill=black] (8.7,-1.8) circle (0.07);
\draw[fill=black] (7.7,-2.8) circle (0.07);
\end{tikzpicture}
\caption{Graphical representation of the quantum Yang-Baxter equation governing scattering of three particles. The equation states, that S-matrix does not depend on the mutual position of the lines, in particular on the position of the black line ($\#3$) at the picture.}
\label{fig:qYBE}
\end{figure}
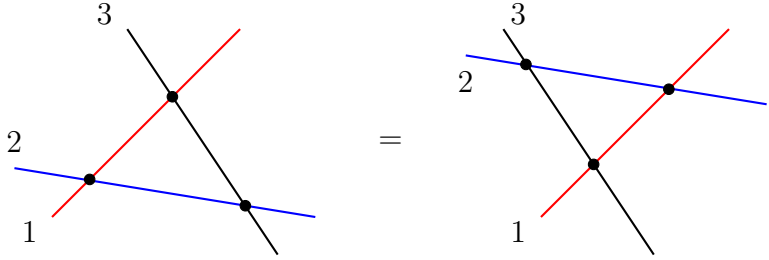

For scattering processes integrability of a quantum system means that there is no particle production. For theories in dimension $d>2$ Coleman-Mandula theorem states that S-matrix is trivial $S=1$ if there is even a single charge that is a second or higher order tensor. In contrast in dimension $d=1+1$ S-matrix remains non-trivial although pretty much restricted:
\begin{itemize}
    \item no particle production;
    \item the initial set of momenta $\{p_i\}_{in} = \{p_i\}_{out}$ is the same as the final set;
    \item scattering factorizes.
\end{itemize}
Factorized S-matrices as exact solutions of $1+1$-dimensional quantum field theories has been first considered in \cite{Zamolodchikov:1978xm} and then used to develop the method of thermodynamic Bethe ansatz in \cite{Zamolodchikov:1989cf}. The factorization property of S-matrix means that S-matrix of $n$ particles decomposes into a product of S-matrices for all pairs of particles. In general such decomposition can be performed in multiple ways, all of which must be equivalent, that leads to consistency constraints. For 3-to-3 particle scatter this can be illustrated by picture on Fig.\ref{fig:qYBE},
where scattering of red, blue and black particles labeled 1, 2 and 3 respectively can be factorized into pair interactions in two ways. Setting the time direction to run from the left to the right at  Fig.\ref{fig:qYBE} and denoting S-matrix $R_{ij}(u)$ for particles with labels $i$ and $j$ and mutual rapidity $u$ we obtain the following equation
\begin{equation}
\label{qYBE}
R_{12}(u-v)R_{13}(u)R_{23}(v)=R_{23}(v)R_{13}(u)R_{12}(u-v).
\end{equation}
This is quantum Yang-Baxter equation first introduced in \cite{Yang:1967bm} to solve eigenvalue problem for an N-particle system using algebraic Bethe ansatz and independently in \cite{Baxter:1972hz} to compute partition function for a certain lattice matrix model.  For more details see e.g. \cite{Bombardelli:2016scq,Ryan:2022ybk}.

Suppose, each particle in addition to rapidities is described by a linear space $V$ of its states, then   R-matrices  $R_{ij}\in { \rm End}(V_i\otimes V_j)$ act on the vector space of states of two particles encoding their interaction. The most known example would be to set $V$ to be a two-dimensional set of states of a particle with spin $\hbar/2$. Then $r$-matrix would take values in the product of groups $\rmSL(2,\mathbb{C})$. Such R-matrices and their generalization to $\rmSL(n,\mathbb{C})$ and to quantum groups have found a wide variety of applications in knot theory and braid groups \cite{Wu:1993ab,Turaev:1988tju,Kaufmann:1989bk}. For us quantum Yang-Baxter equation will be relevant in two aspects: 
\begin{itemize}
    \item it describes scattering of the superstring states on AdS$_5\times \SS^5$,
    \item its quasiclassical limit gives classical Yang-Baxter equation discussed above and relevant to integrable deformations.
\end{itemize}
Keeping the former outside of the scope of the review let us consider the latter in more details.

First, one should be careful with interpretation of qYBE in terms of matrices, since each $r$-matrix acts only at the product of two vector spaces and it is convenient do define an operator $\hat{R}\in {\rm End}(V\otimes V)$. Hence $\hat{R}_{12}(u)\in {\rm End}(V^{(1)}\otimes V^{(2)})$ provides interaction between particles 1 and 2 with quantum states encoded by the spaces $V^{(1)}$ and $V^{(2)}$ scattered at mutual rapidity $u$. Note, that as linear spaces $V^{(1)}=V^{(2)}$ and the numbers are there just to explicitly distinguish the particles. The same can be done just by keeping track of the place at which the operator stands. In what follows we adopt the latter notation. Hence, one writes
\begin{equation}
R_{12}(u)=\hat{R}_{12}(u)\otimes \mathbbm{1},
\end{equation}
where $\mathbbm{1}$ is the identity operator on $V$. For matrix notations one chooses a basis in $V$ as $\{e_{\a}\}={\rm bas}\, V$ and writes
\begin{equation}
\begin{aligned}
R_{12}(u)(e_\a\otimes e_\b \otimes e_\g)&= R_{12}(u)_\a^\d{}_{\b}^{\e}{}_\g^\f e_\d\otimes e_\e\otimes e_\f\\
R_{12}(u)_\a^\d{}_{\b}^{\e}{}_\g^\f&=R(u)_\a^\d{}_{\b}^{\e} \delta_\g^\f.
\end{aligned}\end{equation}
In matrix notations the equation \eqref{qYBE} is an equation for the matrix with 4 indices $R(u)_\a^\d{}_{\b}^{\e}$ depending on the spectral parameter $u$.

Usually, the full quantum Yang-Baxter equation is very complicated to analyze and to solve, and one proceeds with taking a quasi-classical limit. For this one first notices that $R(u)=\id\otimes \id \otimes \id$ trivially solves qYBE and hence it is natural to expand around this point in the space of solutions
\begin{equation}
R_{12}(u)=\id+\hbar \, r_{12}(u).
\end{equation}
Here $\hbar$ is the expansion parameter and $r_{12}(u)$ is constructed from the algebra $\mathfrak{g}$ of the group $G={\rm End}(V)$. Substituting this expansion into the initial equation one finds that at the orders $\hbar^0$ and $\hbar^1$ all terms cancel and the equation is satisfied trivially. Hence, the first non-trivial equation one encounters at level $\hbar^2$ which reads
\begin{equation}
\label{CYBEs}
[r_{12}(u-v),r_{13}(u)]+[r_{13}(u),r_{23}(v)]+[r_{12}(u-v),r_{23}(v)]=0.
\end{equation}
Here, $[x,y]$ is understood as $[x,y]=x\cdot y-y\cdot x$, hence one should be careful with definition of $r_{12}(u-v)$. Indeed, since $R_{12}(u)\in {\rm End}(V\otimes V\otimes V) \equiv G\times G\times \id$, it is natural to define the matrix $r_{12}(u)$ as $ r_{12}(u) \in \f(\mathfrak{g})\otimes \f(\mathfrak{g})\otimes \mathbf{1}$. Here 
\begin{equation}
\f: \mathfrak{g} \to A
\end{equation}
is a map to an associative algebra  $A$ with unit $\mathbf{1}$ defined such that
\begin{equation}
\f(a)\f(b) - \f(b) \f(a) = \f([a,b]), 
\end{equation}
where $[a,b]$ is the Lie bracket in the algebra $\mathfrak{g}$. Hence, the equation \eqref{CYBEs} is understood as an equation on $A\otimes A\otimes A$, which however can be consistently restricted to $\mathfrak{g}\otimes \mathfrak{g} \otimes \mathfrak{g}$ and hence does not depend on the choice of $A$ \cite{BelDri82}. In what follows for definiteness we will choose the algebra $A$ to be the universal enveloping algebra $A=U(\mathfrak{g})$ and drop $\f$ for clarity of notations.

Solutions of the classical Yang-Baxter equation for $r$-matrix with non-trivial spectral parameter can be propagated by a simple shift of a given solution $r_{12}(u)$ as $r'_{12}(u)=r_{12}(u)+r_{12}$ (and similarly for other spaces), where $r_{12}$ must satisfy the constant classical Yang-Baxter equation \cite{BelDri82}
\begin{equation}
\label{CYBE}
[r_{12},r_{13}]+[r_{13},r_{23}]+[r_{12},r_{23}]=0.
\end{equation}
In this work we always refer to this equation when mention CYBE. Each element $r_{12},r_{23}, r_{13}$ above can be decomposed w.r.t. the basis of the algebra $\{t_a\}={\rm bas}\,\mathfrak{g}$
\begin{equation}
\begin{aligned}
r_{12}&=r^{ab}t_a\otimes t_b\otimes \mathbf{1},\\
r_{13}&=r^{ab}t_a\otimes \mathbf{1}\otimes t_b,\\
r_{23}&=r^{ab} \mathbf{1} \otimes t_a\otimes t_b.
\end{aligned}
\end{equation}
In what follows we assume $r^{ab}=-r^{ba}$. Substituting these decompositions back into CYBE \eqref{CYBE} one obtains
\begin{equation}
r^{ab}r^{cd}\Big([t_a,t_c]\otimes t_b\otimes t_d+t_a \otimes t_c \otimes [t_b,t_d]+t_a\otimes [t_b,t_c]\otimes t_d\Big)=0.
\end{equation}
Replacing $[t_a,t_b]=f_{ab}{}^ct_c$ and properly relabelling indices one obtains
\begin{equation}
e_{[a}\otimes e_b\otimes e_{c]} r^{ae}r^{bf}f_{ef}{}^c=0,
\end{equation}
which boils down to the CYBE recovered in deformations of the Type IIA/B supergravity
\begin{equation}
r^{e[a}r^{b|f|}f_{ef}{}^{c]}=0.
\end{equation}
As mentioned above the equation is indeed only on the $\mathfrak{g}\otimes \mathfrak{g} \otimes \mathfrak{g}$ inside $A\otimes A\otimes A$ and does not depend on the chosen algebra $A$ and the precise form of the map. To rouse the curiosity the same is not true for tetrahedron equation, that is a direct analogue of qYBE for scattering of straight strings. Moreover a well defined quasi-classical limit does not exist in this case. We will discuss this more in Section \ref{sec:tetra}.

\subsection{Volume preserving flows and the action principle}
\label{sec:volume2}

Integrability of a Hamiltonian system is tightly related to preservation of its phase volume under evolution. For a given function $G$ a Hamiltonian $H$ defines a flow, that in the infinitesimal form can be written as
\begin{equation}
    G \to G+ \{H,G\}.
\end{equation}
Consider phase space distribution of the system $dw = \r(p,q)d^n p d^n q$, where the distribution function $\r(p,q)$ is the probability to have the system in the phase volume $d^n p d^n q$ at the point $(p,q)$. Liouville equation states that the distribution function is constant
\begin{equation}
    \fr{d \r}{d t} = \fr{\dt \r }{\dt t} + \fr{\dt \r}{\dt q^i}\dot q^i + \fr{\dt f}{\dt p_i}\dot p_i=0.
\end{equation}
Equivalently we can write this as the evolution equation for the density function: $\dot \r = \{H,\r\}$. Let us now show that for Hamiltonian systems phase volume is preserved. For that define a phase vector at arbitrary  time $t$:
\begin{equation}
    x_t = \big(p_1(t),\dots,p_n(t), q^1(t),\dots , q^n(t)\big).
\end{equation}
This vector is related to the vector $x_0$ at $t=0$ by a coordinate transformation in the phase space, whose Jacobian is given by
\begin{equation}
    J = \det \fr{\dt x^I_t}{\dt x^J_0}=\det M^I{}_J,
\end{equation}
where $I=1,\dots,2n$. Hence, preservation of phase volume under evolution is equivalent to such defined Jacobian being independent of time. For that we calculate:
\begin{equation}
    \begin{aligned}
        \fr{dJ}{dt} = \Tr\Big(M^{-1}\fr{dM}{dt}\Big)J = J \fr{\dt x^I_0}{\dt x^J_t}\fr{\dt\dot{x}^J_t}{\dt x^I_0} = J \fr{\dt \dot x^I_t}{\dt x^I_t}.
    \end{aligned}
\end{equation}
For Hamiltonian systems
\begin{equation}
    \fr{\dt \dot x^I_t}{\dt x^I_t} =  -\fr{\dt}{\dt p_i}\fr{\dt H}{\dt q^i} + \fr{\dt}{\dt q^i}\fr{\dt H}{\dt p_i} =0,
\end{equation}
hence the flow preserves phase space volume. In what follows we show that the same is true for Nambu mechanical systems, i.e. defined in terms of tri-brackets.

To define action of the system we start with a vector field $\tilde{L}$ that corresponds to the Hamiltonian evolution flow:
\begin{equation}
    \fr{d f}{dt} =\fr{\dt f}{\dt t}+ \{H,f\} = \tilde{L}(f).
\end{equation}
The vector field can be represented as $\tilde{L} = \dt_t + L$ and components of $L$ read
\begin{equation}
    L^p = -\fr{\dt H}{\dt q}, \quad L^q = \fr{\dt H}{\dt p}.
\end{equation}
This vector field is the line field  $d \w$  that is a derivative of the so-called Poincare-Cartan  1-form $\w$ on the phase space that defines the action.  For a 1+1-dimensional system the 1-form $\w$ can be  written as 
\begin{equation}
    \w = p dq  - H dt.
\end{equation}
To show that $\tilde{L}$ is indeed the line field, i.e.  $\iota_{\tilde{L}}(d\w)=0$, we simply write 
\begin{equation}
    \begin{aligned}
        d \w& = dp \wedge dq - \fr{\dt H}{d p}dp \wedge dt - \fr{\dt H}{\dt q} dq \wedge dt \\
        & = dp \wedge dq - L^q dp \wedge dt + L^p dq \wedge dt.
    \end{aligned}
\end{equation}
The 1-form $\w$ defines an integral invariant $\int_\g \w$ that is usually referred to as the action of the system. This expression is invariant under different choices of 1-chains along the evolution flow. Consider a 1-chain $c$ in the phase space, its image at $t$ under Hamiltonian evolution is given by $g^t(c)$. A tube of phase trajectories is given by a 2-chain
\begin{equation}
    J^tc = \{g^\t(c), 0 \leq \t \leq t\}.
\end{equation}
Simply speaking one takes a closed curve $c$ in the phase space and drags it along the Hamiltonian flow from $0$ to $t$. Given the Stokes theorem and the fact that $\iota_{\tilde{L}}d\w=0$ we have
\begin{equation}
    \int_c \w - \int_{g^t(c)}\w = \int_{J^t(c)}d\w =0,
\end{equation}
where we used that $\dt\big(J^t(c)\big) = c - g^t(c)$. Then extrema of the integral
\begin{equation}
    A(\g) = \int_\g \w = \int_\g p\wedge dq - H dt
\end{equation}
give trajectories of the system (for more details see \cite{Arnold:1989ma}). Interestingly enough, there exists a generalization of the above to Nambu mechanics due to Takhtajan \cite{Takhtajan:1993vr}, that can be uplifted to an action of a membrane. We will return to this later.

\subsection{Integrability in field theory}
\label{sec:ft_int}

For field theories, i.e. when canonical variables depend on a continuous variable, the concept of integrability is more tricky. The phase space of such models is infinitely dimensional and one would have to require a continuous set of integrals of motion to be able to speak about Liouville integrability. For a mechanical system integrability means the possibility to turn to action-angle variables, that allows to explicitly integrate equations of motion. For field theory similarly we will be talking about exact solvability of equations and methods, allowing to construct such solutions. For quantum systems one usually speaks about exact spectrum of operators in the theory, exact energy spectrum of the system, exact S-matrix, which for integrable systems is usually trivial. Let us list few well-known integrable field theories:
\begin{itemize}
    \item Korteweg-de Vries (KdV) equation, that is a mathematical model of waves on shallow water\cite{Helal:1996ab}
        \begin{equation}
            \dot h = 6 h h' - h'''.
        \end{equation}
    \item Non-linear Schr\"odinger equation, which is used to describe propagation of light in a non-linear medium \cite{agrawal2012nonlinear}
        \begin{equation}
            i \dot \y = - \y'' + 2 \k |\y|^2 \y.
        \end{equation}
    \item Sine-Gordon equation, which is used in the theory of crystal dislocation, Bloch-wall motion, magnetic flux in the Josephson effect \cite{Caudrey:1975nc} 
        \begin{equation}
            \ddot \f - \f'' + m^2 \sin \f=0.
        \end{equation}
    \item Principal chiral model on a compact group manifold, that for us is the simplest model of a string. The action is given by
        \begin{equation}
            S = \int \Tr \Big[ (g^{-1}dg) \wedge * (g^{-1}dg)\Big].
        \end{equation}
\end{itemize}
Integrable field equations share in common the property that scattering of their solitonic solutions is factorizable, i.e. after scattering of two or more solitons their shape is restored. This property is described by Yang-Baxter equation, for which reason Lax-Zaharov-Shabat formalism can be repeated for field theories with appropriate changes. 

A 2-dimensional field theory is called integrable if its equations can be written in the form of the flatness condition on the so-called Lax connection $A= A_\a d\s^\a$:
\begin{equation}
    dA + A\wedge A=0.
\end{equation}
In general, the connection may depend on an additional (spectral) parameter $u \in \mathbb{C}$. For example, components of the Lax connection for KdV equation take the form
\begin{equation}
    \begin{aligned}
        A_t = \begin{bmatrix}
            - h' & -4 u^{-1} - 2 h \\
            4 u^{-2} - 2 u^{-1} h + h'' - 2h^2 & h'
        \end{bmatrix}, &&
        A_x =
            \begin{bmatrix}
                0 & -1 \\
                u^{-1} - h & 0
            \end{bmatrix}.
    \end{aligned}
\end{equation}
To construct the Lax pair and to derive an infinite set of conserved currents one has to construct variables, that depend only on time $t$, which is possible due to the flatness condition. Indeed, it guarantees that a parallel transport operator 
\begin{equation}
    U(u;\s_1;\s_0) = \mathrm{Pexp}\bigg[\int_{t_0,x_0}^{t_1,x_1}A(u)\bigg]
\end{equation}
defined as a Wilson line does not depend on continuous variations of the integration path. Endpoints in the integral may be identified with boundaries of the system, say endpoints of a spin chain. Lax pair constructed of such defined parallel transport operator in general depends on boundary conditions. The simplest case is the periodic boundary conditions $\s^1 \sim \s^1+L$, when the Lax pair is defined as
\begin{equation}
    \begin{aligned}
        T(u) &= \mathrm{Pexp}\bigg[\oint A(u)\bigg], \\
        M(u) & = A_t(u)\Big|_{\s^1=0}.
    \end{aligned}
\end{equation}
It is easy to see that these satisfy precisely the desired equations
\begin{equation}
    \dot T(u) = [T(u), M(u)].
\end{equation}
Hence, the continuous family of conserved charges is defined as before as 
\begin{equation}
    F_k(u) = \Tr T(u)^k.
\end{equation}
Note that as before classical $r$-matrix can be used to define canonical brackets. 

In what follows we will be interested in generalizing these integrability structures to 3-dimensional theories, motivated by certain similarities between classical Yang-Baxter equation and what we call generalized Yang-Baxter equation. It is straightforward to assume that in this case the 1-form Lax connection $A$ must be replaced by either a 2-form (gerbe) connection or by connection on a loop space, which is actually equivalent given the transgression map. Both these possibilities can be justified to some extent from string/M-theory point of view, however, in any case it is pretty clear that the mechanics behind these structures must be defined by Nambu brackets. Before proceeding with discussion of Nambu mechanics and the extent to which integrable structures can be generalized, let us discuss another representation of integrable field theories, that is integrable Lax hierarchies.

Note, however, that to have higher dimensional integrable system does not necessarily require turning to higher Nambu mechanics. Example of a 1+2-dimensional integrable system is the so-called Kadomtsev-Petviashvilli (KP) equation
\begin{equation}
    \big(-4 \dot u + u_{xxx}+ 3 u u_x\big)_x+3 u_{yy}=0,
\end{equation}
where subscript denotes derivative. This is a two dimensional generalization of the KdV model of waves on shallow water. It appears, that this theory belongs to a larger (actually, infinite) family of integrable equations, each defined by its own Hamiltonian flow. Such system of commuting integrable Hamiltonian flows is referred to as integrable hierarchy (see \cite{Aratyn:1995qf,Vladimirov:2004cc} for more detailed review). In practice, integrable hierarchies are highly symmetric, infinite sets of nonlinear evolution equations of the Lax type for infinitely many functions $u_i$ of infinitely many variables $t_n$, $n = 1, 2, \dots$. The Lax  equations have the following form
\begin{equation}
    \label{eq:lax_hierarchy}
    \fr{\dt L}{\dt t_m}=[B_m,L], \quad m=1,2,\dots,
\end{equation}
where $L$ and $B_{m}$ are some pseudo-differential operators depending on the variables $u_i$. The Lax equations \eqref{eq:lax_hierarchy} can be written  in the form of zero-curvature condition
\begin{equation}
    \fr{\dt B_m}{\dt t_n} - \fr{\dt B_n}{\dt t_m} + [B_m,B_n] = 0,
\end{equation}
that is usually referred to as Zakaharov-Shabat equation.

Let us illustrate the formalism on the example of KP hierarchy. In this case
\begin{equation}
    \begin{aligned}
        L & = \dt + \sum_{i=1}^{\infty}u_i \dt^{-i} = \dt + u_1 \dt^{-1} + u_2\dt^{-2}+\dots,\\
        B_n & = (L^n)_{\geq 0},
    \end{aligned}
\end{equation}
where $\dt=\dt/\dt x$ is a differential operator, $\dt^{-1}$ is formal integration, the subscript $\geq 0$ in the definition of $B_m$ means that only non-negative powers of $\dt$ must be kept. Let us go through the first few levels of the hierarchy. For $n=1$ we have $B_1 = \dt$ and
\begin{equation}
    [B_1,L]= \sum_{i=1}^{\infty} \dt_x u_i \,\dt^{-i}.
\end{equation}
The equation is the simply $\dt u_i/\dt t_1 = \dt_x u_i$, that means $t_1=x$. Zakharov-Shabat equation when $m=1$ then becomes
\begin{equation}
    \dt_x B_n = [\dt, B_n],
\end{equation}
that is simply the action of momentum operator.

The actual KP equation can be derived from Zakharov-Shabat equation when $m=2,n=3$. For that we calculate
\begin{equation}
    \begin{aligned}
        B_2 & = \dt^2 + 2 u_1,\\
        B_3 & = \dt^3 + 3 u_1' + 3 u_1 \dt + 3 u_2,
    \end{aligned}
\end{equation}
where prime denotes derivative w.r.t. the variable $x$. Zakharov-Shabat equation has terms proportional to $\dt^0$ and $\dt^1$ that leads to two equations, which denoting $y=t_2, t=t_3, u=2u_1,v= u_2 $ can be written as
\begin{equation}
    \begin{aligned}
        \dot u_x - \fr32 u_{xxy} - 3 v_{yx} + \fr12 u_{xxxx} + 3 v_{xxx} - \fr32 (u u_x)_x &=0,\\
        -\fr34 u_y + 3 v_x + \fr34 u_{xx}&=0.
    \end{aligned}
\end{equation}
Now taking derivatives $\dt_{y}$ and $\dt_{xx}$ of the second equation we rewrite the first equation in the following form
\begin{equation}
    \dot u_x - \fr34 u_{yy} - \fr14 u_{xxxx} - \fr32 (u u_x)_x=0.
\end{equation}
This is precisely the KP equation.

Similarly, an integrable hierarchy can be constructed for KdV equation, which itself is a part of an integrable hierarchy. For that we define
\begin{equation}
    \begin{aligned}
        L & = \dt^n + u_{n-2}\dt^{n-2} + \dots + u_1 \dt + u_0,\\
        B_m & = (L^{\fr mn})_{\geq 0}.
    \end{aligned}
\end{equation}
For $n=2$ at level $m=3$ one recovers KdV equation, for $n=3$ at level $m=2$ one recovers the so-called Boussinesq equation
\begin{equation}
    3 \ddot u = - u''' - 4(u u')'.
\end{equation}

To certain extent these structures can also be generalized to the case with more than two Lax operators, which is one of natural ways to generalize the Lax-Zakharov-Shabat approach to Nambu systems. In particular, KP equation becomes a part of hierarchy constructed using Lax triples, however it seems to be not quite integrable.

\section{Nambu mechanics} 
\label{sec:nambuall}

Hamiltonian mechanics described in terms of Poisson brackets as in the previous sections appears to be a particular case of more general Nambu mechanics. Dynamics of a Nambu system is determined in terms of a flow generated by $n-1$ Hamiltonians, correspondingly Poisson bracket is replaced by Nambu bracket, that takes $n$ entries. For the purposes of the present review we are interested in algebraic structures relevant to M-theory, that appear to be 3-brackets, when  speaking about M2-branes, and 5-brackets, when speaking about M5-branes. The history of employing higher algebraic structures, such as $n$-algebras, to describe dynamics of membranes starts with the work of Basu and Harvey \cite{Basu:2004ed} where a generalization of Nahm equation that describes $N$ M2-branes ending on M5-branes has been proposed. In string theory $k$ D1-branes ending on D3-branes from the point of view of the 4-dimensional world-volume theory manifest themselves as infinite spikes \cite{Howe:1997ue,Callan:1997kz}. On the other hand this $k$ monopole system satisfies Bogomolnyi equation that turns out to be the Nahm equation for the moduli space of monopoles in the gauge theory \cite{Nahm:1979yw,Diaconescu:1996rk}. The generalization proposed by Basu and Harvey involves a Nambu 3-bracket\footnote{Strictly speaking, a 4-bracket, but one entry is always fixed} instead of the usual Lie 2-bracket in Nahm's equation. Based on these results in \cite{Bagger:2006sk,Gustavsson:2007vu} a world-volume theory of multiple M2-branes has been proposed. This is a Chern-Simons-like theory based on a Nambu 3-bracket. Although this theory, known as BLG, has later been rewritten in the form of a more conventional gauge theory in \cite{Aharony:2008ug} that does not involve 3-algebras, it is clear that a theory of brane dynamics must be formulated in terms of this kind of higher algebraic structures. We will return to a more detailed discussion of these structures later in Section \ref{sec:11d}, while here we proceed with a review of Nambu mechanics and approaches to generalizations of integrability structures for such systems.

\subsection{Nambu structure}
\label{sec:nambu0}

A generalization of Poisson mechanics to a three-dimensional phase space with evolution defined by two Hamiltonians was proposed by Nambu in \cite{Nambu:1973qe}. Later a detailed investigation of the geometry behind Nambu mechanical system has been performed by Takhtajan in \cite{Takhtajan:1993vr}. In particular, it appears that Nambu systems are much more rigid than Poisson system, that in terms of M-theoretical degrees of freedom manifests itself in the fact that BLG theory describes a stack of two M2-branes, rather than of an arbitrary number.

Dynamics of a Nambu system is defined by the following equation of motion
\begin{equation}
    \fr{d f}{dt} = \{H_1,\dots,H_{n-1},f\},
\end{equation}
where $H_1,\dots,H_{n-1}$ denote Hamiltonians of the system, and $\{\dots\}$ is an $n$-bracket that satisfies Nambu fundamental identities
\begin{equation}
    \begin{aligned}
        &\{f_1,\dots, f_{n},\{f_{n+1},\dots,f_{2n-1}\} + \{f_n,\{f_1,\dots,f_{n+1}\},f_{n+2},\dots,f_{2n-1}\} \\
        &+ \dots+ \{ f_n,\dots,f_{2n-2},\{f_1\dots,f_{2n-1}\}\} = \{f_1,\dots,f_{n-1},\{f_n\dots f_{2n-1}\}\}. 
    \end{aligned}
\end{equation}
These ensure that  given each of $n$ functions $f_i$ satisfies Nambu equation the bracket $\{f_1,\dots,f_n\}$ also satisfies the same equation. Like in the case of Poisson mechanics Nambu bracket can be realized in terms of an $n$-vector $\W \in  \G(\wedge^n TM)$, where $M$ is the configuration space:
\begin{equation}
    \W(df_1,\dots,df_n) = \{f_1,\dots, f_n\}.
\end{equation}
In coordinate notations we have
\begin{equation}
    \W^{m_1\dots m_n}\dt_{m_1}f_1 \dots \dt_{m_n}f_n = \{f_1,\dots,f_n\}.
\end{equation}
A manifold $M$ endowed globally with such $n$-vector is called Nambu-Possion manifold and $\W$ is referred to as Nambu-Possion structure. Equivalently, one says that  $M$ is a Nambu-Possion manifold if an $\RR$-multilinear map
\begin{equation}
    \{\dots\}:\big[C^{\infty}(M)\big]^{\otimes n} \to C^{\infty}(M) 
\end{equation}
is defined on the algebra of (infinitely differentiable) functions $C^{\infty}(M)$.  Given $n-1$ Hamiltonians $H_1,\dots,H_{n-1}$ the $n$-bracket defines evolution of a function $f$, or the so-called Nambu-Hamiltonian flow $g^t$.

Here comes the crucial difference between the Nambu and Poisson structure on a manifold, that is a much stronger set of constraints imposed on the $n$-vector by the fundamental identity. Acting by the $n$-vector $\W$ twice and imposing the fundamental identity we have constraints following from terms with second and first derivatives of functions separately. For the former we have an algebraic constraint
\begin{equation}
    N_{MN} + P(N_{MN})=0,
\end{equation}
where $M=\{m_1,\dots,m_n\}$ and $N=\{n_1,\dots n_n\}$ represent multiindices, the tensor $N_{MN}$ is defined as
\begin{equation}
    \begin{aligned}
     N_{m_1\dots m_n,n_1\dots n_n}  = &\ \W^{m_1\dots m_n}\W^{n_1\dots n_n} + \W^{n_n m_1 m_3\dots m_n}\W^{n_1\dots n_{n-1}m_2}\\
     &+ \dots + \W^{n_n m_2 \dots m_{n-1}m_1}\W^{n_1\dots n_{n-1}m_n} - \W^{n_n m_2\dots m_n} \W^{n_1\dots n_{n-1}m_1},
    \end{aligned}
\end{equation}
and $P$ interchanges the first and $n+1$-th indices, i.e. $m_1$ and $n_1$. One immediately notices that for $n=2$ the condition is identically satisfied, while for $n\geq 3$ it is non-trivial.

The condition descending from terms linear in derivatives of functions reads
\begin{equation}
    \begin{aligned}
    \sum_{l=1}^n \Big(\W^{lm_2 \dots m_n}\dt_l \W^{n_1 \dots n_n} + \W^{n_n l m_3 \dots m_n}\dt_l \W^{n_1 \dots n_{n-1}m_2} + \dots + \W^{n_n m_2 \dots m_{n-1} l}\dt_l \W^{n_1 \dots n_{n-1}m_n}\Big)=0.
    \end{aligned}
\end{equation}
For $n=2$ this is simply
\begin{equation}
    \W^{l[m}\dt_l\W^{nk]}=0.
\end{equation}
Hence, one concludes that in contrast to Poisson manifolds not any totally antisymmetric globally defined tensor $\W^{m_1\dots m_n}$ on a manifold is capable of defining a Nambu structure. On top of the usual differential constraints one has to satisfy algebraic constraints.

An observable $F \in C^{\infty}(M)$ is called an integral of motion if 
\begin{equation}
    \{H_1,\dots,H_{n-1},F\}=0.
\end{equation}
First $n-1$ integrals of motion are the Hamiltonians, the fundamental identity ensures that Nambu bracket of integrals of motion is again an integral of motion. Naively one can extend the notion of Liouville integrability to Nambu systems defining an integrable Nambu system as having $n$ integrals of motion each in involution w.r.t. the Nambu bracket. However, the analogy does not go much further since it is not evident how the action-angle variables can be introduced to completely integrate equations of motion. The same holds for the naive generalization of the proof that Nambu flow preserves phase space volume, although, for some cases it can be shown explicitly. 

\subsection{Examples of Nambu systems}
\label{sec:nambuex}

Although the construction of Nambu mechanics might seem rather exotic it describes mechanical systems many of which are familiar and even integrable in the usual sense. As the first example consider the $n$-dimensional harmonic oscillator with the Hamiltonian
\begin{equation}
    H = \fr12\sum_{i=1}^n (p_i^2 + x_i^2).
\end{equation}
According to  \cite{Chatterjee:1995iq} this system can be written as a Nambu system using other integrals of motion as additional Hamiltonians. Consider for example the case $n=2$, that is the harmonic oscillator in two dimensions. We choose the following set of integrals
\begin{equation}
    \begin{aligned}
        H_1& = \fr{1}{2}(p_1^2+x_1^2),\\
        H_2& = \fr{1}{2}(p_2^2+x_2^2),\\
        H_3& = x_1 p_2 - x_2 p_1.
    \end{aligned}
\end{equation}
Nambu bracket describing the system can then be chosen as
\begin{equation}
    \{H_1,H_2,H_3,f\} = \fr{1}{p_1p_2+x_1x_2}\fr{\dt(H_1,H_2,H_3,f)}{\dt(p_1,p_2,x_1,x_2)}.
\end{equation}
Simple check shows that the above reproduces equations of motion of the two-dimensional oscillator, and the bracket satisfies all the necessary conditions. This system is integrable in the usual sense.

Consider now the example presented by Nambu in the original paper, that describes rotational dynamics of a rigid body with principle axes of inertia $I_i$ and angular momenta $L_i$ where $i=1,2,3$. This system is commonly referred to as the Euler asymmetric top. Equations of motion are
\begin{equation}
    \begin{aligned}
        \fr{d L_1}{dt} & = \fr{I_3 - I_2}{I_2 I_3} L_2 L_3, && \fr{d L_2}{dt} & =\fr{I_1 - I_3}{I_1  I_3}  L_1 L_3, && \fr{d L_3}{dt} & =\fr{I_2 - I_1}{I_1 I_2}  L_1 L_2.
    \end{aligned}
\end{equation}
These can be written in terms of Nambu equations for a system with the following two Hamiltonians:
\begin{equation}
    \begin{aligned}
        H_1 = \fr{L_1^2}{2I_1}+\fr{L_2^2}{2I_2}+\fr{L_3^2}{2I_3}, &&
        H_2 = \fr{1}{2}\big(L_1^2+L_2^2+L_3^2\big).
    \end{aligned}
\end{equation}
These are the full energy and the full momentum of the top and the equations of motion can be written as
\begin{equation}
    \fr{d L_i}{dt} = \e^{ijk}\dt_j H_1 \dt_k H_2,
\end{equation}
where $\dt_i = \dt/\dt L_i$. This suggests the following definition for the Nambu bracket
\begin{equation}
    \{H_1,H_2,f\} = \e^{ijk}\dt_i H_1 \dt_j H_2 \dt_k f,
\end{equation}
that is the most natural choice for a three-dimensional system. 

Equations of motion for the asymmetric Euler top have SU(2) symmetry and are also known under a different name, the Nahm system, when arise in the theory of static monopoles. As we discuss in further sections such equations naturally appear in the description of branes ending on branes as world-volume spike-like monopoles. Its generalization, known as the Basu-Harvey equation, underlies the so-called BLG theory describing dynamics of two M2-branes \cite{Bagger:2006sk,Gustavsson:2007vu}. The Nahm system is usually written as the following set of equations of motion
\begin{equation}
    \begin{aligned}
        \fr{d x_1}{dt} = x_2 x_3, && \fr{d x_2}{dt} = x_1x_3, && \fr{dx_3}{dt}=x_1x_2,
    \end{aligned}
\end{equation}
that can be expressed in the Nambu form given $H_1=x_1^2 - x_2^2$, $H_2 = x_1^2 - x_3^2$. This system is also integrable in the usual sense.

The least action principle can be extended to Nambu mechanics leading to an action presumably describing movement of an open membrane boundaries. Following Takhtadjan \cite{Takhtajan:1993vr} we define 
\begin{equation}
    \w_2 = x^1 dx^2\wedge dx^3 - H_1 dH_2 \wedge dt,
\end{equation}
that is the Poincare-Cartan integral invariant 2-form for Nambu mechanics on the phase space $\tilde{X}$ parametrized by coordinates $\{x^1,x^2,x^3,t\}$. We now follow the same lines as in Section \ref{sec:volume2} where an invariant action for Poisson system was constructed. Define vector field $\tilde{L} = \dt_t +L$ with
\begin{equation}
    \begin{aligned}
    L = L^i \dt_i, && L^i = \fr12 \e^{ijk}\fr{\dt(H_1,H_2)}{\dt(x^j,x^k)}.
    \end{aligned}
\end{equation}
Nambu equations then simply become $\dot f = \tilde{L}(f)$. The vector field $\tilde{L}$ is a line field of the 3-form $d\w_2$, i.e. $\iota_{\tilde{L}}d\w_2=0$, that simply follows from the explicit expression for the derivative
\begin{equation}
    d\w_2 = dx^1 \wedge dx^2 \wedge dx^3 - dH_1 \wedge dH_2 \wedge dt.
\end{equation}
Now for a given 2-chain $c$ in $\tilde{X}$ denote $g^t(c)$ its Nambu-Hamiltonian phase flow, then a tube of phase trajectories will be given by $J^tc = \{g^\t(c), 0\leq \t \leq t\}$. Finally, since $\dt J^tc = c- g^t(c)$ and given $\iota_{\tilde{L}}d\w_2=0$ we have
\begin{equation}
    \int_c \w_2 - \int_{g^t(c)}\w_2 = \int_{J^tc}d\w_2 = 0,
\end{equation}
that basically demonstrates invariance of the integrated two-form. Hence, we call the action $A$ the following integral
\begin{equation}
    A[c] = \int_c \w_2 = \int_c x^1 \wedge dx^2 \wedge dx^3 - H_1 \wedge dH_2 \wedge dt. 
\end{equation}

The first term of the Takhtadjan's action above has precisely the form of the Wess-Zumino term of the action of an M2-brane ending on M5-brane. In the Nambu-Goto form and dropping all possible world-volume gauge fields we can write for the M2-brane
\begin{equation}
    S_{M2}= \int_\S d^3 \x \det(-G) + \int_\S C_3,
\end{equation}
where $G$ denotes the metric pull-back and $C_3=C_{ijk}dx^i\wedge dx^j \wedge dx^k$ denotes the 3-form living in the target space. Supposing the Wess-Zumino term dominates, and that the 3-form varies slowly along the boundary $\dt \S$ of the M2-brane, we write for the boundary action
\begin{equation}
    S_{\dt M2} \propto C_{123}\int_{\dt \S} x^1 \wedge dx^2 \wedge dx^3,
\end{equation}
that is precisely the expression above. This provides yet another evidence that Nambu mechanics with all its structures must be relevant to membranes dynamics in M-theory. More detailed discussion of the above narrative can be found in \cite{Ho:2016hob}. We will return to description of membrane dynamics in further sections.

\subsection{Lax pair and generalized Yang-Baxter equation } 
\label{sec:laxgencybe}

In the previous section we have seen that many of integrable in the usual sense dynamical systems possess Nambu structure. This makes it natural that integrability structures, such as the Lax pair, can be reformulated in terms of Nambu brackets, giving, probably, a criterion for three-dimensional integrability. To our knowledge the program of defining integrability for Nambu dynamical systems has not been completed at least to the level of understanding we have for Poisson systems, however certain progress has been made. The overall aim of this review is to collect observations that give hints at integrability in the theory of membranes, or more generally, in three dimensional systems. Hence, start with a Nambu system with Hamiltonians $H_1$ and $H_2$ and equations of motion given by
\begin{equation}
    \fr{d f}{dt } = \{H_1,H_2,f\},
\end{equation}
and try to generalize the Lax pair construction. Naturally, the very first attempt would be to introduce a Nambu tri-bracket and to rewrite the above in terms of a Lax triple 
\begin{equation}
    \dot{L}=[L,M,N],
\end{equation}
which is indeed a useful construction for defining Nambu hierarchies. We will discuss these in a moment as it is suggestive to start with a different generalization that has more transparent bounds to M-theory. 

Consider a Lax-pair, that is a pair of matrices $L,M \in \mathfrak{g}$, where $\frg$ is an algebra, such that
\begin{equation}
    \dot{L} = [L,M].
\end{equation}
Given a tensor $\r_{123} \in \frg\wedge \frg \wedge \frg$ define a 3-bracket
\begin{equation}
    \{L_1,L_2,L_3\} = [\r_{123},L_1] + [\r_{123},L_2] + [\r_{123},L_3],
\end{equation}
where we denote
\begin{equation}
    \begin{aligned}
    L_1 = L\otimes \id\otimes \id, && L_2 =\id \otimes L\otimes \id, && L_3 = \id\otimes \id \otimes L. 
    \end{aligned}
\end{equation}
Impose the fundamental identity on such defined 3-bracket, that is, turn it into a Nambu structure. This restricts $\r_{123}$ to satisfy a condition similar to classical Yang-Baxter equation. Let $\{T_a\} = \bas\, \frg$ denote basis of the algebra, $f_{ab}{}^c$ denote its structure constants and $\r_{123} = \r^{abc}T_a\wedge T_b\wedge T_c$. Then the condition, that is often referred to as generalized Yang-Baxter equation, can be written in the component form as
\begin{equation}
    \label{eq:genCYBE0}
         {\rho}^{a_1 [a_2 |a_6|} \rho^{a_3 a_4 |a_5|} f_{a_5 a_6}{}^{a_7]}- {\rho}^{a_2 [a_1 |a_6|} \rho^{a_3 a_4 |a_5|} f_{a_5 a_6}{}^{a_7]}=0
\end{equation}
The 3-bracket then defines a Nambu system, whose integrals of motion can be expressed in the usual form $F_k = \Tr L^k$. It is a simple calculation to check, that these are in involution w.r.t. such constructed Nambu bracket
\begin{equation}
    \{F_i,F_j,F_k\}=0.
\end{equation}
If there was a procedure allowing to introduce action-angle variables and completely solve equations of motion using these integrals of motion, one would say that such constructed system is integrable. However, the authors are not aware if this kind of constructions for Nambu systems exist. 

The equations \eqref{eq:genCYBE0} are fascinating in a different respect: it first has been derived when investigating U-dualities of M-theory in \cite{Malek:2019xrf,Sakatani:2019zrs} (and earlier in \cite{Bakhmatov:2019dow} in the form of vanishing of R-flux). To be more precise, equations of 11-dimensional supergravity are known to be symmetric under a set of particular transformations, called Nambu-Lie U-dualities, whose underlying algebraic structure is formulated in terms of the so-called exceptional Drinfeld algebras. A particular subset of such generalized U-dualities can be parametrized by tensor $\r^{abc}$, that has precisely the same meaning as above. The condition for such a deformation to satisfy equations of 11d supergravity is \eqref{eq:genCYBE0}. We will return to this in more detail later, while here it is important to mention that these are mainly generalizations of similar structures in string theory ruled by the usual classical Yang-Baxter equation (see the summary section of \cite{Sakatani:2019zrs}). The most important observation is that deformations parametrized by the matrix $r^{ab}$ satisfying classical Yang-Baxter equation preserve integrability of the string sigma model. Hence, starting from different phenomena involving integrability and generalizing them in the more or less the same way, we arrive at the same equation \eqref{eq:genCYBE0}, which allows to speculate further on these matters. We will do so in Section~\ref{sec:spec}.

\subsection{Lax triples and volume preserving flows}
\label{sec:laxtriple}

As it has already been mentioned above a more straightforward generalization of the Lax construction to Nambu systems is to introduce a Lax triple (multiple). It is convenient to proceed with a construction of integrable hierarchies based on the Lax triple and a Nambu 3-bracket. As in previous explicit examples we will find that familiar systems, the hierarchy KP in this case, can be formulated in terms of such generalized structures. The approach we will be following here is the one advocated by Guha in \cite{Guha:1998ab} and further applied to various examples in 
\cite{Li:2020dfy,Wang:2015ab}. The idea is to generalize the method of \cite{Takasaki:1994xh,Takasaki:1991hn} developed to study area preserving diffeomorphic KP equation. In turn their approach originated from studying self dual Einstein equations. Generalized flows of \cite{Guha:1998ab} briefly reviewed below are integrable in the same sense as the SDiff(2) flows of \cite{Takasaki:1994xh,Takasaki:1991hn}, i.e. in the sense of the nonlinear graviton construction.

Consider a triple of generalized Lax operators $L,M,N$ that are Laurent series in a spectral parameter $\l$ with coefficients being functions of some variables $p,q$. By definition volume preserving integrable hierarchy is given by
\begin{equation}
    \begin{aligned}
        \fr{\dt L}{\dt t_n} &= [B_{1n}, B_{2n},L], \\
        \fr{\dt M}{\dt t_n} &= [B_{1n}, B_{2n},M], \\
        \fr{\dt N}{\dt t_n} &= [B_{1n}, B_{2n},N],
    \end{aligned}
\end{equation}
with the additional involution constraint $[L,M,N]=0$ ensuring volume preservation. As in the case of ordinary integrable hierarchies we restrict the operators $B_{1n}$ and $B_{2n}$ to have  only positive values of $L,M$:
\begin{equation}
    B_{1n} = (L^n)\Big|_{n\geq 0}, \quad B_{2n} = (M^n)\Big|_{n\geq 0}. 
\end{equation}
The condition for the flows to commute boils down to an analogue of Zakharov-Shabat equation
\begin{equation}
    \begin{aligned}
        &[\dt_{m}B_{1n}, B_{2n},\bullet] - [\dt_nB_{1m}, B_{2m},\bullet] + [B_{1n}, \dt_m B_{2n},\bullet] - [B_{1m},\dt_n B_{2m},\bullet] \\
        &=[[B_{1n},B_{2n},B_{2m}],B_{2m},\bullet]-[[B_{1n},B_{2n},B_{1m}],B_{2m},\bullet].
    \end{aligned}
\end{equation}
Important remark here is that for SDiff(2) area preserving flows the above equation is simply the zero-curvature condition. The same we have already observed for Poisson integrable hierarchies, where Zakharov-Shabat equation contained the zero-curvature condition for the Lax connection 1-form. Hence, one would expect that the above contains a three-dimensional analogue of the zero-curvature condition of an analogue of Lax connection, given by a 2-form.

From the geometric point of view, self-duality is simply a Ricci flat K\"ahler geometry, hence, SDiff(2) flow are naturally expressed in terms of a K\"ahler-like 2-form. The analogue here is a 3-form
\begin{equation}
    \begin{aligned}
        \W  &= \sum_{n=1}^\infty dB_{1n}\wedge dB_{2n} \wedge dt_n    = d\l \wedge dp \wedge dq + \sum_{n=2}^\infty dB_{1n}\wedge dB_{2n} \wedge dt_n,
    \end{aligned} 
\end{equation}
where we used the following notations $t_1 = \l$, $B_{11}=p$, $B_{21}=q$. Given the flow equations above the 3-form can be expressed simply as
\begin{equation}
    \W = dL \wedge dM \wedge dN.
\end{equation}
The 3-form $\W$ can be verified to be closed $d\W=0$, that gives
\begin{equation}
    d\Big(M\wedge dL \wedge dN + \sum_{n=1}^\infty B_{1n}\wedge dB_{2n}\wedge dt_n \Big)=0.
\end{equation}
Hence, the expression in brackets at least locally can be written as an exact form 
\begin{equation}
    dQ = M\wedge dL \wedge dN + \sum_{n=1}^\infty B_{1n}\wedge dB_{2n}\wedge dt_n ,
\end{equation}
that is an analogue of Krichever potential, i.e. contains the action.

Let us now consider an example of hierarchy generated by volume preserving Lax triple equations. Here we follow \cite{Wang:2015ab} where KP hierarchy was first written in terms of Lax triples. The hierarchy is defined as
\begin{equation}
    \begin{aligned}
        \fr{d L}{d t_{mn}} &= [B_{m}, B_{n},L], \\
        L& = \dt + \sum_{i = 0}^{\infty} v_i(t)\dt^{-i-1}, \\
        B_n & = (L^{n})_{\geq 0}, \quad n\geq 0,\\
        B_0 & = 1,
    \end{aligned}
\end{equation}
where as before the subscript $\geq 0$ means that only operators with positive powers of $\dt$ are kept. The standard KP hierarchy is recovered from the above when $m=0$:
\begin{equation}
    \fr{d L}{d t_{0n}} = [B_0,B_n,L] \equiv [B_n,L].
\end{equation}

The most interesting is the question of whether integrable hierarchies can be derived for other choices of $B_m$ with $m\neq 0$. According to \cite{Wang:2015ab} it has affirmative answer: at least for certain given pairs $(B_m,B_n)$ one obtains integrable equations that are already present in the KP hierarchy. It is tempting to claim that integrable KP hierarchy can be equivalently written in terms of the usual Lax equation or in terms of generalized equation for Lax triples. However, a subtlety caving this interpretation has been also observed in \cite{Wang:2015ab} when analyzing the hierarchy further for larger values of $(m,n)$: the hierarchy contains equations that do not pass the Painlev\'e integrability test. Hence, not all equations of the generalized Lax triple hierarchy seem to be integrable, however, those that are not, contain soliton solutions.

To conclude, we observe that at least some of the steps in the standard path for construction of integrability structures can be repeated for Nambu systems. In particular, one may introduce infinitely many conserved charges, a Lax operator generating them, involution condition, volume preserving flows and hierarchies based on Lax triples. Moreover, Nambu bracket of a dynamical system can be generated by an analogue $\r$ of the classical $r$-matrix, which is no longer a matrix, however naturally appears in the context of U-duality symmetries of M-theory. The same object is in principle expected to appear in a quasi-classical limit of tetrahedron equation describing factorization of scattering process of straight strings. Combining all these observations together it is tempting to conclude that structures reviewed above must be relevant to description of integrability of 2+1-dimensional systems, i.e. membranes. As we will discuss in more details further in Sections \ref{sec:11d} one indeed finds similar constructions when approaching from the M-theory and supergravity side. In particular, dynamics of membranes naturally leads to tri-brackets via Basu-Harvey equation, the object $\r$ naturally appears in the open membrane metric and an analogue of evolution operator can naturally be constructed using loop algebras. In turn loop algebras appear in the analysis of M5-branes holding boundaries of M2-branes, that scatter precisely as strings in the 6-dimensional world-volume. 

\section{10d supergravity and strings}
\label{sec:10d}

Methods for investigating integrability structures for two-dimensional systems briefly reviewed above can well be applied to dynamics of the fundamental string propagating on a background defined by a solution to supergravity equations. As we discuss in more details below dynamics of the string on certain backgrounds defined in terms of two-dimensional non-linear sigma model (NLSM) is classically integrable, i.e. a Lax connection can be constructed. Among other similar results, integrability of the string on certain backgrounds is of particular interest e.g. in the context of holography. Indeed, holographic correspondence basically tells that the same system can be described equivalently in terms of very different variables. For example, in the case of AdS/CFT correspondence equivalence of descriptions in terms of closed and open strings of the near horizon region of a D3-brane results in the correspondence between 10d supergravity on AdS${}_5\times \SS^5$ and $\mc{N}=4$ $d=4$ super Yang-Mills theory. Now, since the string on AdS${}_5\times \SS^5$  is known to be integrable \cite{Bena:2003wd} (see also \cite{vanTongeren:2013gva} for a review) the same can be claimed for the gauge theory, as this is simply a different way of parametrizing the same dynamics. Indeed, integrability structures of $\mc{N}=4$ $d=4$ have been addressed from different perspectives, that include  e.g. thermodynamic Bethe ansatz, spectral curve. Although intimately related, these approaches stand beyond the scope of our review and hav been nicely covered in various reviews  \cite{Slavnov:2018lec,Schafer-Nameki:2010qho,Staudacher:2010jz,Bajnok:2010ke,Levkovich-Maslyuk:2016kfv}. In this section we focus at integrable non-linear 2d sigma-models on group manifolds and (super-)coset spaces and their continuous Yang-Baxter deformations\footnote{Another examples of integrable strings that we will not focus here on are the $\lambda$ deformations of \cite{Sfetsos:2013wia,Hollowood:2014qma}}. Such deformation, that preserve integrability of a 2d NLSM have been introduced in \cite{Klimcik:2002zj} for a string on a group manifold and further generalized to coset spaces in \cite{Delduc:2013fga}. Their extension to general solutions of supergravity suggested first in \cite{Araujo:2017jap,Araujo:2017jkb} and further developed in \cite{Bakhmatov:2018apn,Bakhmatov:2018bvp} does not seem to have a straightforward relation to integrability, however allows to introduce similar structures for membranes, i.e. 3d NLSM, that will be discussed further in Section~\ref{sec:11d}.

\subsection{Yang-Baxter deformed 2d sigma-models}
\label{sec:YBsigma}

An approach to finding a Lax pair for every 2d principal sigma-model on a simple compact group $G$ based on the inverse scattering method has been suggested by Zakharov and Mikhailov in \cite{Zakharov:1973pp}. A particular example of such model is the model with $G=\rmSU(2)$, that has been found to belong to a continuous family of integrable models by Cherednik in \cite{Cherednik:1981df}. Given $g \in \rmSU(2)$ the model is defined by the following action
\begin{equation}
    S = -\fr12 \int d\t d\s \Tr \Big[\mathrm{Ad}(\dt_+ g \, g^{-1})J\mathrm{Ad}(\dt_- g\, g^{-1})\Big],
\end{equation}
where the diagonal matrix $J=\mathrm{diag}[J_1,J_2,J_3]$ stands for a deformation of the Killing form. In \cite{Klimcik:2002zj} it has been shown that the above model can be understood in terms of Yang-Baxter sigma-models, i.e. sigma-models deformed by a classical $r$-matrix $R$, that satisfies (a modified) classical Yang-Baxter equation
\begin{equation}
[RM, RN] - R\bigl( [RM, N] + [M, RN] \bigr) = c [M, N],
\end{equation}
where $M,N\in \su(2)$. It is worth to mention, that for historical reason classical $r$-matrix defining deformations of 2d sigma-models is denoted by a capital letter $R$, that in mathematical literature is reserved for quantum $r$-matrix, i.e. the one solving quantum Yang-Baxter equation. To keep notations correlated with the rest of the string theory literature we follow this historical rule, which however should not cause much confusion. 

The deformation procedure used in \cite{Cherednik:1981df} is specific for the SU(2) group manifold and cannot be directly generalized to any group manifold taken as a target space. The approach of \cite{Klimcik:2002zj} suggests to consider the following action
\begin{equation}
    S = \int \Big\langle g^{-1}\dt_+ g , \big(\mathbbm{1} + \e R\big)^{-1} g^{-1}\dt_- g \Big\rangle,
\end{equation}
where angle brackets denote Killing form on the Lie algebra $\frg$ of a simple compact Lie group $G$. Further in \cite{Klimcik:2008eq} this model has been shown to be integrable and the corresponding Lax connection can be written as follows:
\begin{equation}
    \label{eq:laxklimcik}
        A_\pm(\l) = \bigg( \e^2 \mp \e R - \fr{1+ e^2}{1\pm \l} \bigg)\big(\mathbbm{1} \pm \e R\big)^{-1} g^{-1}\dt_\pm g,
\end{equation}
where $\l \in \mathbb{C}$ is a complex spectral parameter. When $\e=0$ the above reproduces precisely the Lax connection introduced by Zahkharov and Mikhailov.

The procedure of deforming principal sigma-models preserving integrability has been generalized to sigma-models on coset spaces in \cite{Delduc:2013fga}. The action for the deformed sigma-model on a coset space $G/F$ now involves the so-called dressed $r$-matrix $R_g = \mathrm{Ad} g^{-1} \, R \, \mathrm{Ad} g$:
\begin{equation}
    S = \int \Big\langle (g^{-1}\dt_+ g)^{(1)} , \fr{1+\h^2}{\mathbbm{1} - \h R_g \, P_1} (g^{-1}\dt_- g)^{(1)} \Big\rangle.
\end{equation}
Here $P_1$ is the projection onto the subspace  $\frg^{(1)}$ of the Lie algebra $\frg$ of $G$, that corresponds to the value $\s=+1$ of an order-2 automorphism $\s:\frg \to \frg$. Using this approach an integrable deformation of the AdS$_5\times \SS^5$ superstring in the Metsaev-Tseytlin formalism has been constructed in \cite{Delduc:2013qra}, which we will discuss in a moment. Two important observations have been made  concerning such deformed superstring: i) the deformed sigma-model can be understood as a string propagating on a metric background (ABF) that does not satisfy supergravity equations \cite{Arutyunov:2015qva}; ii) kappa symmetry of the GS superstring still holds \cite{Wulff:2016tju}. Remarkably kappa-symmetry of the GS superstring has been shown to imply a slight generalization of supergravity equations \cite{Arutyunov:2015mqj}, which are precisely the ones solved by the ABF background. Hence, one concludes that the space of consistent vacua, at least for the GS superstring, is wider than the space of solutions to supergravity equations, and moreover certain points in this space are connected by Yang-Baxter deformations\footnote{See however the discussion concerning consistency of generalized supergravity backgrounds on which the string is only Weyl invariant or the corresponding FT counterterm seems to be non-local \cite{Sakamoto:2017wor,Fernandez-Melgarejo:2018wpg,Muck:2019pwj}}. Later in Section \ref{sec:11d} we will see that the same picture holds for 11d supergravity, although several loose ends must be tied up, such as kappa invariance of the membrane on similar deformations.

To the moment huge progress has been made in understanding of Yang-Baxter deformed sigma-models on group manifolds and coset spaces and in finding new examples. Let us mention some of the most compelling results. A slight generalization of the q-deformation of \cite{Delduc:2013qra} has been suggested in \cite{Kawaguchi:2014qwa} that contains twists of the R-operator allowing to perform partial deformations affecting only of the sphere part of the AdS$_5\times \SS^5$ superstring. Recall that the $r$-matrix describing the standard q-deformations is of the so-called Drinfeld-Jimbo type:
\begin{equation}
    R_{DJ} = \alpha \sum_{M}\fr{1}{\Tr[e_M f_M]} e_M \wedge f_M,
\end{equation}
where the index $M$ labels positive $e_M$ and negative $f_M$ roots of the isometry algebra. Jordanian matrices are constructed by a linear twist of $R_{DJ}$ by an arbitrary (bosonic) root. A general class of integrable deformations of sigma models on coset spaces whose Poisson brackets are related to those of \cite{Delduc:2013qra} by an analytic continuation has been found in \cite{Hollowood:2014rla}. This generalizes earlier works \cite{Balog:1993es,Evans:1994hi,Sfetsos:2013wia} where integrable deformations have been constructed as interpolations between exact WZW CFTs. For more details see e.g. the reviews \cite{Magro:2010jx,Thompson:2019ipl}, the PhD thesis \cite{Driezen:2019yjv} and references therein. Integrable deformations of the string on AdS$_n\times \SS^n$ have been intensively investigated in \cite{Hoare:2014pna,Lunin:2014tsa,Sfetsos:2015nya,Hoare:2015gda}. Given the discussion in the beginning of this section it is also worth to mention the works \cite{vanTongeren:2015uha,Araujo:2017jap,Araujo:2017jkb} where a gauge theory interpretation of integrable deformations have been presented using the formalism of Drinfeld twists. This generalizes the known interpretation of abelian deformations, such as the U(1)$\times$U(1) deformation of Lunin and Maldacena as twisting of fields product, to the non-abelian case (for more details on the abelian case see e.g. \cite{Imeroni:2008cr}). The result is a non-commutative Yang-Mills theory, that is expected since the generators of the deformations are taken along the AdS space.

\subsection{Integrable deformation of the \texorpdfstring{AdS$_5\times \SS^5$}{AdS5xS5} superstring}
\label{sec:ads5}

Let us illustrate the formalism of integrable Yang-Baxter deformations by the example of $\eta$-deformation of the AdS$_5\times \SS^5$ superstring following \cite{Delduc:2013fga}. We start with recalling the construction of Lax connection for Metsaev-Tseytlin superstring \cite{Arutyunov:2009ga}. The Type IIB background AdS$_5\times \SS^5$ is supported by  the non-vanishing self-dual RR five-form flux and hence the superstring on such background can conveniently by described using the Green-Schwarz formalism. Such superstring lives in the following symmetric  super-space
\begin{equation}\label{AdScoset}
    \frac{PSU(2,2|4)}{SO(4,1) \times SO(5)} \supset \frac{SU(2,2) \times SU(4)}{SO(4,1) \times SO(5)} \cong \frac{SO(4,2) \times SO(6)}{SO(4,1) \times SO(5)} = AdS_5\times \SS^5.
\end{equation}
The corresponding GS sigma model is formulated in terms of a 1-form $A \in \mathfrak{su}(2,2|4)$ built out of a supergroup element $\mathfrak{g} \in SU(2,2|4)$ as
\begin{equation}\label{defA}
    A = -  \mathfrak{g}^{-1} d \mathfrak{g} = A^{(0)} + A^{(1)} + A^{(2)} + A^{(3)}.
\end{equation}
Here the decomposition is due to $\mathbb{Z}_4$-grading of $\mathfrak{su}(2,2|4)$ induced by a certain order 4 automorphism. Such defined 1-form $A$  is flat:
\begin{equation}\label{Aflatness}
   \dt_\a A_{\beta} - \dt_{\beta} A_{\alpha} - [A_{\alpha}, A_{\beta}] = 0.
\end{equation}
The action of the superstring on AdS$_5\times \SS^5$ then takes the form  of the so-called Metsaev-Tseytlin superstring \cite{Metsaev:1998it}:
\begin{equation}\label{GSAdS}
    S_{MT} = - \frac{g}{2} \int d\tau d\sigma [\gamma^{\alpha \beta} str(A^{(2)}_{\alpha} A^{(2)}_{\beta}) + \kappa \epsilon^{\alpha \beta} str(A^{(1)}_{\alpha} A^{(3)}_{\beta})], \quad \sigma \in (-r,r),
\end{equation}
with $g = \frac{R^2}{2 \pi \alpha'}$, where $R$ is $\SS^{5}$ radius and $\alpha'$ is string slope. $\gamma^{\alpha \beta}$ is worldvolume metric and $\epsilon^{\alpha \beta}$ is worldvolume totally antisymmetric tensor, $\epsilon^{\tau \sigma} = 1$.

For the further discussion it is convenient to introduce tensors
\begin{equation}
    P_{\pm}^{\a \b} = \frac{1}{2} (
\gamma^{\a\b} \pm \kappa \epsilon^{\a\b} ),
\end{equation}
which are orthogonal projectors in cases $\kappa = \pm 1$, and four projectors $P_{k}$ on to the corresponding subspaces of $\mathfrak{su}(2,2|4)$ with grade $k=0,\dots,3$, such that $A^{(k)} = P_{k} A$. Also we will use the following conventions for projected vectors $V^{\alpha}_{\pm} = P^{\alpha \beta}_{\pm} V_{\beta}$. In these notations 
\begin{equation}\label{GSAdSP}
    S_{GS} = - \frac{g}{2} \int d\tau d\sigma P_{-}^{\alpha \beta} str(A_{\alpha} [P_{1} + 2P_{2} - P_{3}] A_{\beta}).
\end{equation}

The GS action (\ref{GSAdSP}) must obey a local fermionic symmetry, called $\kappa$-symmetry, that provides the space-time supersymmetry of the physical spectrum. Its transformation acts on $A$ as
\begin{equation}
    \delta_{\epsilon} A = - d \epsilon + [A, \epsilon], \quad \epsilon = \epsilon^{(1)} + \epsilon^{(3)},
\end{equation}
with
\begin{equation}
    \begin{aligned}
        \epsilon^{(1)} &= A^{(2)}_{\a,-} \kappa^{(1),\a}_{+} + \kappa^{(1),\a}_{+} A^{(2)}_{\a,-}\, , \\ 
        \epsilon^{(3)} &= A^{(2)}_{\a,+} \kappa^{(3),\a}_{-} + \kappa^{(3),\a}_{-} A^{(2)}_{\a,+}.
    \end{aligned}
\end{equation}
Interesting fact is that $\kappa$-invariance of the action requires $\kappa = \pm 1$ and hence $P_\pm^{\a\b}$ are indeed orthogonal projectors.

Equations of motion for (\ref{GSAdSP}) can be written in the following compact form
\begin{equation}\label{AdSeom}
    \begin{aligned}
        0 = & \, \dt_{\a}(\gamma^{\a\b}A_{\b}^{(2)}) - \gamma^{\a\b} [A_{\a}^{(0)},A_{\b}^{(2)}] +\frac{1}{2} \kappa \epsilon^{\a \beta} \big([A_{\a}^{(1)},A_{\b}^{(1)}] - [A_{\a}^{(3)}, A_{\b}^{(3)}]\big) \, ,\\
        0 = & \, P_{-}^{\a\b}[A_{\a}^{(2)},A_{\b}^{(3)}]\, ,\\
        0 = & \, P_{+}^{\a\b}[A_{\a}^{(2)},A_{\b}^{(1)}]\, .
    \end{aligned}
\end{equation}
Global $PSU(2,2|4)$ symmetry of the sigma model corresponds to conservation of the following Noether's current
\begin{equation}
J^{\alpha} = g \mathfrak{g} \Big[ \gamma^{\a\b} A^{(2)}_{\b}
- \frac{1}{2}\kappa\, \epsilon^{\a\beta} (A^{(1)}_{\beta} - A^{(3)}_{\beta}) \Big] \mathfrak{g}^{-1}, \quad \dt_{\alpha} J^{\alpha} = 0.
\end{equation}

This model is classically integrable meaning that the equations of motion (\ref{AdSeom}) together with the flatness condition for $A$ (\ref{Aflatness}) are equivalent to zero curvature condition 
\begin{equation}
   \dt_{\a} L_{\beta}-\dt_{\b} L_{\a}- [L_{\a},L_{\b}] = 0, 
\end{equation}
for a Lax connection $L_{\a}$ defined by
\begin{equation}
    L_{\a} = \ell_0 A_{\a}^{(0)} + \ell_1 A_{\a}^{(2)} + \ell_2 \gamma_{\a\b} \epsilon^{\beta\rho} A_{\rho}^{(2)} + \ell_3 A_{\a}^{(1)} + \ell_4 A_{\a}^{(3)}.
\end{equation}
The prefactors must be chosen as 
\begin{equation}
    \ell_0=1\, ,\quad
\ell_1 = \frac{1}{2}\Big(z^2+\frac{1}{z^2}\Big)\, , \quad
\ell_2 = -\frac{1}{2\kappa}\Big(z^2-\frac{1}{z^2}\Big)\, \quad
\ell_3 = z\, , \quad \ell_4 = \frac{1}{z}\, ,
\end{equation}
where $z$ is spectral parameter and $\kappa = \pm 1$. This  means that  the requirement of integrability automatically leads to the same constraints as $\kappa$-invariance. 

Let us now briefly discuss the results of \cite{Delduc:2013qra} where integrability of the superstring on the $\eta$-deformed AdS$_5\times \SS^5$ has been demonstrated. Note that for $A$ we use the convention (\ref{defA}) of \cite{Arutyunov:2009ga} that differs from \cite{Delduc:2013qra} by a ``$-$'' sign. The superstring on the $\eta$-deformed AdS$_5\times \SS^5$ can be written as the following $\eta$-deformation of the Metsaev-Tseytlin superstring (\ref{GSAdSP}):
\begin{equation}\label{etaGSAdSP}
    S^{\eta}_{MT} = - g \int d\tau d\sigma \frac{(1+\eta^2)^2}{2(1-\eta^2)} P_-^{\alpha\beta} str\Big(A_\alpha
    \, P \circ \frac{1}{1-\eta R_\mathfrak{g}\circ P} (A_\beta)\Big),
\end{equation}
where
\begin{equation}
    P = P_1+\frac{2}{1 - \eta^2} P_2 - P_3, \quad \tilde{P} = - P_1 + \frac{2}{1-\eta^2} P_2 + P_3.
\end{equation}
The crucial ingredient here is a skew-symmetric operator on $\mathfrak{su}(2,2|4)$, which acts as $R_\mathfrak{g}=\mbox{Ad}_{\mathfrak{g}}^{-1} \circ R \circ \mbox{Ad}_\mathfrak{g}$ and solves the modified classical Yang-Baxter equation. Specifically, $\forall M, N \in \mathfrak{su}(2,2|4)$:
\begin{equation}
[RM, RN] - R\bigl( [RM, N] + [M, RN] \bigr) = [M, N] \end{equation}
and $str\bigl( M RN\big)=-str\big( RM N\bigr)$.

For $\eta = 0$ the action (\ref{etaGSAdSP}) reproduces (\ref{GSAdSP}). The following vectors
\begin{align}
J_\alpha &=  \frac{1}{1-\eta R_\mathfrak{g} \circ P}(A_\alpha), \\
\tilde{J}_{\alpha} &= \frac{1}{1 + \eta R_\mathfrak{g} \circ \tilde{P}}(A_\alpha),
\end{align}
allow to write equations of motion for (\ref{etaGSAdSP}) in the most convenient way 
\begin{equation}\label{eomdef}
0 = P(\partial_\alpha J_-^\alpha)+
\tilde{P}(\partial_\alpha  \tilde{J}_+^\alpha)
-\ [ \tilde{J}_{+\alpha}, P(J_-^\alpha)] - [J_{-\alpha}, \tilde{P}( \tilde{J}_+^\alpha)].
\end{equation}
Finally, we can define
\begin{equation}
L_+^\alpha = \tilde{J}_+^{\alpha(0)}
+ \lambda \sqrt{1+\eta^2} \tilde{J}_+^{\alpha (1)}
+ \lambda^{-2} \frac{1+\eta^2}{1-\eta^2} \tilde{J}_+^{\alpha(2)} + \lambda^{-1} \sqrt{1+\eta^2} \tilde{J}_+^{\alpha(3)},
\end{equation}
\begin{equation}
M_-^\alpha = J_-^{\alpha(0)}
+ \lambda \sqrt{1+\eta^2} J_-^{\alpha (1)}
+ \lambda^{2} \frac{1+\eta^2}{1-\eta^2} J_-^{\alpha(2)} + \lambda^{-1} \sqrt{1+\eta^2} J_-^{\alpha(3)},
\end{equation}
with the spectral parameter $\lambda$. Then, the whole set of
equations of motion (\ref{eomdef}) and the zero curvature equations (\ref{Aflatness}) are equivalent to
\begin{equation}
\partial_\alpha L_+^\alpha-\partial_\alpha M_-^\alpha - 
[M_{-\alpha},L_+^\alpha]=0.
\end{equation}
Introducing $\mL_\alpha=L_{+\alpha}+M_{-\alpha}$, we obtain the standard Lax equation
\begin{equation}
\partial_\alpha \mL_\beta-\partial_\beta \mL_\alpha
 - [\mL_\alpha,\mL_\beta]=0.
\end{equation}
This confirms integrability of the  $\eta$-deformed sigma model. Also, it is worth mentioning that the action (\ref{etaGSAdSP}) is $\kappa$-invariant.

\subsection{Poisson-Lie T-duality}
\label{sec:PL}

Yang-Baxter deformations appear to be a particular example of Poisson-Lie T-duality and in particular can be represented as a non-abelian T-duality with an additional parameter, whose inverse is precisely the deformation parameter \cite{Borsato:2018idb}. While a detailed review of Poisson-Lie T-dualities is not really necessary to define Yang-Baxter equations, this part of the story is still important for the purposes of the present review. The reason is that there are two known ways to arrive to a 3d generalization of the classical Yang-Baxter equation: to use algebraic arguments based on generalizations of the Drinfeld double construction \cite{Malek:2019xrf,Sakatani:2019zrs} or to address deformations from the supergravity side \cite{Bakhmatov:2019dow,Gubarev:2020ydf}. The former are based on generalization of the U-duality symmetry of the membrane to the so-called Nambu-Lie symmetry along the same lines that lead from the ordinary T-duality to Poisson-Lie duality. This approach is restricted to only group manifolds, while for a general 11d background one follows the latter approach that is based on exceptional field theory and eventually again on U-duality, now understood as a local symmetry of a specially extended space. For this reason we find it useful to show the relation between Yang-Baxter deformations and Poisson-Lie T-dualities to further exploit the logic in the 11d case.

Poisson-Lie T-duality transformations has been suggested in 
\cite{Klimcik:1995dy} to answer the question whether an inverse of a non-abelian T-duality transformation can be constructed. The crucial observation here is that the standard (abelian) T-duality defined by Buscher rules preserves the U(1) isometries on which it is constructed. To generalize these dualities one may start with a non-abelian group of symmetries $G$ of a background rather than the abelian group of a torus \cite{delaOssa:1992vci}. These non-abelian T-duality transformation in general break the initial isometries and it is very non-obvious how an inverse transformation can be constructed. To T-dualise backgrounds without isometries in the usual sense in \cite{Klimcik:1995dy} an algebraic approach has been suggested based on the notion of Drinfeld doubles, where the isometry actually exists and is hidden inside the algebraic structure of the double. Let us give more details of the construction focusing primarily on Yang-Baxter deformations inside Drinfeld double. For a review of non-abelian T-dualities and their applications as a solution generating technique see e.g. \cite{Sfetsos:2011jw,Thompson:2019ipl,Thompson:2015lzd}, for more details on Poisson-Lie T-dualities including their realization in double field theory see the e.g. \cite{Petr:2010nzh,Hassler:2017yza,Blumenhagen:2017noc,Demulder:2019bha}. For a recent discussion related to Poisson-Lie and non-abelian T-duality symmetry for the quantum superstring see \cite{Borsato:2020wwk,Hassler:2020tvz}, for a general approach to solution generating techniques see \cite{Borsato:2021gma,Borsato:2021vfy}.

To go beyond T-dualization along isometries defined by conserved charges in \cite{Klimcik:1995dy} a notion of the non-commutative conservation law has been introduced. For a sigma-model model on a group manifold $G$ the non-commutative conservation law is defined as 
\begin{equation}
    \label{eq:nclaw}
    d J_a = \fr12 \tf_a{}^{bc}J_b \wedge J_c,
\end{equation}
where the currents defined by 1-forms $J_a$ correspond to the standard action of the group $G$ on itself. In coordinates the action can be written as  $\d x^i=v_a{}^i \e^a$. Under such coordinate shifts the action of the 2d sigma-model transforms as 
\begin{equation}
    \d S=\int d^2 \s \e^a \mL_{v_a}(E_{ij})\dt x^i 
    \bar{\dt} x^j + \int d \e^a \wedge J_a
\end{equation}
where $E=G+B$. Integrating the last term by parts and assuming proper boundary conditions we see that the action stays invariant under the transformation when holds either the usual conservation law $dJ_a = 0$ or the non-commutative conservation law \eqref{eq:nclaw} together with
\begin{equation}
    \mL_{v_a}(E_{ij})=\tf_a{}^{bc}v_a{}^k v_b{}^l E_{ki}E_{jl}.
\end{equation}
Integrability of this constraint implies the following relation between the quantities $\tf_{a}{}^{bc}$ and structure constants $f_{ab}{}^c$ of the Lie algebra $\frg$ of the isometry group $G$
\begin{equation}
\label{compaftf}
    4 \tf_{[a}{}^{a[c}f_{b]e}{}^{d]}-\tf_{e}{}^{cd}f_{ab}{}^e=0.
\end{equation}
Together with the integrability condition for \eqref{eq:nclaw} that is $\tf_{e}{}^{g[a}\tf_{g}{}^{bc]}=0$ and Jacobi identity for $f_{ab}{}^c$ this has the form of the compatibility condition for the structure of a Lie bi-algebra on  $\frg$. In \cite{Klimcik:1995dy} it has been shown that given the algebras $\frg$ defined by $f_{ab}{}^c$ and $\tgg$ defined by $\tf_a{}^{bc}$ for a Drinfeld double $\mc{D}$ (to be defined below) sigma-models on backgrounds realizing $\frg$ and $\tgg$ are equivalent. Equivalence here is meant in the sense that both sigma models can be obtained translating a $d$-dimensional linear space $\mc{E}=T_e D$ tangent to Drinfeld group $D$ at unity $e$ by either $\exp\frg$ or $\exp\tgg$. In more physical terms: equations of motion are the same.

Let us give more details on the Drinfeld algebra construction however avoiding the categorical language of commutative diagrams since working with explicit backgrounds one always has to chose a specific basis. Hence, let $\{T_{a}\}=bas\,\frg$ and $\{\tT^{a}\}=bas\,\tgg$ with $a=1,\dots,d$, then Drinfeld double can be realized as a Manin triple $(T_a,\tT^a,\h)$, where $\h$ is a non-degenerate quadratic form defined as
\begin{equation}
    \h(T_a,\tT^b) = \d_a{}^b.
\end{equation}
Commutation relations in this basis read
\begin{equation}
    \begin{aligned}{}
        [T_a,T_b] &= f_{ab}{}^c T_c,\\
        [T_a,\tT^b] & = \tf_a{}^{bc}T_c - f_{ac}{}^b \tT^c,\\
        [\tT^a,\tT^b]& = \tf_c{}^{ab}\tT^c.
    \end{aligned}
\end{equation}
To define Poisson-Lie T-dualities in these terms it is convenient to denote the whole basis by $\{T_A\} = \{T_a,\tT^a\}$ and structure constants by $F_{AB}{}^C$, i.e. $[T_A,T_B] = F_{AB}{}^C T_C$. The quadratic form is then given by the invariant tensor $\h_{AB}$ of the O$(d,d)$ group
\begin{equation}
    \h_{AB}= 
        \begin{bmatrix}
            0 & \d_a{}^b \\
            \d_c{}^d & 0
        \end{bmatrix}.
\end{equation}
Poisson-Lie T-duality transformations are then such O$(d,d)$ rotations of the basis 
\begin{equation}
    T'_{A} = C_A{}^B T_B,
\end{equation}
that preserve the Drinfeld double. To construct a geometric realization one takes the so-called geometric subgroup, that is by definition the one generated by $\frg$ and constructs right-invariant 1-form $r = g^{-1}d g$, where $g\in G$. The dual background is then constructed as geometric realization of the transformed geometric subgroup generated by $\frg'$. For a more detailed description of this algorithm see \cite{Malek:2019xrf,Malek:2020hpo,Musaev:2020nrt}. Search for such a matrix $C_A{}^B \in \rmO(d,d)$ preserving a Drinfeld double is the most complicated task in constructing Poisson-Lie T-dual backgrounds. Certain classification results for lower dimensional Lie algebras are available in the literature \cite{Hlavaty:2001fb,Snobl:2002kq,Hlavaty:2002kp}. For special matrices $C_A{}^B$ corresponding to inner automorphisms of the $\rmO(d,d)$ group (factorized T-dualities) one has discrete transformations that are guaranteed to preserve a given Drinfeld double. An example of such transformation is switching $\frg \leftrightarrow \tgg$, that is a different way of saying that a given Drinfeld algebra can be decomposed into two Manin triples. Moreover in \cite{VonUnge:2002xjf} examples of Drinfeld doubles have been found that can be decomposed into more than three Manin triples, that has been called Poisson-Lie T-plurality.

After this long introduction we are finally at the point to define Yang-Baxter deformations in terms of Drinfeld doubles and Poisson-Lie symmetries. Consider a continuous family of deformations of a given Drinfeld algebra $\tf_{a}{}^{bc} = r^{d[b}f_{ad}{}^{c]}$, that corresponds to deformation of the Drinfeld algebra with $\tf_a{}^{bc}=0$ by the following matrix
\begin{equation}
    C_A{}^B=
        \begin{bmatrix}
            \d_a{}^b & r^{ac} \\
            0 & \d_d{}^c
        \end{bmatrix}.
\end{equation}
Such defined dual structure constants $\tf_a{}^{bc}$ satisfy all compatibility conditions if $r^{ab}$ satisfies classical Yang-Baxter equation
\begin{equation}
    r^{e[a}r^{|f|b}f_{ef}{}^{d]}=0.
\end{equation}
In the language of double field theory to be discussed below such matrix corresponds simply to a special case of generalized diffeomorphisms of extended space \cite{Sakamoto:2017cpu,Catal-Ozer:2019tmm}. Since such Yang-Baxter deformation changes the initial Drinfeld algebra it is strictly speaking not a duality in the usual sense of a relation between two different descriptions of the same physics. However, geometric realization of the deformed Drinfeld double solves supergravity equations, given the initial background is a solution and the so-called unimodularity constraint $r^{ab}f_{ab}{}^c=0$ holds, which is the best seen in the formalism of double field theory which we now turn to.

\subsection{Bi-vector deformations of 10d supergravity backgrounds}
\label{sec:bivector}

As it has already been  mentioned the construction above is restricted to group manifolds and coset spaces (in the case of NATD \cite{Borsato:2018idb}). The reason can be seen in the fact that it is very algebraic in its nature and severely relies on the usage of right-invariant forms as target-space vielbeins. A more field theoretic approach to Yang-Baxter deformations has been suggested in \cite{Araujo:2017jap,Araujo:2017jkb} and further developed in \cite{Bakhmatov:2018apn,Bakhmatov:2018bvp,Catal-Ozer:2019tmm,Borsato:2020bqo}.  The approach of \cite{Araujo:2017jap,Araujo:2017jkb} was based on noticing that Yang-Baxter deformations of a background given by metric $G$ can be  represented in the form of the open-closed string map
\begin{equation}
    (G^{-1}+\b)^{-1} = g+b,
\end{equation}
where $g$ and $b$ are the deformed metric and deformed 2-form Kalb-Ramon field, and the deformation parameter $\b = r^{ab}k_a\wedge k_b$ is defined in terms of Killing vectors $k_a=k_a{}^m\dt_m$ of the initial background. Let us note here that the deformation parameter $\b^{mn}$ enters equations very similar to the non-commutativity parameter of Seiberg and Witten \cite{Seiberg:1999vs}. Although the deep meaning of this is not clear, exactly one observes the same in 11 dimensions. There the deformation parameter has 3 indices $\W^{mnk}$ precisely as the membrane non-commutativity parameter, and generalized Yang-Baxter deformation rules have precisely the form of the open-closed membrane map to be discussed in Section \ref{sec:loopnc}. 

Since we are not able to comment more on this very intriguing relation, we prefer to formulate Yang-Baxter deformations in the $\rmO(10,10)$ covariant language, that is ready to generalize to 11 dimensions. This is the language of double field theory, where all supergravity fields depend on a doubled set of coordinates $\{\XX^M\}=\{x^m,\tx_m\}$ subject to the so-called section constraint
\begin{equation}
    \h^{MN}\dt_M \bullet \dt_N \bullet =0,
\end{equation}
where the bullets stand for any of the fields of the theory and their combinations. Basically, the section constraint removes dependence on a half of the coordinates, e.g. on the so-called non-geometric ones $\{\tx_m\}$. In what follows we  will always assume this choice of the section. The idea of doubling of the coordinates follows from the early work by Fradkin and Tseytlin \cite{Fradkin:1984ai} where right and left moving modes of the closed string on a torus are considered independently, hence the alternative reference for $\tx_m$ as winding coordinates. The notion of the section condition and generalized Lie derivative has been introduced in \cite{Siegel:1993th,Siegel:1993xq}. Full formulation of double field theory  has been developed in \cite{Hohm:2010pp} for the NS-NS sector, in \cite{Hohm:2011dv} for the full bosonic field content of supergravity and in 
\cite{Jeon:2012hp} to include supersymmetry. For the purposes of this review double field theory simply provides a convenient choice of parametrization of fields for which Yang-Baxter deformations become a linear $\rmO(10,10)$ transformation. Hence, we will provide only the necessary bits of the formalism and for a more detailed review the reader is referred to \cite{Thompson:2011uw,Hohm:2013bwa,Aldazabal:2013sca}.

In the covariant formalism the metric and the B-field of supergravity are packed into the so-called generalized metric $\mH_{MN} \in \rmO(10,10)/\rmO(1,9)\times \rmO(9,1)$. For our purposes it is more convenient to introduce a generalized vielbein $\mH_{MN}=E_M{}^A E_N{}^B \mH_{AB}$, where $\mH_{AB}$ is a constant unity matrix. The generalized vielbein in the upper triangular form can be defined by exponentiating the space-time vielbein $e_m{}^a$ and $b_{mn}$ with certain generators of $\rmO(10,10)$. For that purpose decompose the generators w.r.t the action of the geometric GL(10) subgroup, i.e. parameterize the generators as follows $\{T_{ab}, T_a{}^b, T^{ab}\} = \mathrm{bas}\,\o(10,10)$. Then the generalized vielbein is defined as
\begin{equation}
    E_M{}^A = \exp \big[e_m{}^a T_a{}^m\big] \exp \big[b_{ab}T^{ab}\big] = 
    \begin{bmatrix}
        e_m{}^a & b_{m k}e_b{}^k \\
        0 & e_b{}^n
    \end{bmatrix}.
\end{equation}
Consider now an $\rmO(10,10)$ transformation of the form
\begin{equation}
    E'{}_M{}^A = O_M{}^N E_N{}^A, \quad O_M{}^N = \exp \big[\b^{mn}T_{mn}\big] =
        \begin{bmatrix}
            \d_m{}^n & 0 \\
            \b^{nk} & \d_l{}^k
        \end{bmatrix}.
\end{equation}
The generalized metric then transforms linearly by conjugations $\mH' = O^{-1}\mH O$ and in terms of the space-time fields $g,b$ the transformation has precisely the form of the open-closed string map. Following \cite{Araujo:2017jap,Araujo:2017jkb} suppose the bi-Killing ansatz for the bivector 
\begin{equation}
    \b^{mn} = r^{ab}k_a{}^m k_b{}^n,
\end{equation}
where $r^{ab}= - r^{ba}$ is a constant matrix and $k_a{}^m$ are Killing vectors of the initial background $g,b$. Now the advantage of the covariant language is that to ensure that such transformed background is still a solution to supergravity equations it is enough to check that the so-called generalized fluxes stay invariant \cite{Borsato:2018idb}. The latter are defined as a generalization of anholonomy coefficients
\begin{equation}{}
    \begin{aligned}
        \mL_{E_A}{E_B} &= \mF_{AB}{}^C E_C,\\
        \mL_{E_A} d & = \mF_A,
    \end{aligned}
\end{equation}
where $\mL_{E_A}$ denote generalized Lie derivative along the vielbein $E_A{}^M$ and $d$ is the invariant dilaton. In general the fluxes $\mF_{AB}{}^C$ and $\mF_A$ are some combinations of the fields $g,b,\phi$  and their derivatives, and become constant in the case of group manifolds. Then these are precisely the structure constants $F_{AB}{}^C$ of the corresponding Drinfeld double. Crucial for the discussion feature of double field theory is that pretty much like general relativity its action and field equations can be written completely in terms of fluxes and their derivatives \cite{Geissbuhler:2013uka}. Hence, if the deformation does not change generalized fluxes, it is a solution generating transformation. This boils down to a condition on the matrix $r^{ab}$ that remarkably is the classical Yang-Baxter equation together with the unimodularity constraint:
\begin{equation}
    \begin{aligned}
        r^{e[a}r^{|f|b}f_{ef}{}^{c]}=0, && r^{ab}f_{ab}{}^c=0.
    \end{aligned}
\end{equation}

To conclude this section recall that from the point of view of sigma-model Yang-Baxter deformations are the ones that preserve integrability. In terms of T-dualities, these act as a special case of the Poisson-Lie symmetry deforming a given Drinfeld double. Beyond group manifolds these act as solution generating transformations preserving generalized fluxes of double field theory. At the moment the latter two of these approaches to deformations have been generalized to 11 dimensions, apparently without reference to integrability of the membrane.

\section{11d supergravity and membranes}
\label{sec:11d}

Classical integrability of a two-dimensional system (say, the fundamental string) implies that its equations of motion can be recast in the form of the Lax-Zakharov-Shabat equation requiring a 1-form $A$, the Lax connection, to be flat. In turn this implies that an evolution operator can be constructed as a Wilson line, that does not depend on the choice of the path. Taking the flatness condition as a starting point one may use $r$-matrix satisfying classical Yang-Baxter equation to generate Poisson brackets of the system. This equation is a quasi-classical limit of the quantum Yang-Baxter equation, whose solution $R$ defines S-matrix of the system. This factorization property of S-matrix means that the theory is integrable. Given an integrable superstring on a background classical $r$-matrix $r\in \frg\wedge \frg$ can be used to define its integrable deformations. For a general case of a solution to 10d supergravity equations such (unimodular) Yang-Baxter deformations  work as solution generating transformations. The algebraic structure behind these symmetries is provided by classical Drinfeld double and Poisson-Lie T-dualities.

In Section \ref{sec:laxtriple} we have seen that when replacing Poisson bracket by Nambu bracket one naturally arrives at an action structurally similar to that of the membrane. Although integrability in three dimensions is not a well defined concept, certain generalization of the structures responsible for integrability in 2d can be made. Consider as before a Lax pair, then a Nambu bracket can be generated by making use of the so-called $\r$-tensor $\r \in \frg\wedge \frg\wedge \frg$, that satisfies certain generalization of the classical Yang-Baxter equation. Such $\r$-tensor can be used to deform the so-called exceptional Drinfeld algebra, that is a generalization of classical Drinfeld double standing behind Nambu-Lie U-duality. The condition for the deformation to preserve the algebraic structure is the same generalized Yang-Baxter equation. Beyond group manifolds one finds generalized Yang-Baxter deformations as transformations generating families of solutions to 11d supergravity equations. Unfortunately, at this moment one is not able to claim that these deformations preserve integrability of the membrane due to the lack of its clear description. Formally following the 2d constructions one faces the absence of the proper ordering of points on a 2d surface generalizing Wilson line. The way out might be in turning to loop algebra variables that seem to be more natural for describing membrane dynamics. In this section we give more detailed review of Nambu-Lie U-duality, deformations of exceptional Drinfeld algebras leading to generalized Yang-Baxter equation and generalization of the construction beyond group manifolds; in more details describe the arguments for Nambu brackets and loop algebraic variables to appear naturally in M-theory; describe the so-called transgression map relating the two; and finally provide some vague considerations following possible definitions of Wilson surface and quasi-classical limits of tetrahedron equations, that seems to be the proper 3d analogue of quantum YB equation.

\subsection{Nambu-Lie U-duality}
\label{sec:NL}

M-theory can be understood as the theory whose weak coupling approximation is the perturbative string theory. This statement is supported by the double dimensional reduction of the membrane when it wraps the compact cycle, that gives the fundamental string. Given that one may think of string theory as of a theory of various membranes on 11-dimensional background space-time, that can be described in terms of Polyakov string at certain points of the moduli space of vacua where the coupling $g_s$ is small. Since in the double dimensional reduction $g_s$ is determined by size of the compact cycle dimension of the background space at these points is effectively 10. A more detailed discussion of M-theory from this point of view can be found in \cite{Townsend:1996xj}. Since $g_s$ does not play the role of a coupling constant in M-theory various branes whose tensions differ by its powers can be mapped into each other by a symmetry that enhances T-duality and includes S-duality. This so-called U-duality (for unity or unified) was first observed in 11d supergravity compactified on a 7-torus in \cite{Cremmer:1978ds}, where the resulting 4d equations of motion have been shown to be invariant under $\rmE_{7(7)}$ transformations. In general compactifying 11d supergravity on a $d$-torus one ends up with a theory invariant under $\rmE_{d(d)}$ symmetry, where $\rmE_{3(3)}=\rmSL(2)\times \rmSL(3)$, $\rmE_{4(4)} = \rmSL(5)$, $\rmE_{5(5)} = \rmSO(5,5)$ and for $d\geq9$ the symmetry algebra becomes infinite and special constructions are needed. Taking into account quantum effects breaks this symmetry to $\rmE_{d(d)}(\ZZ)$ as it has been shown in \cite{Hull:1994ys}. A highly detailed discussion of exceptional symmetries of toroidal compactifications of 11d supergravity can be found in \cite{Cremmer:1997ct,Cremmer:1998px} and of U-duality symmetries of M-theory can be found in \cite{Obers:1998fb}.

Although a construction similar to that of Buscher for the string does not seem to exist for membranes, simply due to the fact that one must simultaneously consider M2 and M5 branes to properly define duality transformations, extension of the standard abelian U-duality symmetry to non-abelian Nambu-Lie U-duality is possible\footnote{Note however the work \cite{Duff:1990hn} where invariance of membrane equations of motion have been shown for the specific case of a 4-dimensional target space where the M5-brane cannot fit. Dual coordinates then correspond to windings of the M2-brane only.}. For that one has to construct a generalization of classical Drinfeld double, called exceptional Drinfeld algebra (EDA), where the $\rmO(d,d)$ symmetry is replaced by one of the U-duality symmetry groups. This has been done in \cite{Sakatani:2019zrs,Malek:2019xrf,Malek:2020hpo,Sakatani:2020iad} and the step-by-step algorithm of constructing a Nambu-Lie dual can be found in \cite{Musaev:2020nrt}. Let us sketch the construction without going into much details. Generators of exceptional Drinfeld algebra belong to $\mathbf{10},\,\mathbf{16}_s,\, \overline{\mathbf{27}}$ for groups $\rmSL(5),\, \rmSO(5,5),\, \rmE_{6(6)}$ respectively, so in general one has for the basis:
\begin{equation}
    T_{A} = \{T_a, T^{a_1a_2}, T^{a_1\dots a_5}\}.
\end{equation}
The generators $\{T_a\}$ form a basis of the so-called geometric subalgebra $\frg$, while the others can be understood as corresponding to windings of the M2- and M5-brane, pretty much like in the classical Drinfeld double case generators $\tT^a$ defined the dual algebra. Algebraic structure is defined by the following multiplication table
\begin{equation}
    T_A \circ T_B = F_{A,B}{}^C T_C,
\end{equation}
where $F_{A,B}{}^C$ are to generalized fluxes of exceptional field theory as structure constants of the classical Drinfeld double are to generalized fluxes of double field theory. More concretely the multiplication table can be represented in the following form
\begin{equation}
\label{eq:eda6}
\begin{aligned}
 T_a \circ T_b =&\ f_{ab}{}^c\,T_c \,,
\\
 T_a \circ T^{b_1b_2} &= f_a{}^{b_1b_2c}\,T_c + 2\,f_{ac}{}^{[b_1}\,T^{b_2]c}
 +3\,Z_a\,T^{b_1b_2}\,,
\\
 T_a \circ T^{b_1\cdots b_5} &= -f_a{}^{b_1\cdots b_5c}\,T_c - 10\,f_{a}{}^{[b_1b_2b_3}\,T^{b_4b_5]} - 5\,f_{ac}{}^{[b_1}\,T^{b_2\cdots b_5]c} 
 +6\,Z_a\,T^{b_1\cdots b_5}\,,
\\
T^{a_1a_2} \circ T_b &= -f_b{}^{a_1a_2c}\,T_c + 3\,f_{[c_1c_2}{}^{[a_1}\,\delta^{a_2]}_{b]}\,T^{c_1c_2} -9\,Z_c\,\delta_b^{[c}\,T^{a_1a_2]}\,,\\
 T^{a_1a_2} \circ T^{b_1b_2} &= -2\, f_c{}^{a_1a_2[b_1}\, T^{b_2]c} - f_{c_1c_2}{}^{[a_1}\,T^{a_2]b_1b_2c_1c_2} +3\,Z_c\,T^{a_1a_2b_1b_2c}\,,
\\
 T^{a_1a_2} \circ T^{b_1\cdots b_5} &= 5\,f_c{}^{a_1a_2[b_1}\, T^{b_2\cdots b_5]c} \,,
\\
 T^{a_1\cdots a_5} \circ T_b &= f_b{}^{a_1\cdots a_5c}\,T_c + 10\,f_b{}^{[a_1a_2a_3}\,T^{a_4a_5]}  + 20\,f_c{}^{[a_1a_2a_3}\,\delta_b^{a_4}\,T^{a_5]c} 
 + 5\,f_{bc}{}^{[a_1}\,T^{a_2\cdots a_5]c} \\
 &\quad+ 10\,f_{c_1c_2}{}^{[a_1}\,\delta^{a_2}_b\,T^{a_3a_4a_5]c_1c_2} 
 -36\,Z_c\,\delta_b^{[c}\,T^{a_1\cdots a_5]}\,,
\\
 T^{a_1\cdots a_5} \circ T^{b_1b_2} &= 2\,f_c{}^{a_1\cdots a_5[b_1}\,T^{b_2]c} - 10\,f_c{}^{[a_1a_2a_3}\, T^{a_4a_5]b_1b_2c}\,,
\\
 T^{a_1\cdots a_5} \circ T^{b_1\cdots b_5} &= -5\,f_c{}^{a_1\cdots a_5[b_1}\, T^{b_2\cdots b_5]c} \,.
\end{aligned}
\end{equation}
In contrast to classical Drinfeld double exceptional Drinfeld algebra is a Leibniz algebra as structure constants $F_{A,B}{}^{C}$ are not antisymmetric in lower indices. Consistency requires quadratic relations on the constants $f_{ab}{}^c, f_a{}^{bcd}, Z_a$ that written in terms of the covariant object $F_{A,B}{}^C$ repeat quadratic constraints of maximal gauged supergravity \cite{Samtleben:2008pe}. Nambu-Lie U-duality is defined as such an $\rmE_{d(d)}$ transformation $T_A \to C_A{}^BT_B$ that preserves the algebra. One immediately notices the absence of the natural duality swapping the geometric algebra $\frg$ spanned by $T_a$ in the chosen basis and its dual spanned by the rest, since dimensions are different and the rest of generators do not form an algebra. However, an analogue of such swapping has been suggested in 
\cite{Musaev:2020bwm} based on outer automorphisms of $\mathfrak{e}_d$, that allowed to generate several examples of non-abelian U-dual backgrounds in \cite{Musaev:2020nrt}. 

We are interested here in deformations of exceptional Drinfeld algebras consistent with their structure and defined in analogy with Yang-Baxter deformations of classical Drinfeld double as follows:
\begin{equation}
    f_a{}^{bcd} = \r^{e[bc}f_{ea}{}^{d]}, \quad f_a{}^{a_1\dots a_6} = \r^{e[a_1\dots a_5}f_{ea}{}^{a_6]}, 
\end{equation}
where $f_{a}{}^{a_1\dots a_6} = \e^{a_1\dots a_6}Z_a$. Such deformations of exceptional Drinfeld algebras have been introduced in \cite{Malek:2019xrf,Sakatani:2019zrs}. In the context of deformations of supergravity backgrounds these have been considered earlier in \cite{Bakhmatov:2019dow} as a generalization of the open-closed string map to the case of 11d supergravity fields. The condition for the deformation to preserve exceptional Drinfeld algebra structure is called generalized Yang-Baxter equations and reads
\begin{equation}
    \label{eq:genCYBE}
    \begin{aligned}
         {\rho}^{a_1 [a_2 |a_6|} \rho^{a_3 a_4 |a_5|} f_{a_5 a_6}{}^{a_7]}- {\rho}^{a_2 [a_1 |a_6|} \rho^{a_3 a_4 |a_5|} f_{a_5 a_6}{}^{a_7]} -3 {f}_{a_5 a_6}\,^{[a_1}\rho^{a_2] a_3 a_4 a_5 a_6 a_7}&=0,\\
         \rho^{a_1 a_2 [a_8} \rho^{a_3 a_4 a_9} \rho^{a_5 a_6 a_7]} f_{a_8 a_9}{}^{a_{10}}- 18 {\rho}^{a_1 a_2 [a_8} {\rho}^{a_3 a_4 a_5 a_6 a_7 a_9]} f_{a_8 a_9}{}^{a_{10}}&=0.
    \end{aligned}
\end{equation}
When restricted to the $\rmSL(5)$ EDA, i.e. four geometric generators, the above condition is precisely the one of \cite{Bakhmatov:2019dow} obtained from the vanishing R-flux condition (to be discussed in the next section). Note that generalized Yang-Baxter equations in the first line above are exactly the conditions  \eqref{eq:genCYBE0} that ensure that the tri-bracket defined for Lax operators in terms of the $\rho$-tensor $\rho^{abc}$ is a Nambu bracket, i.e. satisfies the fundamental identity. To our knowledge in the present context first this  has been observed in \cite{Malek:2020hpo} (Section 4 there) and a candidate for an equation whose quasiclassical limit gives \eqref{eq:genCYBE} has been suggested. Although the details are not completely clear, the suggested quantum generalized Yang-Baxter equation looks very similar to tetrahedron equation in the form of decorated Yang-Baxter equation (see Section \ref{sec:tetra}). If proven, this would be a strong hint towards integrability of the membrane.

\subsection{Polyvector deformations}
\label{sec:polyvector}

As in the case of Poisson-Lie T-duality and Yang-Baxter deformations generalized Yang-Baxter deformations discussed above in the context of exceptional Drinfeld algebras can be extended beyond group manifolds. We will follow here the construction of \cite{Bakhmatov:2020kul,Gubarev:2020ydf} where generalized Yang-Baxter deformations of a general supergravity background with at least three Killing vectors are given by a certain $\rmE_{d(d)}$ transformation. The transformation acts on fields of exceptional field theory, that is a covariant formulation of supergravity with field transforming in irreducible representations of $\mathfrak{e}_{d}$ and in general depend on coordinates $X^M$ parametrizing space-time extended by membrane winding modes. Similarly to double field theory consistency of local symmetries requires section condition, which we will assume to be solved by keeping only geometric coordinates $x^m$.  We will not go into details of the construction for which the interested reader is referred to plenty of detailed literature on the subject \cite{Baguet:2015xha,Baguet:2017rmz,Hohm:2019bba,Musaev:2019zcr,Berman:2020tqn}. 

Since the local symmetry group of exceptional theories changes with dimension $d$ of the so-called internal space entering the split $11=D+d$, explicit expressions for generalized Yang-Baxter deformations also  significantly change. To illustrate the main idea we take the simplest SL(5) theory, whose deformations might be trivial in the sense discussed below, however all main features present. As in double field theory we focus only at the generalized viebein $E_M{}^A$ and the corresponding generalized metric $m_{MN}\in SL(5)/SO(5)\times \RR^+$. Note however, that in contrast to the full double field theory exceptional field theory includes gauge fields, that transform non-trivially under U-duality. To restrict the discussion to the generalized vielbein and a dilaton-like field $\phi$ corresponding to determinant of the external $D$-dimensional metric, a specific truncation must be performed \cite{Bakhmatov:2020kul}. Leaving the details aside we note that generators of SL(5) when decomposed under the action of its GL(4) subgroup follow the same labeling pattern as generators of the SL(5) EDA
\begin{equation}
    \mathrm{bas} \, \mathfrak{sl}(5) = \{T_A\} = \{T_{abc}, T_a{}^b, T^{abc}\}.    
\end{equation}
As before the last 10, i.e. generators of non-negative level w.r.t. the action of the GL(1) subgroup of GL(4), define the generalized veilbein itself
\begin{equation}
    E_M{}^A = e^{\phi T}\exp\big[e_m{}^a T_a{}^m\big]\exp\big[C_{abc}T^{abc}\big],
\end{equation}
where $T$ is the generator of $\RR^+$. Deformation map is then defined by negative level generators and has the folllwing form
\begin{equation}
    \label{eq:defmap}
    E'{}_M{}^A = O_M{}^N E_N{}^A, \quad O_M{}^N = \exp\big[\W^{mnk}T_{mnk}\big] = 
        \begin{bmatrix}
            \d_m{}^n & 0 \\
            \e_{npqr}\W^{pqr} & 1
        \end{bmatrix}.
\end{equation}
As before imposing tri-Killing ansatz 
\begin{equation}
    \W^{mnk} = \r^{abc}k_a{}^m k_b{}^n k_c{}^k,
\end{equation}
where $k_a{}^m$ denote Killing vectors of the initial background, and requiring that the deformed background is a solution to supergravity equation in the exceptional field theory form we arrive at the condition on $\rho^{abc}$ that is precisely the generalized Yang-Baxter equation \eqref{eq:genCYBE} together with the unimodularity constraint
\begin{equation}
    \rho^{abc}f_{ab}{}^d=0.
\end{equation}
To have 6-vector deformations one has to go to a larger symmetry group. The triviality of tri-vector deformations inside the SL(5) theory mentioned above  comes from the fact that to ensure invariance of generalized fluxes, equivalently satisfaction of supergravity equations, the unimodularity condition is enough. This is the same to the O(3,3) double field theory and is related to dimension of the internal space, which renders (generalized) Yang-Baxter equation in the form of the vanishing of R-flux to be equivalent to the unimodularity condition. One however is able to consider Yang-Baxter deformations that are non-unimodular and hence do not solve equations of the standard supergravity, instead leading to their generalization \cite{Bakhmatov:2022rjn,Bakhmatov:2022lin}.

To recap, a generalization of Poisson-Lie T-duality symmetries to an algebra that includes abelian U-duality transformations naturally leads to the notion of exceptional Drinfeld algebra that underlies the Nambu-Lie U-duality symmetry.  As the geometric realization of classical Drinfeld double in terms of generalized vielbein leads to a bi-vector defining Poisson structure, geometric realization of exceptional Drinfeld algebra leads to a 3-vector and a 6-vector defining a Nambu structure. In the context of generalized Yang-Baxter deformations these define a deformed background and the condition for it to satisfy supergravity field equations is precisely the equations \eqref{eq:genCYBE0}. The latter appear as the condition for a 3-bracket of a system defined by Lax pair and the tri-vector to satisfy the fundamental identity. 

Precisely as in the stringy case the map \eqref{eq:defmap} has the form of the open-closed membrane map of \cite{Seiberg:1999vs}. To see that one should start with a background with no $C$-field, then the  deformed background in this language will be precisely the background seen by the open membrane. The tri-vector $\W^{mnk}$ is then one of the generalized theta-parameters in the notations of \cite{Berman:2001rka} and defines open membrane non-commutativity. As we discuss below the non-commutativity relations are naturally written in terms of loop variables, that suggests to do the same for generalized Yang-Baxter equations.

\subsection{Membranes ending on membranes}
\label{sec:basu}

In Section \ref{sec:nambuex} we have seen that the Nahm system can be equivalently described in terms of Poisson and Nambu structures. In the latter case one has to introduce two Hamiltonians, one of which is simply one of the conserved charges in the standard Poisson formulation. For the narrative the Nahm system is of interest due to its tight relation to dynamics of systems of Dp-branes, speaking more concretely: boundary conditions of the D1-D3-brane system can be described in terms of Nahm equations. When uplifted to M-theory this becomes the system of M2 and M5 branes, where the former ends on the latter, and the Nahm equation becomes the so-called Basu-Harvey equation. This procedure has first been considered in \cite{Basu:2004ed}, where a generalization of Nahm equation has been proposed, that naturally involves a tri-bracket, and hence possesses a Nambu structure.

\begin{figure}[http]
    \centering
    \includegraphics[scale=0.4]{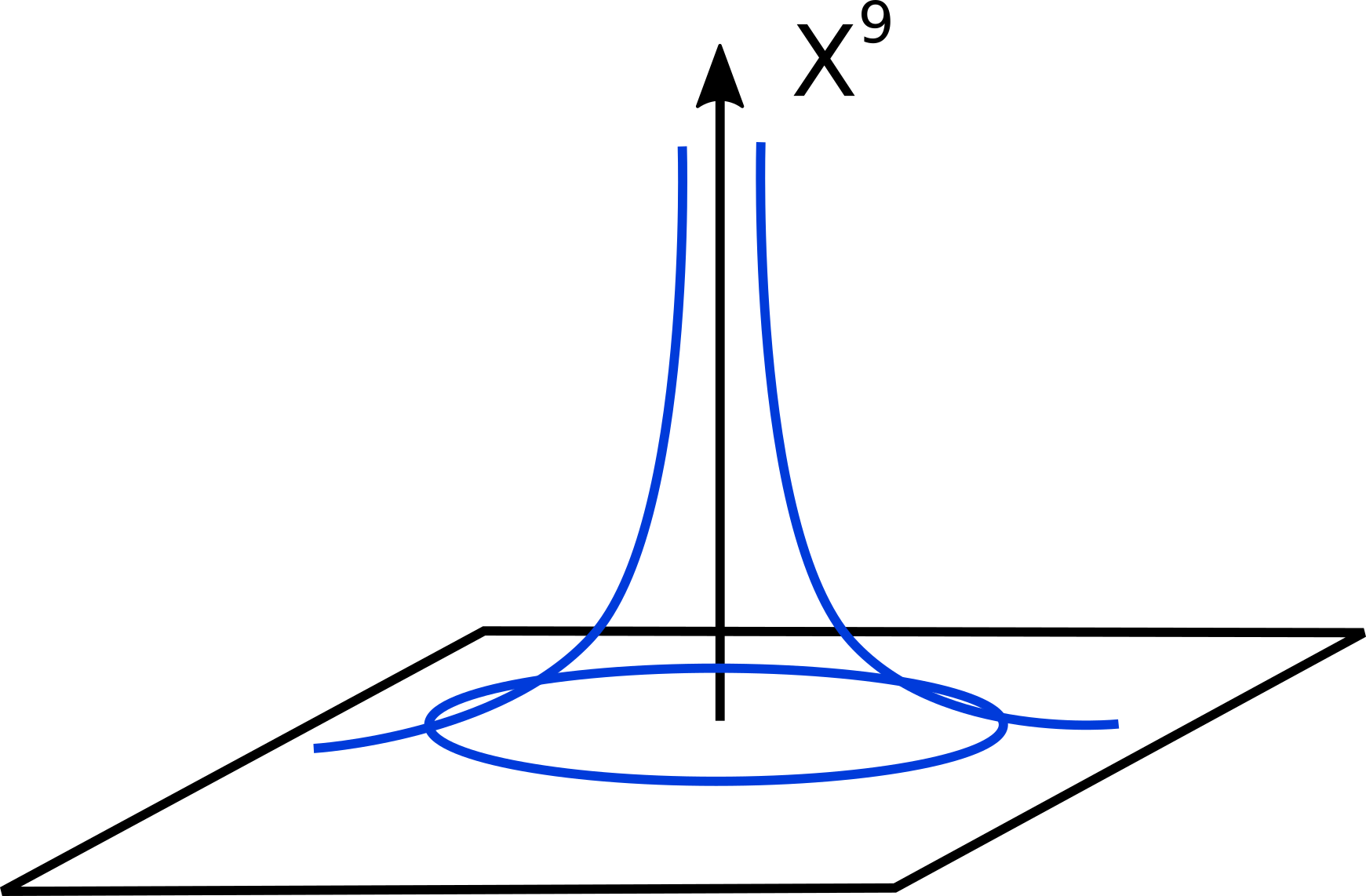}
    \hspace{1cm}
    \includegraphics[scale=0.4]{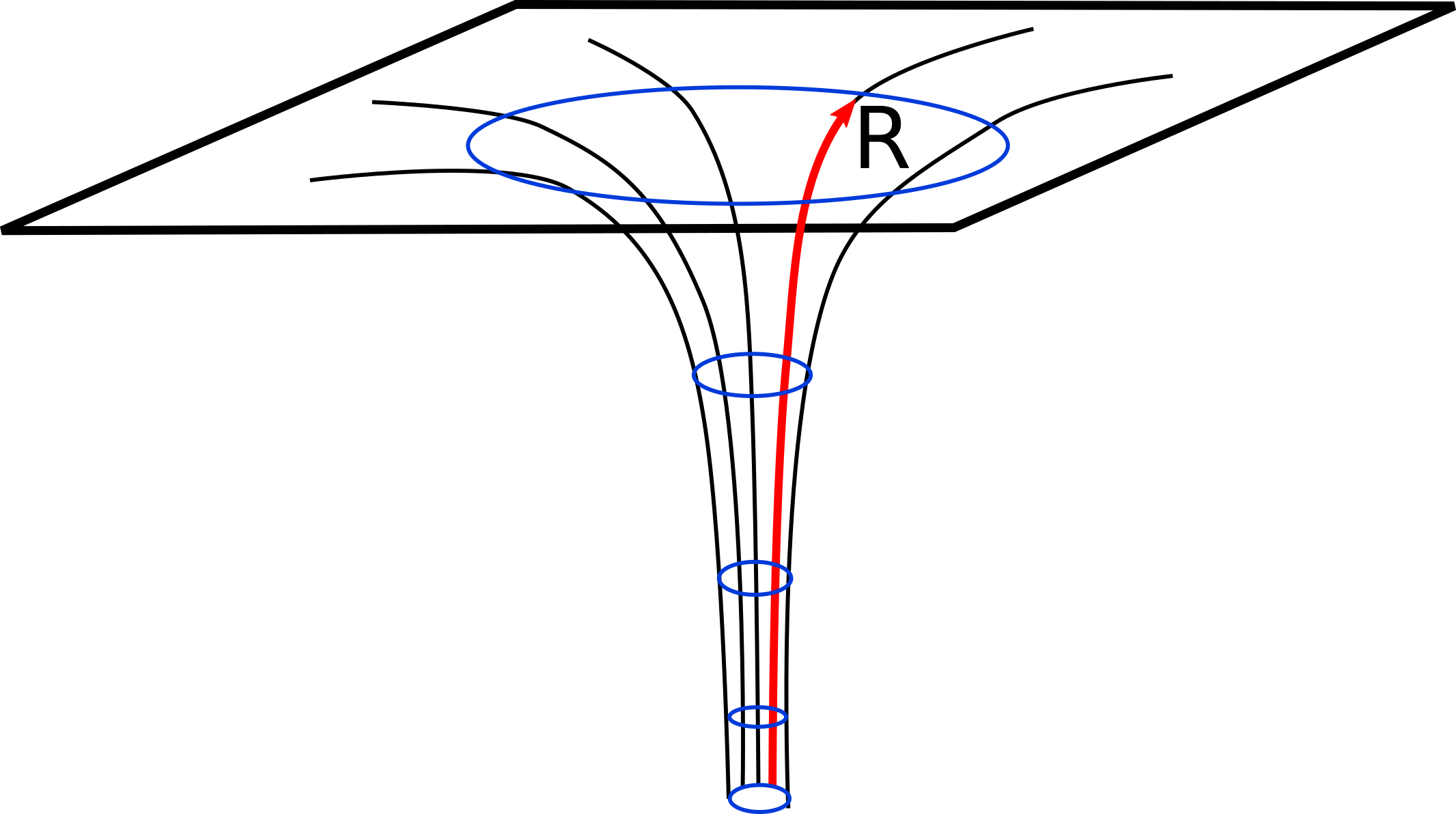}
    \caption{A D1-brane ending on a D3-brane from different points of view. On the left: as a soliton solution of the D3-brane worldvolume theory. On the right: as a throat representing fuzzy sphere geometry around the D1-brane. $X^9$ denotes the coordinate along which the soliton field descends. $R$ denotes physical radius of the fuzzy sphere.}
    \label{fig:DD}
\end{figure}

Let us first look at the D1-D3-brane system. The starting point here is to notice that the brane intersection locus can be equivalently described by i) fuzzy funnel non-commutative geometry interpolating between D1 and D3 brane geometries, ii) geometric engineering of Yang-Mills monopoles on the D3-brane. It is worth to mention, that this is also true for a more general intersection of the Dp and D(p+2) branes. These two pictures basically correspond to considering the intersection from the point of view of the D1-brane and of the D3-brane respectively.

\textbf{D1-brane from the D3-brane point of view.} For the second picture we start with the worldvolume action of an infinitely large D3-brane:
\begin{equation}
    \begin{aligned}
        S^{D3} &= \int d^4 \x e^{-\f}\sqrt{\det \big(g+\mF\big)} + \int C_4 + C_2 \wedge B_2 + \fr12 C_0 \wedge B_2\wedge B_2,\\
        \mF & = dA_1 + B_2, \quad \t = C_0 + i e^{-\f}.
    \end{aligned}
\end{equation}
This action describes a D3-brane that interacts electrically with the fundamental string F1, that can be seen from the $A_1$ field in the determinant coupled to open string ends. To replace F1 by D1 and hence to describe interaction with the D1 brane one performs S-duality:
\begin{equation}
    \begin{aligned}
        \t & \to -\fr1{\t} = -\fr{C_0 - i e^{-\f}}{C_0^2 + e^{-2\f}}, \quad  g_{\m\n} \to |\t|\, g_{\m\n},\\
        B_2 & \to - C_2, \quad A_1 \to -c_1.
    \end{aligned}
\end{equation}
Assuming the background is generated purely by the D-branes, i.e. $B_2 = 0$, we have
\begin{equation}
    S'{}^{D3} = \int d^4 \x e^{-\f}\sqrt{\det \big(g - |\t|^{-1} dc_1 - |\t|^{-1}C_2\big)} + \int C_4 - \fr12 |\t|^{-2}C_0 \wedge \, C_2 \wedge C_2.
\end{equation}
This action describes interaction between D3-brane and D1-brane in the sense, that D1-brane endpoints are charged w.r.t. the world-volume field $c_1$. From the point of view of the world-volume theory D1-branes are seen as spikes of energy, corresponding to Yang-Mills monopoles carrying magnetic charge. Let us choose $(X^4,\dots,X^9) = \vec{X}_{\perp}$ to be transverse directions, and chose spherical coordinates on the brane: $(X^1,X^2,X^3) = (r,\q,\f)$. This theory has classical monopole solution
\begin{equation}
    \begin{aligned}
        X^9 & = \fr{N}{2 r}, && F_{\q \f} = -r^{-2}\dt_r X^9,
    \end{aligned}
\end{equation}
whose charge is given by
\begin{equation}
    Q_m = \fr{1}{2\p}\int d\W F_{\q\f} = N.
\end{equation}
This is interpreted as $N$ D1-branes ending on the D3-brane at the point $r=0$ in the chosen coordinates and stretching along $X^9$. Schematically this is depicted on the left of Fig.\ref{fig:DD}.

\textbf{D3-brane from the D1-brane point of view.} The opposite picture describing D3-brane from the point of view of the D1-brane is slightly more subtle and involves non-abelian Yang-Mills theories. Let us briefly describe the idea here and send the interested reader for details to \cite{Diaconescu:1996rk}. One starts with the description of $N$ D1-branes stretching along the $X^9$ direction in terms of $N\times N$ real matrices $(\XX^1,\dots, \XX^8)$ (see e.g. \cite{Myers:2000in}). The worldvolume gauge field also becomes represented by a matrix $\mathbb{A}_i$ and we fix the gauge choice to be $\mathbb{A}_9=0$. Now, we are looking for a (supersymmetric) solution to equations of the non-abelian Yang-Mills theory with $\XX^A=(\XX^4,\dots,\XX^8)=0$, that corresponds to the position of the D3-brane (see Fig.\ref{fig:DD} right). Equations of motion together with supersymmetry render \cite{Constable:2002yn}:
\begin{equation}
    \fr{\dt \XX^i}{\dt x^9} = \pm \fr i 2 \e^{ijk}[\XX^j,\XX^k],
\end{equation}
where $i,j,k=1,2,3$, these are Nahm equations. The following solution to this system of equations precisely reproduces the monopole profile obtained before in the opposite approach:
\begin{equation}
    \XX^i = \pm \fr{1}{2 x^9} \s^i,
\end{equation}
where $\s^i$ are the standard Pauli matrices. 

The ``coordinates'' $\XX^i$ are used to measure the physical radius of the sphere around the D1-brane on the surface $\S_\perp$ defined by $\XX^A=0$:
\begin{equation}
    R^2 = \fr{2 \p \a'}{N}\Tr\bigg[\sum_{i=1}^3 \XX^i \XX^i\bigg] = \fr{\p \a' (N^2-1)}{(x^9)^2}. 
\end{equation}
We see, that the space near the intersection has the geometry of an infinite throat, that at large $N$ indeed matches the previous result.

The above picture has been uplifted to M-theory in \cite{Basu:2004ed} to describe M2-M5 brane junctions. The overall idea is basically the same: from the point of view of the M5-brane theory boundary of the M2-brane is described by a string-like BPS soliton in the $\mc{N}=(2,0)$  supersymmetric gauge theory in $d=6$. Scalar fields of the theory that correspond to embedding functions of the membrane belong to the supermultiplet that contains a self-dual 3-form, that makes writing a Lagrangian to be a hard task. Equations of motion for fields of the gauge theory have been obtained in \cite{Howe:1996yn,Howe:1997fb} in the so-called superembedding formalism, where the supermanifold describing the M5-brane world-volume is embedded into another supermanifold whose bosonic part is the target space-time. These equations are in the Green-Schwarz form and the string-like soliton solution has been obtained in \cite{Howe:1997ue}. Although we cannot provide a detailed review of the formalism without extending the text well beyond its scope, it is wise to give some more details and sketch the main results following \cite{Basu:2004ed}. First, we note that all equations below are written in the so-called static gauge, where 16 out of 32 fermionic fields are set to zero. In search for classical solutions we set the remaining fermionic fields to be zero as well and write the following bosonic equations:
\begin{equation}
    \begin{aligned}
        G^{\m\n}\nabla_{\m}\nabla_{\n}X^{a'}&=0,\\
        G^{\m\n}\nabla_{\m}H_{\n\r\s} & = 0.
    \end{aligned}
\end{equation}
Conventions for the indices are the following: $\m,\n,\k,\dots = 0,\dots,5$ and $a,b,c,\dots = 0,\dots,5$ are curved and flat world-volume indices; $a',b',c',\dots=1',\dots,5'$ label transverse directions. In what follows  we split $\m=(0,1,m)$ and $a=(0,1,\a)$ with $i=2,3,4,5$ and $\a=2,3,4,5$ labeling directions transverse to the soliton (curved and flat respectively). The covariant derivative $\nabla_{m}=\nabla_{m}[g]$ is constructed on the metric written in terms of the standard world-volume vielbein $g_{\m\n} = e_\m{}^a e_\n{}^b \h_{ab}$. The remaining fields are defined as
\begin{equation}
    \begin{aligned}
        G_{\m\n} & = E_\m{}^a E_\n{}^b \h_{ab}, \\
        E_\m{}^a & = e_\m{}^b (m^{-1})_b{}^a, \\
        m_a{}^b & = \d_a{}^b - 2 h_{a c d}h^{b c d}, \\
        H_{\m\n\r} & = E_\m{}^a E_\n{}^b E_\r{}^c m_b{}^d m_c{}^e h_{a d e},
    \end{aligned}
\end{equation}
where $h_{abc}$ is the world-volume self-dual 3-form. Note, that the field $H_{\m\n\k}$ is not self-dual and moreover it can be written as $H_{\m\n\k} = 3 \dt_{[\m}B_{\n\k]}$. 

Now, we are looking for a string-like solution that lies in the $(0,1)$-plane, for which we introduce the following ansatz
\begin{equation}
    \begin{aligned}
        X^{5'}& = \f, \\
        h_{01\a} & = v_\a, \\
        h_{\a\b\g} & = \e_{\a\b\g\d}v^\d.
    \end{aligned}
\end{equation}
Denoting $H_{01m}=V_m$ one is able to write equations in the following form: 
\begin{equation}
    \begin{aligned}
        \delta^{mn}\dt_m\dt_n \phi&= 0,\\
        \delta^{mn}\nabla_{m}V_n & = 0.
    \end{aligned}
\end{equation}
The solution describing $N$ string-like BPS solitons then becomes
\begin{equation}
    \begin{aligned}
        H_{01m} & = \pm \fr14 \dt_m \phi, \\
        H_{mnp} & = \pm \fr14 \e_{mnpq}\d^{qr}\dt_r \phi, \\
        \phi & = \phi_0 + \sum_{I=0}^{N-1}\fr{2 Q_0}{|x-y_I|^2}.
    \end{aligned}
\end{equation}
Note, that there is no need in a source term, hence the solution is indeed solitonic, and due to the self-duality it possesses both electric and magnetic charges w.r.t. $H_{mnk}$ both equal to $\pm Q_0$. 

The string soliton solution above has a non-trivial profile of the $X^{5'}$ field stretching along $X^1$ and corresponds to M2-brane stretched along $(015')$ directions ending on the M5-brane stretching along $(012345)$ directions. To arrive at a generalization of the Nahm's equation for the M2-M5-brane system we will proceed as before: describe the junction in terms of the fuzzy sphere construction and write a matrix equation, whose solution gives the string soliton profile. For that we need an equation that has SO(4) symmetry rather than the SO(3) symmetry of the Nahm's equation, that is basically the technical reason for the Nambu bracket to appear. The fuzzy 2-sphere describing the D1-D3-brane junction must be generalized to the fuzzy 3-sphere construction, that has been presented in \cite{Guralnik:2000pb}. The space is described by four $N\times N$ matrices $G^i$ where 
\begin{equation}
    N= \fr{(n+1)(n+3)}{2}, \quad n= 2k+1, \quad k \in \mathbb{Z}.
\end{equation}
For $n=1$ these matrices are simply the standard gamma-matrices in four dimensions. The matrices are given explicitly as
\begin{equation}
    \begin{aligned}
        G^i & = \mc{P}_{\mR_+}\sum_{s=1}^N \r_s \big(\G^i P_-\big) \mc{P}_{\mR_-} + \mc{P}_{\mR_-}\sum_{s=1}^N \r_s \big(\G^i P_+\big) \mc{P}_{\mR_+},\\
        \sum_{s=1}{}^n\r_s \big(\G^i\big) & = \Big(\G^i \otimes \dots \otimes 1 + \dots + 1\otimes \dots \otimes \G^i\Big)_{\mathrm{sym}},
    \end{aligned}
\end{equation}
where the ``sym'' subscript denotes complete symmetrization of the tensor product. The projectors $P_\pm = 1/2(1\pm \G^5)$, where $\G^5$ is the standard gamma-matrix. The projectors $\mc{P}_{\mR_\pm}$ project on the irreducible representations
\begin{equation}
    \begin{aligned}
        \mR_+ = \left(\fr{n+1}{4}, \fr{n-1}{4}\right), && \mR_- = \left(\fr{n-1}{4}, \fr{n+1}{4}\right)        
    \end{aligned}
\end{equation}
of the Spin(4)=SU(2)$\times$SU(2) group. Finally, we denote $G_5 = \mc{P}_{\mR_+} + \mc{P}_{\mR_-}$.

Now, taking $\XX^i \in \mathrm{Mat}_{N}(\mathbb{C})$ the proper generalization of the Nahm's equation can be written as
\begin{equation}
    \fr{d \XX^i}{ds} + \fr{\l M_{11}^3}{8 \p} \e_{ijkl}\big[G_5,\XX^j,\XX^k,\XX^l\big] = 0,
\end{equation}
where the Nambu bracket is given by the following sum over permutations
\begin{equation}
    [A_1,A_2,A_3,A_4] = \sum_\s \mathrm{sgn}(\s)A_{\s_1}A_{\s_2}A_{\s_3}A_{\s_4}.
\end{equation}
This equation can be interpreted as the Bogomolnyi equation for the membrane theory. As in the Nahm case its solution can be written in terms of the matrices defining the fuzzy 3-sphere, that in the large N limit takes the following form
\begin{equation}
    \XX^i(s) = \fr{\sqrt{2\p}i}{\sqrt{\l(n+2)s}M_{11}^{3/2}} G^{i}.
\end{equation}
To see the string soliton profile we first introduce the physical radius of the fuzzy 3-sphere
\begin{equation}
    R^2 = \fr{1}{N}\left|\Tr\sum_i (\XX^i)^2\right|.
\end{equation}
Taking $s= X^{5'}$ the above gives the desired result
\begin{equation}
    X^{5'} = \fr{2\pi N}{\l(n+2)M_{11}^{3/2}R^2}.
\end{equation}

Hence, we conclude that the attempt to generalize Nahm's equation describing D1-D3-brane junctions to branes of higher dimensions, Nambu bracket naturally appears and the generalization is commonly referred to as the Basu-Harvey equation. Later in \cite{Bagger:2006sk,Gustavsson:2007vu} based on this observation an action for a stack (of 2) M2-branes has been suggested, that essentially includes Nambu bracket of world-volume fields. Although later in \cite{Aharony:2008ug} an alternative formulation of the world-volume theory (ABJM) has been suggested that requires no Nambu brackets, there seem still be traces of this structure inside. We are referring here to the U(1)${}^3$ deformation of the AdS$_3\times \SS^7$ background first addressed by Lunin and Maldacena in \cite{Lunin:2005jy}, that is holographically dual to the $\b$-deformation of ABJM theory. From the supergravity point of view this corresponds to an SL(2) transformation of the parameter
\begin{equation}
    \t = C_{123} + i \sqrt{G},
\end{equation}
where $(123)$ are three $\SS^1$ directions of the $\SS^7$. Alternatively, this is simply a 3-vector deformation described by generalized Yang-Baxter equation as has been discussed above. We have already seen, that this appears naturally when a Lax triple is introduced for a Nambu system. 

On the ABJM theory side $\b$-deformations of Lunin-Maldacena correspond to introducing certain phase factors for fields entering the superpotential \cite{Imeroni:2008cr}
\begin{equation}
    W \to W_\b = \fr{4 \pi}{ k} \Tr\left(e^{-i\fr{\p\b}{2}}A_1B_1A_2B_2  - e^{\fr{i\p\b}{2}}A_1B_2A_2B_1\right). 
\end{equation}
It is suggestive here to consider a similar deformation of the $\mc{N}=4$ SYM theory that is the U(1)$^2$ $\b$-deformation. On the gravity side this is an abelian bivector deformation along two of three abelian Killing vectors of the dual AdS$_5\times \SS^5$ solution. On the gauge theory side things get much more interesting and the deformation corresponds to deforming product of fields
\begin{equation}
    f\, g \to f * g = \exp\left[i\p \b \left(p^f_1 p^g_2 - p^f_2 p^g_1\right)\right]f g,
\end{equation}
where $p_1$ and $p_2$ denote generators of the two U(1) isometries and the superscript denotes whether the generator acts on $f$ or $g$. Now, if both isometries are along the AdS space we end up with a Moyal product and a non-commutative deformation of SYM \cite{Imeroni:2008cr}, if one isometry is along AdS and one is along the sphere we get the so-called dipole deformations, when both isometries are along the sphere we get the $\b$-deformation of Lunin-Maldacena. In this latter case the generators act simply by multiplication by weight of the operator and the deformed superpotential becomes (see \cite{Imeroni:2008cr} for more details and for the corresponding brane picture)
\begin{equation}
    W \to W_\b = \Tr\left(e^{i\pi \b}\F_1 \F_2 \F_3 - e^{-i\p \b} \F_1 \F_3 \F_2\right). 
\end{equation}
The most intriguing here is that the bi-vector deformation effectively introduces a non-trivial bracket of the operators
\begin{equation}
    [f,g] = \b.
\end{equation}
When both isometries are non-compact, the non-commutative parameter on the RHS becomes literally the bi-vector deformation parameter naturally appearing from double field theory. Now, for the expectional case we have a tri-vector, which presumably must correspond to a non-associativity parameter
\begin{equation}
    [x^m,x^n,x^k] = \Omega^{mnk}
\end{equation}
or to define a Moyal-like tri-product, whose explicit form has not been really established in the literature (see however \cite{Gustavsson:2010nc} for a definition of star-three product in relation to ABJM theory). Precise definition of such tri-product as well as a controllable formulation of a non-associative gauge theory stands among the fascinating directions of research to deepen the understanding of membrane dynamics.

\subsection{Open membranes and loop variables}
\label{sec:loopnc}

Since boundary of an M2-brane ending on an M5-brane is string-like one could expect that natural variables to define world-volume theory on a membrane are those taking values in a loop algebra. To our knowledge the first  mention of loop algebras in the context of membrane dynamics appears  in \cite{Bergshoeff:2000jn} where analysis of the canonical Dirac bracket for membrane world-volume fields has been performed. The observation was that pretty similar to the way how non-commutativity of open string ends appears for D-branes on a constant NS 2-form field background, loop space non-commutativity appears for the case of membranes. The authors define star product of fields $X^i(s)$ with $s$ parametrizing the boundary loop. The fascinating observation relating this approach to polyvector deformations is that the $\W^{mnk}$ tensor parametrizing deformations defines the open membrane metric.

Let us start with a brief reminder of the expressions related to the open-string non-commutati\-vity, for more details the reader is addressed to the original work \cite{Seiberg:1999vs}. Consider the theory of an open string on a background with non-trivial B-field with the standard second order action
\begin{equation}
    S_{F1} = T \int d^2\s (\sqrt{h} g_{mn}h^{\a\b} + b_{mn}\e^{\a\b})\dt_\a X^m \dt_\b X^n, 
\end{equation}
where the integration is taken over the string world-sheet $\S$ with induced metric $h_{\a\b}$ and coordinates $\s^\a = (\t,\s)$. Given the metric and the B-field are constant boundary conditions along a Dp-brane world-volume take the form
\begin{equation}
    g_{mn}\dt_{\vec n} X^n + b_{mn}\dt_t X^n\Big|_{\dt \S}=0,
\end{equation}
where $\dt_{\vec n}$ denotes derivative along a normal vector to $\dt \S$ and $\dt_t$ denotes tangent derivative. One is interested in quantum properties of such two-dimensional field theory with a boundary, in particular in two-point correlation functions $\langle X^m(\t,\s)X^n(\t',\s')\rangle$ that define the propagator of the theory. Restricting the surface $\S$ to be a disk for simplicity and introducing complex coordinates $(z,\bar{z})$ as usual one arrives at the following expression
\begin{equation}
    \begin{aligned}
        \langle X^m(z) X^n(z')\rangle  = &\ g^{mn}\log|z-\bar{z}| - g^{mn}\log|z-z'| \\
         & - G^{mn} \log|z-\bar{z}'|^2 - \Q^{mn}\log \fr{z-\bar{z}'}{\bar{z} - z'} + D^{mn},
    \end{aligned}
\end{equation}
where $D^{mn}$ does not depend on world-volume coordinates. The matrices $G^{mn}$ and $\Q^{mn}$ are defined as symmetric and antisymmetric parts of the matrix $(g+b)^{-1}$ respectively.

Now, the fascinating observation is the following. If both of the points $z$ and $z'$ are inside the world-volume and almost coincident, then the correlator behaves as the usual propagator of a two-dimensional CFT of scalar fields $X^m$. The matrix $g_{mn}$ is then the proper metric for these fields and we refer to it as the closed string metric. For the open string however vertex operators must be inserted on the boundary, i.e. when both $z$ and $z'$ are at the edge of the disk, i.e $z=\t$, $z'=\t'$. Then we have
\begin{equation}
    \label{eq:corr_string}
    \langle X^m(\t)X^n(\t')\rangle = -G^{mn}\log(\t - \t')^2 + \fr i 2 \Q^{mn}\e(\t-\t'),
\end{equation}
where $D^{mn}$ has been set to a convenient value and $\e(\t)$ is the function that is $+1$ for positive argument and $-1$ for negative. We see, that the matrix $G^{mn}$ can now be interpreted as the metric seen by open string ends, since it provides the correct behavior of the propagator for close points. The object $\Q^{mn}$ is the non-commutativity parameter of the open string ends, meaning, that the effective field theory on a Dp-brane on a background with non-vanishing B-field must be described by non-commutative Yang-Mills theory. In \cite{Seiberg:1999vs} it has been shown that it is indeed the case.

The relation between open and closed string parameters can be recast in the following form
\begin{equation}
    (g+b)^{-1} = G+\Q, 
\end{equation}
that is precisely the bi-vector deformation rule, when $\Q$ is understood as the deformation tensor and $G$ as the initial undeformed background. Although it is not completely clear what is the deep reason behind this similarity, it can not be merely a coincidence, as precisely the same match one observes between open-closed membrane relations of \cite{Berman:2001rka} and tri-vector transformation rules. Before turning to the case of the membrane it is suggestive to mention another origin of the open string metric $G^{mn}$, that is the Dp-brane action. Schematically it has the form
\begin{equation}
    S_{Dp} = T_p \int d^{p+1}\x e^{\phi}\sqrt{\det(g+\mF)} + \int \mC_{(p+1)} + \dots,
\end{equation}
where $\mF = dA + b$ with $A=A_\a d\x^\a$ denoting the Born-Infeld world-volume vector field interacting with the open string endpoints, $g$ and $b$ denoting pullbacks of the target space fields, $\mC_{(p+1)}$ denotes the top RR form interacting with the Dp-brane and ellipses denote the remaining terms altogether rendering the action gauge invariant. Varying the action with respect to the scalar fields $X^m$ we obtain equation
\begin{equation}
    \Box[G]X^m + \dots = 0,
\end{equation}
where the box denotes the world-volume d'Alambertian constructed of the pullback of the open string metric $G_{mn}$ and ellipses denote various terms containing only linear derivatives $\dt_\a X^m$. Hence, we see that the open string metric appears as the natural metric for dynamics of the scalar fields, that are nothing but open string excitations transverse to the Dp-brane.

Let us now turn to a three-dimensional sigma-model interacting with target space-time metric $g_{mn}$, a 3-form field $C_{mnk}$, that will be a model of the M2-brane of M-theory. The action of the model can be written as follows
\begin{equation}
    S = \fr{1}{2 l_p^2} \int_\S d^3 \s \sqrt{-h} g_{mn} h^{\a\b}\dt_\a X^m \dt_\b X^n + \int_\S C_{(3)} + \int_{\dt \S} B_{(2)},
\end{equation}
where $\S$ denotes world-volume of the model and $\dt\S$ denotes its space-like boundary. Similarly to open string ends, whose dynamics is effectively described by Dp-brane, boundary of the M2-brane is described in terms of the M5-brane world-volume theory. Since this theory is non-Lagrangian, in the sense that its proper Lagrangian description is not known, the task to write an analogue of the DBI action becomes really tough. On the other hand, the standard CFT methods used above to obtain correlations on the boundary of the 2d disk fail here since the theory is three-dimensional. To circumvent this difficulties in \cite{Bergshoeff:2000ai} an elegant approach has been suggested: i) impose a special decoupling limit to freeze out bulk modes keeping the M5-brane world-volume theory non-degenerate, ii) using primary constraint of the resulting theory to construct a Dirac bracket, that actually encodes loop non-commutativity of the boundary fields. 

To comment on the first step let us first notice following \cite{Bergshoeff:2000ai} that in M-theory there is no sense in which tension of the probe M2-brane is much smaller than that of the background M5-brane. This is in contrast to the open string theory where the small string coupling $g_s \ll 1$ results in large Dp-brane tension $T_{Dp} \sim g_s^{-1}$, while the fundamental string tension $T_{F1}\sim 1$. To prevent the probe M2-brane of deforming the background suppose the latter is generated by a large stack of $N_5$ M5-branes:
\begin{equation}
    ds^2 = H^{-\fr13}dx_{||}^2 + H^{\fr23} dx_{\perp}^2, \quad H=1 + \fr{N_5 l_p^3}{x_\perp^3}, 
\end{equation}
where $\vec{x}_{||}$ and $\vec{x}_\perp$ denote directions parallel and transverse to the M5-branes respectively. The 4-form field strength $F_4$ is proportional to the volume form in the transverse space. Now detach a single M5-brane of the stack and shift it to the position $r_0$ close to the initial stack. If $N_5 \gg 1$ and $r_0$ is small enough interaction between this M5-brane and the remaining stack makes it effectively frozen such that the M2-brane can probe it without deformation. Introducing $\e \to 0$ we can write the decoupling limit as
\begin{equation}
    \begin{aligned}
        l_p & \sim \e l_p, \\
        \fr{N_5}{r_0^3} &\sim \e^{-3}\fr{N_5}{r_0^3},
    \end{aligned}
\end{equation}
such that the product $h l_p^3$ remains finite. Here $h$ enters the self-dual world-volume field strength  $\mH_{\m\n\r}$ as
\begin{equation}
    \begin{aligned}
        \mH_{012} & = - \fr{h}{\sqrt{1+l_p^6 h^2}},\\
        \mH_{345} & = h.
    \end{aligned}
\end{equation}
In this limit the action to quantize becomes simply
\begin{equation}
    S = \fr13 \int_{\dt \S} d^2\s \mH_{\m\n\r} X^\m \dot{X}^\n X'{}^\r,
\end{equation}
where dot and prime denote derivatives w.r.t. the coordinates $(\t,\s)$ on the boundary $\dt \S$. The equal time Dirac brackets between the fields $X^a = \{X^3,X^4,X^5\}$ then becomes
\begin{equation}
    \big[X^a(\t,\s),X^b(\t,\s')\big] = - \fr{1}{h}\fr{\e^{abc}X'{}^c(\s)}{|X^a(\s)|^2}\d(\s-\s').
\end{equation}
We see that the variables $X^a(\s)$ parametrized by the world-volume boundary coordinate $\s$ are indeed non-commutative. The equation above can be understood as commutation relation for loop algebra variables $X^a(\s)$. The same observation will be made in the next section based on the membrane ADHM construction.

Comparing the above commutator to the antisymmetric part of \eqref{eq:corr_string}, we would expect the RHS to be interpreted in terms of the metric seen by the open membrane. That is precisely the case as it can be learned from the work \cite{Berman:2001rka}, where parameters $\Q^{\m_1\dots \m_p}$ have been introduced, called there generalized theta parameters. Instead of copying the relevant expressions from this paper let us illustrate the idea in terms of exceptional field theory parametrization. The open-closed string map $G+\Q = (g+b)^{-1}$ is nothing but two different ways of writing the same $\rmO(d,d)/\rmO(d)\times \rmO(d)$ coset
\begin{equation}
    \begin{bmatrix}
        g- bg^{-1}b & b g^{-1 }\\
        g^{-1} b & g^{-1}
    \end{bmatrix} = \mH =
    \begin{bmatrix}
        G & \Q G \\
        G \Q & G^{-1} - \Q G \Q
    \end{bmatrix}.
\end{equation}
Moreover, decomposing O$(d,d)$ generators under the action of its GL$(d)$ subgroup (upper left and lower right blocks in the matrix notation) as $\{T_{\m\n}, T^\m{}_\n, T^{\m\n}\}$ the matrix $\mH$ can be written as
\begin{equation}
    \mH = \mc{O}^{T}
    \begin{bmatrix}
        G & 0 \\
        0 & G^{-1}
    \end{bmatrix}
    \mc{O}, \quad \mc{O} = \exp\big[\Q^{\m\n}T_{\m\n}\big].
\end{equation}
In other words, adding the non-commutative parameter $\Q^{\m\n}$ can be understood as an O$(d,d)$ transformation. Now, lifting this to exceptional field theory, i.e. replacing the orthogonal group by one of the groups SL(5), SO(5,5) or E${}_{d}$ with $d=6,7,8$ one reproduces precisely the expressions of \cite{Berman:2001rka}. Let us illustrate that by the SL(5) example. Decompose generators under the action of its GL(4) subgroup (upper left block) $\{T_{\m\n\r}, T^\m{}_\n, T^{\m\n\r}\}$ and write
\begin{equation}
    \begin{aligned}
        \mc{U} & = \exp\big[\Q^{\m\n\r}T_{\m\n\r}\big], &&
        \mc{U}^{-1} 
            \begin{bmatrix}
                G & 0 \\
                0 & 1
            \end{bmatrix}
        \mc{U}
        =
        \begin{bmatrix}
            G_{\m\n} & \e_{\m\r_1\r_2\r_3}\Q^{\r_1\r_2\r_3} \\
            \e_{\n\r_1\r_2\r_3}\Q^{\r_1\r_2\r_3} & 1 - \Q_{\r_1\r_2\r_3}\Q^{\r_1\r_2\r_3} 
        \end{bmatrix}.
    \end{aligned}
\end{equation}
The overall prefactor proportional to powers of $1 - \Q_{\r_1\r_2\r_3}\Q^{\r_1\r_2\r_3} $ comes from the proper non-linear realization of the SL(5)/SO(5) coset element in terms of actual space-time metric and the 3-form C-field. 

As we see, three indices of the parameter $\Q^{\m\n\r}$ naturally descent from three space-time dimensions of the M2-brane, suggesting there must be $\Q^{\m_1\dots \m_6}$ accompanying it, that appears to be precisely the case. Reducing this to Type IIA, that is breaking exceptional group w.r.t. its O(d,d) subgroup one generates the parameters $\Q^{\m_1\dots \m_p}$ and all the formulas listed in \cite{Berman:2001rka}. Hence, the ``3-index'' feature of the membrane theory can be either understood as the need of Nambu structure to describe its dynamics in algebraic terms, or loop non-commutativity. The possibility is that these two pictures are completely interchangeable, that stands as an interesting direction of further research.

\subsection{Loop ADHMN construction}
\label{sec:adhmn}

Naturalness of the loop algebra description of membrane dynamics can be seen from the membrane ADHM construction. For that let us return to the Nahm and Basu-Harvey equations and elaborate on results of the works \cite{Gustavsson:2008dy,Saemann:2010cp} (see also the lectures \cite{Samann:2016ksp}), where Basu-Harvey equation has been shown to be a loop-space version of Nahm equation. The approach of the former starts from noticing that $\so(4)=\so(3)\oplus \so(3)$, that allows to understand the fuzzy 3-sphere describing M2-M5-brane junctions as a couple of fuzzy 2-spheres. Construction of the latter work, which we will review in more details below, uses basically the same loop-space variables and is based on the so-called transgression transformation allowing to map a finite dimensional gerbe with a 2-form connection to an infinite dimensional vector bundle with a 1-form connection taking values in loop-space.

To illustrate the construction of \cite{Saemann:2010cp} let us start with the case of the ordinary Nahm equation describing Dirac monopole. In the notations of \cite{Saemann:2010cp} we write
\begin{equation}
    \fr{d}{ds}\XX^i = \fr12 \e_{ijk}[\XX^j,\XX^k],
\end{equation}
where $\XX^i{}^\dagger = - \XX^i$ take values in the algebra $\bu(k)$, hence describing $k$ D1-branes. To construct the Dirac monopole solution define Dirac operator
\begin{equation}
    \slashed{\nabla}_s = - \mathbbm{1} \fr{d}{ds} + \s^i \otimes i\, \XX^i .
\end{equation}
Defining a Laplace operator $\D = \slashed{\nabla}^\dagger \slashed{\nabla}$ we see that the condition $[\D_s, \s^i \otimes \mathbbm{1}]=0$ is equivalent to the condition that $\XX^i$ solves the Nahm equation. Following the standard ADHMN construction \cite{Atiyah:1978ri,Nahm:1979yw} we introduce the following twist of the Diract operator
\begin{equation}
    \slashed{\nabla}_{s,x} = -\mathbbm{1}\fr{d}{ds} + \s^i \otimes \big(i\, \XX^i + x^i \mathbbm{1}\big),
\end{equation}
that preserves the condition $[\D_{s,x}, \s^i \otimes \mathbbm{1}]=0$ on solutions to the Nahm equation. Now orthonormalized zero modes of the twisted Dirac operator
\begin{equation}
    \begin{aligned}
        &\slashed{\nabla}_{s,x}^\dagger \psi_{s,x,\a} = 0 ,\quad \a=1,\dots,N,\\
        &\d_{\a\b} = \int ds \,\psi^\dagger{}_{s,x,\a} \psi_{s,x,\b}
    \end{aligned}
\end{equation}
define gauge and scalar fields of the monopole. Here $x^i$ have the meaning of coordinates in the transverse space and $N$ denotes the total number of D3-branes carrying endpoints of $k$ D1-branes. Gauge potential and the Higgs field then read
\begin{equation}
    \begin{aligned}
        A_i & = \int ds \, \psi^\dagger{}_{s,x} \fr{\dt}{\dt x^i}\psi_{s,x},\\
        \Phi & = - i  \int ds\, \psi^\dagger{}_{s,x} s \,\psi_{s,x},
    \end{aligned}
\end{equation}
 where the indices $\a$ labeling D3-branes are hidden. These fields solve the corresponding Bogomolnyi equations $F_{ij} = \e_{ijk}\dt_k \Phi$, that descent from the higher dimensional Yang-Mills self-duality condition. To be precise, we are working in the setup where all fields of the 10d SYM but $A_i$ and $\Phi^6=\Phi$ vanish.
 
 Let us illustrate this by two most simple examples. Start with $N=k=1$, that corresponds to a single D1-brane ending on a single D3-brane. The D1-brane is stretched along $x^6=s$. In this case the solution to the Nahm equation is $\XX^i=0$, zero modes of the twisted Dirac operator read
\begin{equation}
    \begin{aligned}
        \y_+ = e^{-s R}\fr{\sqrt{R+x^3}}{x^1 - i x^2}
            \begin{bmatrix}
                x^1 - i x^2\\
                R - x^3
            \end{bmatrix}, &&
        \y_- = e^{-s R}\fr{\sqrt{R-x^3}}{x^1 + i x^2}
            \begin{bmatrix}
                R+x^3 \\
                x^1 + i x^2
            \end{bmatrix},
    \end{aligned}
\end{equation}
where $R^2 = x^ix^i$. For the $\y_+$ zero mode one gets the following fields
\begin{equation}
    \begin{aligned}
        \Phi= -\fr{i}{2R}, \quad A_i = \fr{i}{2(x^1 + x^2)^2}
            \begin{bmatrix}
                x^2 \Big(1- \fr{x^3}{R}\Big), & - x^1 \Big(1- \fr{x^3}{R}\Big), & 0 , & 0
            \end{bmatrix},
    \end{aligned}
\end{equation}
that apparently describe a monopole. For the $\y_-$ zero mode one get fields that are related to the above by a gauge transformation everywhere but the points $|x^3| = R$. Similarly for two D1-branes, i.e. when $k=2$, $N=1$, one finds $\XX^i = \fr{i}{2s}\s^i$ and $\Phi = -i/R$, that is precisely the solution we started with in the previous section.

The above approach can be applied to the Basu-Harvey equation and the corresponding Bogomolnyi equation for the self-dual 3-form almost without changes up to the point when a twisting occurs. The equation reads
\begin{equation}
    \fr{d}{ds}\XX^i = \fr{1}{3!}\e^{ijkl}[\XX^j,\XX^k,\XX^l],
\end{equation}
where now $\XX^i$ belongs to a 3-Lie algebra, that is basically a linear space with Nambu bracket. Motivated by T-dualization and further uplift of the D1-D3-brane system to M2-M5-system one writes the following Dirac operator
\begin{equation}
    \slashed{\nabla}_s = - \g_5 \fr{d}{ds} + \fr12 \g^{ij}D(\XX^i,\XX^j),
\end{equation}
where $D(\XX^i,\XX^j) = [\XX^i,\XX^j,\;]$ is the inner derivative and $\g^i$ are the standard Dirac gamma-matrices. For a more detailed discussion on type IIB and type IIA Dirac operators motivating the uplift see the original paper \cite{Saemann:2010cp}. As before the condition that $\XX^i$ satisfies the Basu-Harvey equation can be written in the form $[\D_s,\g^{ij}]=0$, where $\D_s = \slashed{\nabla}_s{}^\dagger\slashed{\nabla}_s$. Next we need to introduce an appropriate twist of such defined Dirac operator, for which we apparently need something of the form $\g^{ij}a_ib_j$, where $a_i \neq \alpha b_i$ in general for some coefficient $\alpha$. Here one uses the fact that vector bundle of a 2-sphere describing Dirac monopole gets replaced by a gerbe over a 3-sphere, that by the transgression map can be understood in terms of loops over $\SS^3$. Hence, we introduce fields $x^i(\t)$ with $\t$ parametrizing the loop, that are restricted by $x^i(\t)x^i(\t) = R^2$. From this it follows $x^i \dot{x}^i=0$ and in addition we impose $\dot{x}^i \dot{x}^i = R^2$. Then the proper twist of the Dirac operator can be written as
\begin{equation}
    \slashed{\nabla}_{s,x(\t)} = -\g_5 \fr{d}{ds} + \g^{ij} \bigg[\fr12 D(\XX^i,\XX^j) - i \, x^i(\t)\dot{x}^j(\t)\bigg].
\end{equation}
We see, that loop space variables naturally enter the twisted Dirac operator, while the construction itself pretty much repeats the conventional ADHMN approach. The next step is to construct gauge and scalar fields, now defined on the loop space, using zero modes of the twisted Dirac operator:
\begin{equation}
    \begin{aligned}
        \Phi(x(\t)) = - i \int ds \psi^\dagger{}_{s,x(\t)} s\, \psi{}_{s,x(\t)}, &&         A_i(x(\t)) =  \int ds \psi^\dagger{}_{s,x(\t)} \dt_i \psi{}_{s,x(\t)}, 
    \end{aligned}
\end{equation}
where the derivative is defined as $\dt_i = \int d\t \fr{\d}{\d x^i(\t)}$. Field strength of the gauge field $A_i(x(\y))$ is then defined as usual as $F_{ij} = 2\dt_{[i}A_{j]}$ and given $\XX^i$ satisfy the Basu-Harvey equation it satisfies 
\begin{equation}
    \label{eq:sd2}
    F_{ij}(x(\t)) = \e_{ijkl} \dot{x}^k \dt_l \Phi(x(\t)).
\end{equation}
The crucial statement here is that while this has schematically the form of the Bogomolnyi equation for SYM theory, its loop structure actually makes it the desired Bogomolnyi equation for the $D=6$ $\mc{N}=(2,0)$ theory. The relation between the self-dual 3-form and such defined 2-form field strength has the following form
\begin{equation}
    F(V_1,V_2) = \int d\t H_{ijk}(x(\t)) \dot{x}^k(\t)V_1{}^i V_2{}^j,
\end{equation}
where $V_{1,2}$ are arbitrary vectors. The equation \eqref{eq:sd2} is then equivalent to self-duality of the 3-form
\begin{equation}
    \begin{aligned}
        H_{05i} = \fr14 \dt_i \Phi, && H_{ijk} = \fr14 \e_{ijkl}\dt_l \Phi.
    \end{aligned}
\end{equation}
As an example let us look at explicit solutions for $N=k=1$, that is a single M2-brane attached to a single M5-brane. Then $\XX^i=0$ and there are eight  zero-modes, of which we will need only four, $\g_5 \y_{s,x(\t)}=\y_{s,x(\t)}$. This is related to doubling of zero modes when going from Pauli matrices to gamma matrices, which in turn is required since SU(2) symmetry of the Nahm equation gets replaced by the SO(4) symmetry of the Basu-Harvey equation. The remaining zero-modes can be arranged as follows:
\begin{equation}
    \y_{s,x(\t)} = e^{-R^2 s}
        \begin{bmatrix}
            i(R^2 + x^2 \dot{x}^1 - x^1 \dot{x}^2 - x^4 \dot{x}^3 + x^3 \dot{x}^4)\\
            x^3 (\dot{x}^1 + i \dot{x}^2  ) + x^4 (\dot{x}^2 - i \dot{x}^1) - (x^1 + x^2)(\dot{x}^3 - i \dot{x}^4)\\
            0\\
            0
        \end{bmatrix}.
\end{equation}
Note the $R^2$ in the power of the exponent, that renders the correct dependence of the scalar field  on the physical distance $\Phi(x) = i/2R^2$. As before for the case $N=1$ $k=2$ we reproduce the previously discussed solution with $\XX^i \propto G^i$.

To summarize, following \cite{Saemann:2010cp} we have observed that turning to fields defined on loop space one is able to apply the standard ADHMN procedure to the Basu-Harvey equation and to describe self-dual string solitons in a way very similar to that of the SYM monopole.  In the process one replaces gerbes, also appearing naturally in membrane theory, by vector bundles, however over a loop space.

\subsection{Speculations on integrability in M-theory}
\label{sec:spec}

In Section \ref{sec:10d} we have briefly reviewed how string theory as a two-dimensional sigma model becomes (classically) integrable on certain backgrounds. This means it is possible to write equations of motion for the string in terms of a Lax connection or to write a quantum Yang-Baxter equation for its S-matrix. At the classical level integrability requires the Lax connection to be flat, that translates to the possibility to define a parallel transport operator, that is basically a Wilson loop calculated on the Lax 1-form. Turning to a theory of two-dimensional membranes and naively generalizing all these structures one would expect to have a two-dimensional analogue of the Wilson loop, which is natural to call a Wilson surface\footnote{Another hint comes from higher gauge theories where Wilson surfaces understood as higher holonomies provide  a set of observables \cite{Ganor:1996nf,Gustavsson:2004gj,Alekseev:2015hda}}. Crucial here is that on the one hand there is no naturally defined ordering on a two-dimensional surface, and on the other hand the 1-form Lax connection should be replaced by a two-form.

Let us first comment on the latter. Overall it is natural to expect a 2-form in the problem since endpoints of a two-dimensional membrane form a string, that naturally interacts with the 2-form. In M-theory the M2-brane ends on an M5-brane, hence the 6-dimensional world-volume theory of the latter is formulated in terms of a 2-form. Supersymmetry requires it to be selfdual, rendering a Lagrangian formulation really hard to construct (for various approaches see \cite{Pasti:1996vs,Bandos:1997ui,Ko:2013dka}). Lifting the notion of a 1-form connection to the 2-form case one naturally ends up with the notion of gerbe connection, that appears when gluing co-cycles close on intersection of four and more charts \cite{Murray:2007ps,Hitchin:1999fh}. Hence, one of the possibility to construct an analogue of the evolution operator is to use a 2-form gerbe connection. 

Although gerbes provide nice geometric background for the problem, it still remains unclear whether there exists a natural ordering on a 2-dimensional surface. One way to parametrise the surface is to swipe it by loops (see e.g. \cite{Alvarez:2009dt}) and hence the Lax connection naturally becomes a 1-form taking values in the loop space. As we have discussed above the idea that loop spaces must be relevant to membrane dynamics is long-standing, and in particular in \cite{Bergshoeff:2000jn} it has been noticed that pretty much like endpoints of an open string become non-commutative, string-like boundaries of the M2-brane become loop-space-non-commutative. Moreover, the metric seen by the open M2-brane boundary is precisely the one, that appears in the exceptional field theory approach to deformations. As we briefly review below the loop-space-non-commutativity of open membrane boundaries naturally appears in the analysis of the Basu-Harvey equations, describing M2-M5 brane junctions.

Finally, let us consider quantum Yang-Baxter equation describing factorization in scattering of point-like particles. Loosely speaking, from the string theory point of view this is related to scattering of endpoints of an open string, that is further motivated by the Wilson loop construction discussed above. Speculating further, one concludes that to describe integrability of a membrane one should be interested in factorization of scattering of strings. Indeed, the corresponding equation has been derived and is known under the names tetrahedron equation, Zamolodchikov equation or Frenkel-Moore equation. We briefly review the progress on relating this structures to 3d integrability and to M-theory below.

\subsubsection{Wilson surfaces and  loop-space connections}
\label{sec:wilson}

Recall, that to discuss integrability of a two-dimensional field theory one introduces Lax connection $A= A_\a d\s^\a$ satisfying flatness condition $F=dA+A\wedge A=0$ and constructs an evolution operator, that is basically a Wilson line. For periodic boundary conditions we write
\begin{equation}
    T = P \exp\Big[\oint_\g A\Big],
\end{equation}
where integration is performed along a line. The flatness condition, that is a different way of writing equations of motion of such a system, then implies
\begin{equation}
    \dot{T} = [T,M],
\end{equation}
where $M=A_0(\s^1=0)$ and possible dependence on spectral parameter is undermined. Then trace of various powers of $T$ give conserved charges. Considering this as a starting point one is able to generate Poisson brackets of the corresponding integrable system by using $r$-matrix $r\in \mathrm{End}(V\otimes V)$, where $V$ represents Hilbert space of the system:
\begin{equation}{}
    \{T_1,T_2\} = [r_{12}, T_1] + [r_{12},T_2].
\end{equation}
Subscripts denote space on which the operator acts, i.e. 
\begin{equation}
    T_1 = T\otimes \mathrm{id}, \quad T_2 = \mathrm{id}\otimes T.
\end{equation}

To generalize these constructions to say a three-dimensional theory one naturally needs a Wilson surface instead of a Wilson line along which a 2-form $B_{\m\n}$ is integrated. While the 1-form $A$ represents a connection on a fibre bundle, the 2-form can be naturally thought of as a connection on a gerbe, which in turn can be mapped to 1-form connections on a loop space \cite{Freund:1981qw,Brylinski:1993ab}. The so-called transgression map has been used in \cite{Papageorgakis:2011xg} to rewrite the $\mc{N}=(2,0)$ theory on the M5-brane formulated in terms of Nambu brackets in \cite{Lambert:2010wm} as a Yang-Mills like theory on a loop space. The map naturally identifies elements of a 3-Lie algebra with elements of an associated Lie algebra (of inner derivatives). This is similar to the construction we have discussed in Section \ref{sec:laxgencybe} where a generalization of the $r$-matrix $\rho \in \mathrm{End}(V\otimes V\otimes V)$ has been used to define Nambu bracket for a system described by Lax pair $\dot{L}=[L,M]$ as
\begin{equation}{}
    \label{eq:namburho}
    \{L_1,L_2,L_3\} = [\r_{123},L_1]+[\r_{123},L_2]+[\r_{123},L_3],
\end{equation}
where the notations are the same as above. The fundamental identity for the 3-bracket is precisely the generalized Yang-Baxter equation \eqref{eq:genCYBE}. Now, on the one hand we have a formulation of the theory on the M5-brane, that is a theory of boundaries of M2-branes ending on it. On the other hand we have a generalization of classical Yang-Baxter equation that presumably describes scattering of straight strings, that could also be understood as boundaries of M2-brane. We will return to the latter point in the Section \ref{sec:tetra}, while now let us describe the construction of \cite{Papageorgakis:2011xg} in more details.

\begin{figure}
    \centering
        \begin{tikzpicture}[overlay]
            \node at (0.9,0.2) (s) {$s$};
            \node at (1.3,-0.1) (t) {$t$};
            \node at (11.7,2) (C) {$C(s)$};
            \node at (8,2) (C) {$C'(s)$};            
        \end{tikzpicture}
        \includegraphics[height=3cm]{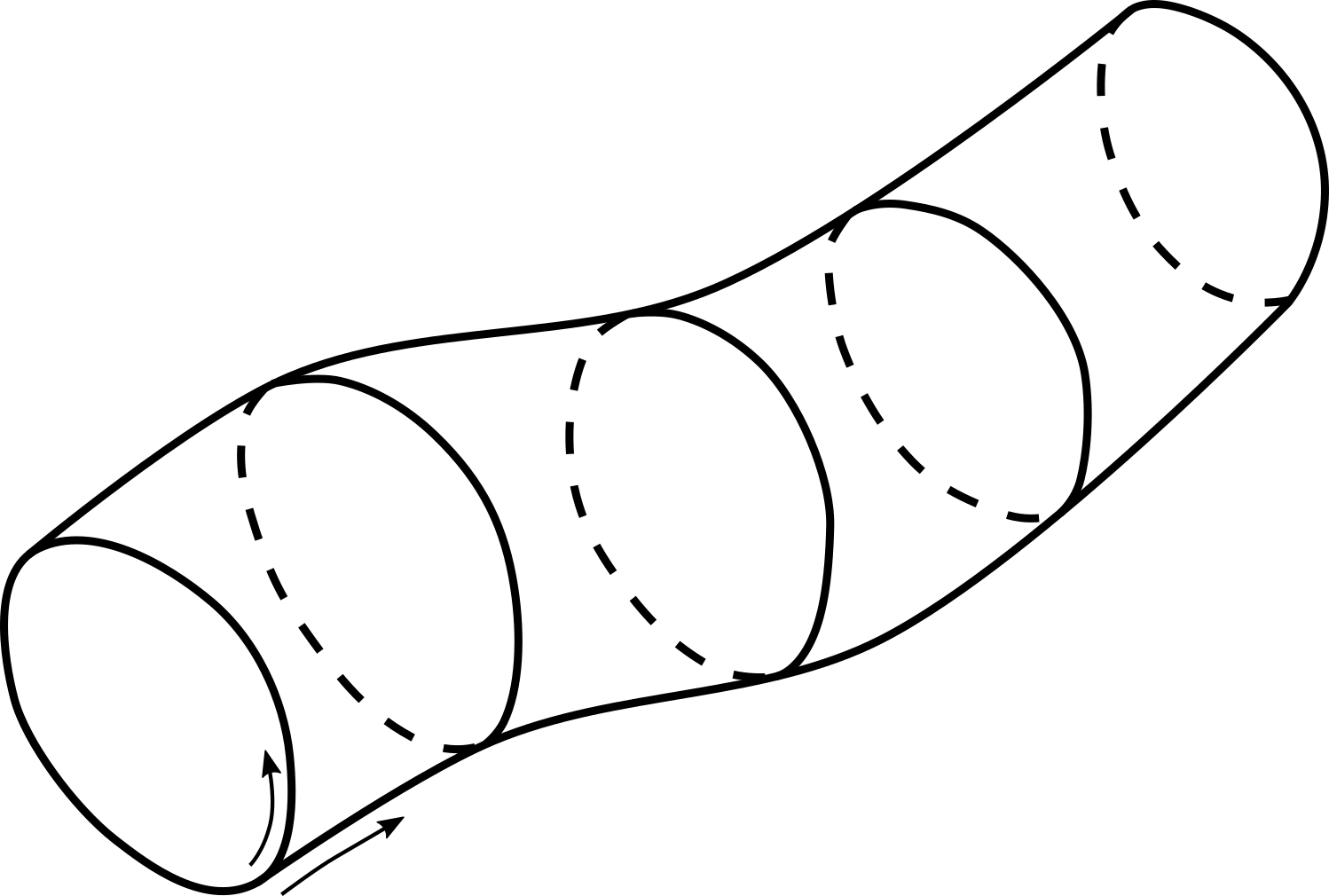}
        \hspace{1cm}
        \includegraphics[height=3cm]{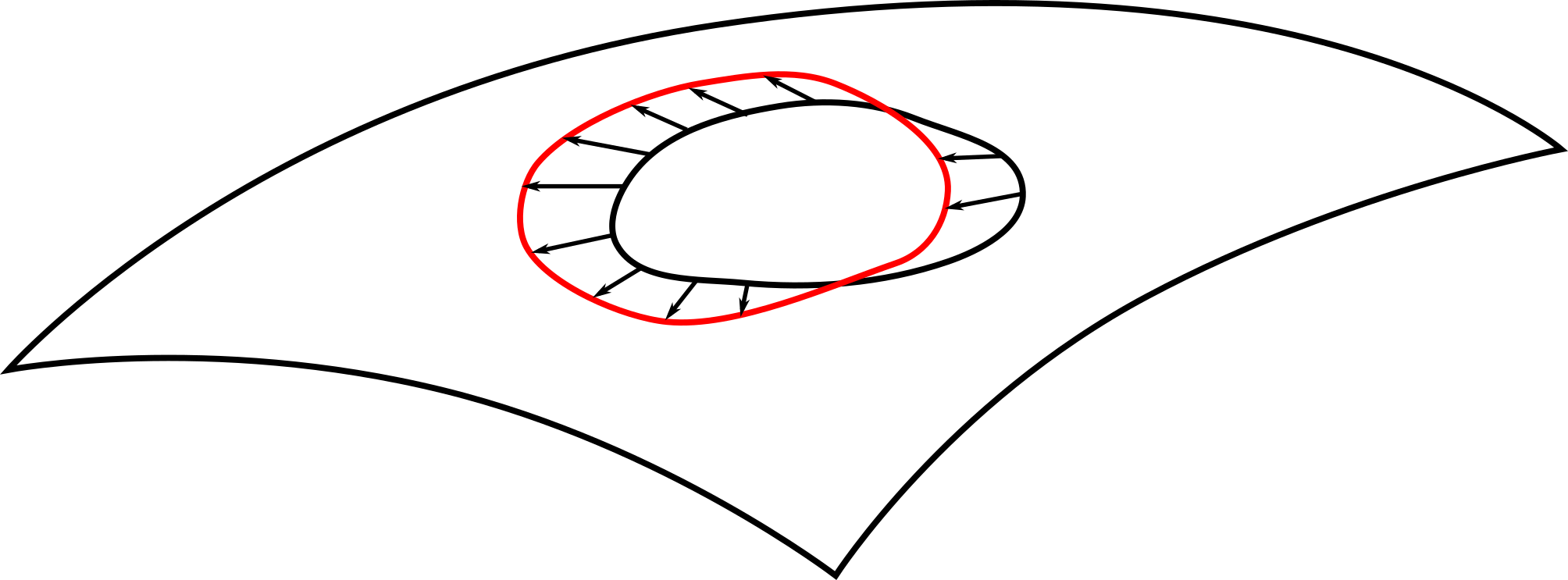}
    \caption{On the left: a path in the space of loops parametrized by a variable $t$. Points along each loop are parametrized by $s\in [0,2\p]$. On the right: a loop $C(s)$ and its deformation $C'(t)$. Vector field tangent to the deformation is depicted by arrows.}
    \label{fig:loops}
\end{figure}
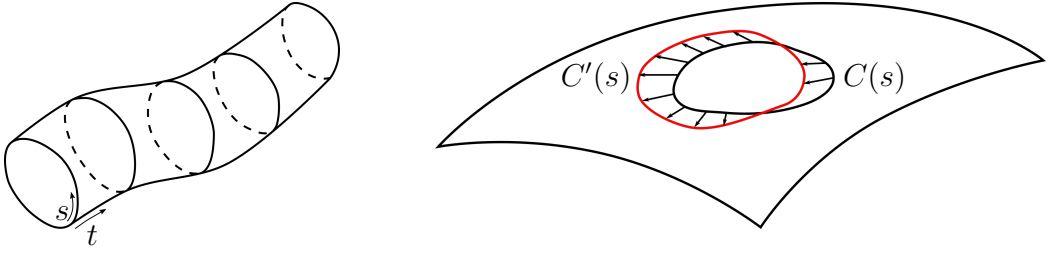

The evolution operator $U = P\exp \int_\g A $ does not depend on the path $\g$ given the connection $A$ on a fibre bundle is flat. When considering a surface integral $\int_\S B$ of a 2-form one faces the problem of absence of a naturally defined ordering of points on a surface. This could be overcome by splitting a cylinder shape surface $\S$ into a collection of loops $C(t)$ parametrized by $t\in[0,1]$ varying along the cylinder (see Fig. \ref{fig:loops}). Hence, instead of a curve on a set of points we consider a curve on a set of loops and instead of the 2-form $B$ we consider a 1-form $\mc{A}$ representing a connection on the loop space. To be more precise, consider a space 
\begin{equation}
    \mL M = \{C:\SS^1 \to M\}
\end{equation}
of all loops on a manifold $M$. In a given patch the curve is defined by coordinate maps $x^\m=x^\m(s)$ with $s\in [0,2\pi]$. 
At each point of the curve one can construct a tangent vector $X^m(s)$, a collection of such tangent vectors for a given curve $C$ we will call a vector tangent to a curve (see Fig. \ref{fig:loops})\footnote{We intentionally keep the discussion more intuitively clear. For more rigor and mathematically formal description of these structures see e.g. \cite{Brylinski:1993ab} and references therein}. Naturally one has a tangent bundle to the space of loops $\mL M$. Denote basis for vector fields as $\d/\d x^\m(s)$ and basis of 1-forms $\d x^\m(s)$, then the usual action of 1-forms reads
\begin{equation}
    \d x^\m(s)\fr{\d}{\d x^\n(s')} = \d^\m{}_\n \d(s-s').
\end{equation}
Following \cite{Brylinski:1993ab,Papageorgakis:2011xg} we construct the transgression map $\mc{T}: \W^{k+1}(M) \to \W^k(\mL M)$ relating $k+1$-forms on the manifold $M$ to $k$-forms on the loop space $\mL M$. In a given basis the map reads
\begin{equation}
    (\mc{T} \w )_C \big(v_1(x),\dots, v_k(x)\big) = \oint_{\SS^1} d s \, \w\big(v_1(s),\dots, v_k(s),\dot{x}(s)\big),
\end{equation}
where $x^\m=x^\m(s)$ represent coordinates on a loop $C$. Hence, on the LHS we have a $k$-form evaluated on $k$ vector fields at a point $C$ of $\mL M$, while on the RHS we have a $k+1$-form evaluated on $k+1$ vector fields at points $x^\m(s)$ and integrated to keep information of the whole loop. As a more explicit example consider the case $k=2$:
\begin{equation}
    \int ds dt (\mc{T} \w)_{\m,s;\n,t} v_1^\m\big(x(s)\big) v_2^\n\big(x(t)\big) = \oint_{\SS^1} dr \,\w_{\m\n\r} v_1^\m\big(x(r)\big)v_2^\n\big(x(r)\big) \fr{d x^\r}{dr}. 
\end{equation}
Note the integrals over $s$ and $t$ on the LHS, that can be understood as an analogue of index contraction when acting by a form on a vector on the loop space, i.e. a form on the loop space has a discrete index $\m,\n,\dots$ and a continuous ``index'' $s,t,\dots$. Inserting Dirac delta-functions on the RHS and dropping two integrals we finally have
\begin{equation}
    (\mc{T} \w)_{\m,s;\n,t} = \int dr\, \w_{\m\n\r}\big(x(r)\big) \fr{d x^\r}{dr} \d(s-r)\d(t-r).
\end{equation}

The next step in relating a 3-Lie algebra variables of the $\mc{N}=(2,0)$ theory on the M5-brane to a Yang-Mills like theory is to construct a Lie algebra out of Nambu brackets. For that we simply consider an associated algebra $\frg_{\mc{A}}$ of inner derivatives, i.e. for any two elements $a,b \in \mc{A}$ of the 3-Lie algebra consider an action
\begin{equation}
    D(a,b)\triangleright x \equiv [a,b,x], \quad x \in \mc{A}.
\end{equation}
The self-dual 3-form $H_{\m\n\r}$ of the theory takes values in the 3-Lie algebra $\mc{A}$ and can be mapped to a 2-form Yang-Mills field strength on the loop space by making use of the transgression map as described above. For that consider a loop $C^m$ taking values in the 3-Lie algebra. Assume, that the algebra and loop variables detach, i.e. 
\begin{equation}
    C^\m(s) = C\, x^\m(s), \quad C \in \mc{A}.
\end{equation}
Formally, this can be ensured by imposing $D(C^\m,C^\n) =0$. The transgression-like map for the self-dual 3-form is then defined as
\begin{equation}
    (F_{\m\n})_C (v_1,v_2) = \oint_{\SS^1}ds \, D\Big(C,H_{\m\n\r}\big(x(s)\big)\dot{x}^\r(s)\Big) v_1^\m(s)v_2^\n(s).
\end{equation}
In the component form action of the 2-form field strength inside the algebra $\frg_{\mc{A}}$ then reads
\begin{equation}
    F_{\m,s;\n,t} \triangleright \bullet= \oint_{\SS^1}dt\, [C, H_{\m\n\r}(r)\dot{x}^\r(r),\bullet] \d(s-r)\d(t-r).
\end{equation}
Given the 3-form is exact $H=dB$ the 2-form $F_{\m\n}$ on the loop space can be represented as a field strength for a 1-form defined naturally as
\begin{equation}
    A_{\m,s} \triangleright \bullet= \oint_{\SS^1}ds\, [C,B_{\m\n}(s)\dot{x}^\n(s),\bullet]\d(s-t).
\end{equation}
Here we are restricted to loops that are covariantly constant w.r.t. such defined 1-form connection:
\begin{equation}
    \begin{aligned}
        0&=\nabla_\m C^\n \equiv \dt_\m \dot{x}^\n(s)\, C + \dot{x}^\n (A_\m \triangleright C)\\
        & = \oint_{\SS^1}ds \fr{\d}{\d x^\m(s)}\dot{x}^\n(s) \,C +\dot{x}^\n (A_\m \triangleright C).
    \end{aligned}
\end{equation}
The integral is apparently zero and vanishing of the remaining term effectively implies
\begin{equation}
    [C,\bullet,C]=0.
\end{equation}
The transgression-like map as above can be extended also to fermionic and scalar fields of the $\mc{N}=(2,0)$ theory, all taking values in the 3-Lie algebra. Hence, the whole formalism including equations of motion and supersymmetry transformation rules gets rewritten in loop variables.

Speculating on these results and results of the previous section one may conclude that loop-space variables are more natural for describing world-volume theory of the M5-brane and hence dynamics of the open M2-brane boundaries. Although the above construction describes the M5-brane world-volume theory, it gives suggestive hints on how integrability structures for the M2-brane can be formulated. To start with, in previous sections we have seen a tight relation between Lax connection for a string on a certain background, classical Yang-Baxter equation for $r$-matrix, and quantum Yang-Baxter equation. The second is simply the quasi-classical limit of the latter, that in turn defines S-matrix of the string on certain backgrounds. Poisson brackets for evolution operators constructed of the former can be defined in terms of the classical $r$-matrix as in the beginning of this subsection. Moreover, classical Yang-Baxter equation appears in relation to bivector deformations given by the open-closed string map as it has been discussed in Section \ref{sec:bivector}. Now we see the same relation between open-closed membrane map and the so-called generalized Yang-Baxter equation, that appears in relation to tri-vector deformations of 11d backgrounds. The generalized theta parameter $\q^{\m\n\r}$ that defines the open-closed membrane map also defines non-commutativity of the membrane on a background with non-vanishing C-field. Moreover, commutation relations are written for loop space variables hence one is talking about loop-space non-commutativity. Similarly to the 2d case generalized Yang-Baxter equation guarantees that the bracket defined using the $\r$-tensor as in \eqref{eq:namburho} is a Nambu bracket, i.e. satisfies the fundamental identity. 

To wrap up the above logic one would like to define an evolution operator using a properly defined Wilson surface or a holonomy in the loop space, that causes the most trouble. A way to define an analogue of the Lax-Zaharov-Shabat construction for higher dimensional theories using loop space variables has been suggested in \cite{Alvarez:1997ma} (for a concise review see \cite{Alvarez:2009dt}). The main idea is to parametrise Wilson surface by a collection of loops, each satisfying a parallel transport equation for a loop space connection $A_\m$. The connection 1-form takes values in a non-abelian algebra and acts non-trivially on loops. Although this formalism is not completely identical to the one described above it is similar enough to be of interest for further investigation.

\subsubsection{Tetrahedron equation}
\label{sec:tetra}

The whole discussion around open-closed string/membrane parameters associated $r$-matrix or $\r$-tensor and Poisson/Nambu brackets has been placed around classical Yang-Baxter equation and its generalized analogue for the $\r$-tensor. As it has been briefly discussed in Section \ref{sec:ads5} integrability of the string means not only the possibility to define a flat Lax connection, but also a possibility to write string S-matrix in terms of the quantum $r$-matrix that solves quantum Yang-Baxter equation (QYBE)
\begin{equation}
    R_{12}R_{23}R_{13}=R_{13}R_{23}R_{12}.
\end{equation}
Here the subscript denotes the Hilbert space on which the $r$-matrix acts at each intersection. Quantum Yang-Baxter equation describes scattering of point-like particles, stating that S-matrix factorizes, and appears to be a particular case of an infinite series of the so-called simplex equations (see e.g. \cite{frenkel1991}). 
\begin{figure}[http]
    \centering
    \includegraphics[height=4cm]{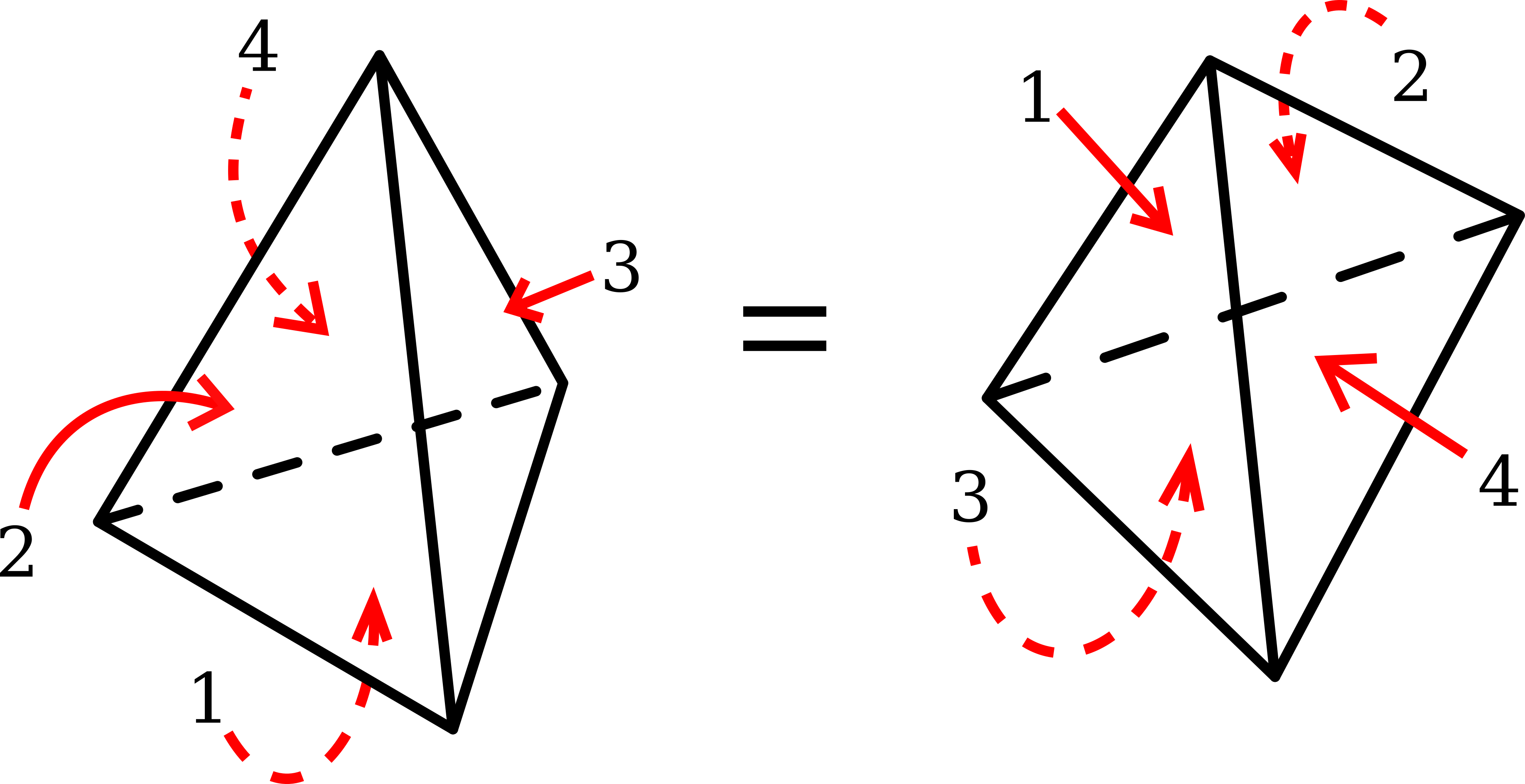}
    \caption{Factorization of scattering of straight strings depicted in the form of tetrahedron equation. Labeling scheme is chosen according to Frenkel and Moore and numbers correspond to faces.}
    \label{fig:tetra}
\end{figure}
An $n$-simplex equation may be understood as describing factorization of S-matrix corresponding to scattering of $n-1$-dimensional objects. Hence, we become naturally interested in $2$-simplex equations, also known as Zamolodchikov tetrahedron  equation (ZTE), introduced in \cite{zamolodchikov1980} and further developed in \cite{zamolodchikov1980z,zamolodchikov1981} as a description of 3d integrable systems. 4-simplex equation appeared in the work \cite{Bazhanov:1981zm}, $n$-simplex equations have been studied for example in \cite{Maillet:1989gg,frenkel1991,carter1996}.

Different labeling schemes can be used to write down the 3-simplex equation: label the states of string segments between vertices \cite{zamolodchikov1980}, label the state of vacua between the strings (faces of tetrahedron) \cite{frenkel1991} or label particles at the edges of the strings \cite{Bazhanov:1981zm}. Let us start with the former and consider scattering of straight strings. $r$-matrix then acts on the space of states $V$ of intersecting strings at each tetrahedron vertex and hence $R\in \mathrm{End}(V\otimes V\otimes V)$. Then the equation reads:
\begin{equation}
R_{123}R_{145}R_{246}R_{356}=R_{356}R_{246}R_{145}R_{123}.
\end{equation}
An alternative labeling scheme uses labels for the tetrahedron faces, which are now four in total and hence the scheme is non-local \cite{Hietarinta:1994ab}. This corresponds to equation on $V^{\otimes 4}$ rather than $V^{\otimes 6}$ and is referred to as Frenkel-Moore equation:
\begin{equation}
\label{FM}
R_{234}R_{134}R_{124}R_{123}=R_{123}R_{124}R_{134}R_{234}.
\end{equation}
Note the important difference to the qYBE and ZTE that indices for each space are contracted more than twice. More details on tetrahedron equation and its relation to other quantum group equations can be found e.g. in  \cite{Talalaev:2021hva,Isaev:2022mrc}.

Given the relations between quasi-classical limit of the quantum Yang-Baxter equation and bi-vector deformations of 10d string backgrounds, it is suggestive to search for similar relations between tetrahedron equation (in any formulation) and the so-called generalized Yang-Baxter equation \eqref{eq:genCYBE}. Unfortunately, this path as not straightforward as it seems since it is not known how to define a quasi-classical limit of tetrahedron equation. One may take the most naive path, for which the Frenkel-Moore equation suits the best and consider $R_{123}=\id\otimes \id \otimes \id+\hbar \, \r_{123}$. Substituting this  into \eqref{FM} we find that the orders $\hbar^0$ and $\hbar^1$ are satisfied trivially, while the order $\hbar^2$ provides
\begin{equation}
\label{cFM}
[\r_{123},\r_{124}]+[\r_{123},\r_{134}]+[\r_{124},\r_{134}]+[\r_{123},\r_{234}]+[\r_{134},\r_{234}]+[\r_{124},\r_{234}]=0.
\end{equation}
This has the form of a nice generalization of the classical Yang-Baxter equation, however, cannot be written in the form \eqref{eq:genCYBE} for a general algebra of endomorphisms. To illustrate that we  decompose $\r$ into a basis $\{t_{a}\}={\rm bas}\,\mathfrak{g}$, say:
\begin{equation}
r_{123}=\r^{abc}\,t_{a}\otimes t_{b}\otimes t_{c}\otimes \mathbf{1}
\end{equation}
assuming $\r^{abc}=\r^{[abc]}$ is completely antisymmetric. Now, it is easy to see, that in each term in the classical 3-simplex equation \eqref{cFM} one obtains expressions of the form 
\begin{equation}
    a\cdot{}b\otimes{}c\cdot d-b\cdot a\otimes d\cdot c,
\end{equation}
where $a\cdot b$ denotes multiplication in the universal enveloping algebra $U(\mathfrak{g})$. This can be transformed into
\begin{equation}
a\cdot{}b\otimes{}c\cdot d-b\cdot a\otimes d\cdot c=[a,b]\otimes \{c,d\}+\{a,b\}\otimes [c,d],
\end{equation}
where $[a,b]$ is the image of the Lie bracket and we formally define $\{a,b\}:=a\cdot b+b\cdot a$. Since we are interested in the algebra of Killing vectors it is not completely clear how to define a symmetric product without introducing a connection. Certainly, this does not prevent from searching solutions for another realizations of the algebra $\frg$, e.g. for Spin$(d)$ the anticommutator is perfectly defined and one may proceed.

Seemingly more fruitful approach is to turn ZTE to the so-called decorated Yang-Baxter equation. For that suppose that the spaces with labels, say $1,2,3$, are considered as additional (color) states. Then two tetrahedra on Fig. \ref{fig:tetra} are simply two triangula with additional lines, decorating them. Introducing labels $\a,\b,\g$ instead of $1,2,3$ we may rewrite ZTE as
\begin{equation}
    R_{\a,45}R_{\b,46}R_{\g,56} = R_{\a\b\g}^{-1}R_{\g,56}R_{\b,46}R_{\a,45}R_{\a\b\g}.
\end{equation}
Hence, we see the familiar structure of quantum Yang-Baxter equation, where i) each $R$-matrix carries additional label (is decorated), ii) the RHS gets twisted by the adjoint action of $R_{\a\b\g}$. It is tempting to think that the additional color label corresponds to having loop-space variables, however, a precise realization of this statement is not clear.

\section{Discussion}
\label{sec:disc}

In this review we have made an attempt to collect methods aimed at investigation of integrability in string theory as a 1+1-dimensional sigma-model, and various observations that give hints on possible generalization of these methods to the theory of membranes. In the main text we briefly discuss  each of the methods and observations in some details to give a general expression of the techniques behind, and provide references to original works and to reviews, lectures and introductory papers. To conclude lets us first recap all these in the form of simple lists. 

Let us start with the list of techniques and observations addressed above that are related to integrability of the string and to symmetries of its space of vacua, i.e. 10d supergravity backgrounds.
\begin{itemize}
    \item A 1+1-dimensional field theory is integrable when its equations of motion can be written in the form of flatness of a connection, the corresponding Wilson line defines Lax operator.
    \item Given a Lax pair the classical $r$-matrix can be used to generate Poisson brackets that define an integrable system.
    \item Integrable deformation of 2d sigma model are generated by $r$-matrix solving classical Yang-Baxter equation.
    \item Classical Yang-Baxter equation is a limit of quantum Yang-Baxter equation describing factorization of S-matrix for scattering of particles.
     \item Yang-Baxter deformations generate families of classical Drinfeld algebras that stand behind Poisson-Lie T-duality symmetries.
    \item At the level of supergravity backgrounds Yang-Baxter deformation are generated by a bi-vector that is the $r$-matrix dressed by Killing vectors.
    \item The deformation map has the same form as the open-closed string map with the bi-vector having the meaning of the non-commutativity parameter.
\end{itemize}
We see here tight connections of classical Yang-Baxter equation to integrability of the string, that is in some sense expected, and to symmetries of the space of solutions of supergravity field equations. Similar connections have been found in 11d supergravity that provides backgrounds for the membranes. The corresponding list of statements can be composed as follows.
\begin{itemize}
    \item Families of exceptional Drinfeld algebras standing behind Nambu-Lie U-duality can be generated by generalized Yang-Baxter deformation defined by $\r$-tensors satisfying generalized Yang-Baxter equation.
    \item At the level of supergravity generalized YB deformations are generated by tri- and six-vectors that are $\rho$-tensors dressed by Killing vectors.
    \item The deformation map has the same form as the open-closed membrane map with the 3-vector having the meaning of the loop non-commutativity parameter.
    \item Given a Lax pair $\r$-tensor can be used to generate Nambu brackets that define a mechanical system.
\end{itemize}

This list intentionally does not mention speculations on integrability. Indeed, there is no construction for a 1+2-dimensional theory similar to Lax-Zakharov-Shabat description of integrability of 1+1-systems. The results collected in the second list above and their similarity to the first list suggest that generalized Yang-Baxter equation must have some relation to integrability properties of the membrane. Certainly, the relation is far from being obvious, however there are some observations concerning algebraic description of membrane dynamics in general and in the context of M-theory that we find particularly useful. They can be combined in the following list.
\begin{itemize}
    \item Quantum version of generalized YB equation is not known, however the index structure of the $\r$-tensor suggests that it must be the tetrahedron equation, that describes factorization of S-matrix for straight strings.
    \item Canonical analysis and ADHMN construction for the membrane suggest that natural variables to describe its dynamics could be functions taking values in the algebra of loops.
    \item Using loops one is able to introduce a natural ordering on Wilson surface presumably defining Lax operator for the corresponding 1+2-dimensional system.
    \item Quasiclassical limit of tetrahedron equation is not known, however it can be written as a set of Yang-Baxter equations on decorated quantum $R$-matrices.
    \item The natural (Takhtajan) action for a system described by a Nambu 3-bracket has the form of the WZ term for the M2-brane ending on an M5-brane.
    \item A generalization of KP hierarchy can be defined using Nambu brackets that has time variables  $t_{m,n}$ parametrized by pairs of indices. Similar variables one finds when defining a generalization of Schur polynomials to the case of 3d Young diagrams presumably describing integrability of 1+2-dimensional systems \cite{Morozov:2022ndt}.
\end{itemize}

The most promising we find the following directions of research. First, one is naturally interested in finding a way to arrive from tetrahedron equation, that naturally describes factorization of S-matrix for straight strings scattering, at generalized Yang-Baxter equation. Second, it would be interesting to construct an analogue of evolution operator in 3d using loop algebra variables, that seem to be natural for describing M2-brane dynamics (or at least dynamics of its endpoints). Hopefully more will be reported on these and other related question in the nearest future.

\section*{Acknowledgments}

This work has been supported by Russian Science Foundation grant RSCF-20-72-10144 and in
part by the Foundation for the Advancement of Theoretical Physics and Mathematics “BASIS”,
grant No 21-1-2-3-1 and by Russian Ministry of Education and Science. We would like to acknowledge insightful discussions with Ilya Bakhmatov, David Berman, Lena Lanina, Nikita Tselousov, Yegor Zenkevich. We are grateful to Riccardo Borsato for his comments on the text.

\bibliography{bib.bib}
\bibliographystyle{utphys.bst}

\end{document}

%% file: Defs.tex
\def\a{\alpha} 
\def\b{\beta} 
\def\g{\gamma} 
\def\d{\delta} 
\def\e{\epsilon}

\def\h{\eta} 

\def\k{\kappa} 
\def\l{\lambda} 
\def\m{\mu}
\def\n{\nu} 
\def\x{\xi} 
\def\p{\pi}
\def\r{\rho}
\def\q{\theta}
\def\s{\sigma} 
\def\t{\tau}  
\def\f{\phi}
\def\y{\psi} 
\def\w{\omega}

\def\G{\Gamma} 
\def\D{\Delta} 

\def\F{\Phi}

\def\Q{\Theta}

\def\S{\Sigma}

\def\W{\Omega}

\def\fr{\frac}  \def\dt{\partial}

\def\mc{\mathcal}

\def\mC{{\mathcal{C}}}

\def\mF{\mathcal{F}}

\def\mH{\mathcal{H}}
\def\mL{\mathcal{L}}
\def\mH{\mathcal{H}}

\def\mR{\mathcal{R}}

\def\tf{\tilde{f}}

\def\tx{\tilde{x}}

\def\tT{\tilde{T}}

\def\XX{\mathbb{X}}
\def\RR{\mathbb{R}}
\def\SS{\mathbb{S}}
\def\ZZ{\mathbb{Z}}

\def\id{\mathbbm{1}}

\def\Tr{\mathrm{Tr}}

\def\frg{\mathfrak{g}}
\def\tgg{\tilde{\mathfrak{g}}}
\def\sl{\mathfrak{sl}}
\def\gl{\mathfrak{gl}}
\def\so{\mathfrak{so}}
\def\o{\mathfrak{o}}
\def\bu{\mathfrak{u}}
\def\su{\mathfrak{su}}

\def\rmSL{\mathrm{SL}}

\def\rmO{\mathrm{O}}
\def\rmSO{\mathrm{SO}}

\def\rmSU{\mathrm{SU}}
\def\rmE{\mathrm{E}}

\def\bas{{\mathrm{bas}}}